\documentclass[aps,floatfix,rmp]{revtex4-1}
\let\it=\textit
\let\rm=\textrm
\usepackage{microtype}
\usepackage{amsmath}
\usepackage{siunitx}
\usepackage{bbm}
\usepackage{fancyhdr}
\fancypagestyle{plain}{ %
  \fancyhf{} % remove everything
}

\def\rck{rqk}

\usepackage{natbib}
\usepackage{breakcites}
\bibpunct{[}{]}{;}{a}{}{;}
\usepackage{url}
\usepackage{epsfig}
\usepackage{amssymb}
\usepackage{times}
\usepackage{bm}

\usepackage{framed}

\def\lvert{\left |}
\def\rvert{\right |}
\newenvironment{myenum}
{\begin{enumerate}
  \setlength{\itemsep}{1pt}
  \setlength{\parskip}{0pt}
  \setlength{\parsep}{0pt}}
{\end{enumerate}}
\let\tilde=\widetilde
\def\bm#1{\boldsymbol{#1}}
\usepackage{braket}
\usepackage{booktabs}
\usepackage{eucal}
\usepackage{soul}
\usepackage{xcolor}
\usepackage{MHequ}
\def\pseudo{\dagger}
\def\T{\rm T}

\def\epsilonb{{\bm \epsilon}}
\def\elastic{\mathcal E}
\def\ks{\kappa_{\rm s}}
\def\kw{\kappa_{\rm w}}

\def\eref#1{Eq.~(\ref{#1})}
\def\fref#1{Fig.~\ref{#1}}
\def\sref#1{Sec.~\ref{#1}}

\def\real{{\mathbb R}}

\let\epsilon=\varepsilon
\let\rho=\varrho
\def\HALF{{\textstyle\frac{1}{2}}}
\def\eg{{\it{e.g.,~}}}
\def\ie{{\it{i.e.,~}}}
\def\thecomma{\ifx,\thenext \else\ifx;\thenext \else\ifx.\thenext
\else\ifx!\thenext \else\ifx:\thenext\else\ifx)\thenext \else \
\fi\fi\fi\fi\fi\fi}
\def\condblank{\futurelet\thenext\thecomma}

\def\lams{{\lambda_\ast}}
\def\ubs{{\ub_{\ast}}}
\def\ubsT{{\ub_{\ast}^\T}}

\def\Bb{{\mathbf{B}}}
\def\cb{{\mathbf{c}}}

\def\sb{{\mathbf{s}}}
\def\Db{{\mathbf{D}}}
\def\cbs{{\cb_{\ast}}}

\def\id{{\mathbf{1}}}

\def\Kb{{\mathbf{K}}}

\def\nb{{\mathbf{n}}}
\def\fb{{\mathbf{f}}}
\def\dfb{\mathbf{\delta{f}}}

\def\rb{{\mathbf{r}}}

\def\vb{{\mathbf{v}}}

\def\ub{{\mathbf{u}}}
\def\xb{{\mathbf{x}}}
\def\rb{{\mathbf{r}}}
\def\dub{\mathbf{\delta{u}}}

\def\vbT{\vb^\T}

\def\CC{{\mathbf{C}}}
\def\DD{{\mathbf{D}}}
\def\UU{{\mathbf{U}}}
\def\SS{{\mathbf{S}}}
\def\Ck{{\CC_k}}
\def\Uk{{\UU_k}}
\def\Sk{{\SS_k}}
\def\MM{{\mathbf{M}}}
\def\AAA{{\mathbf{A}}}
\def\BB{{\mathbf{B}}}
\def\QQ{{\mathbf{Q}}}
\def\EE{E}

\def\epij{{e_{i,j}}}
\def\epijk{{e_{i,j,k}}}
\def\HH{\mathbf{H}}
\def\DH{\delta{\mathbf{H}}}
\def\DHi{\delta{\mathbf{H}_i}}
\def\DHj{\delta{\mathbf{H}_j}}
\def\DHij{\delta{\mathbf{H}_{i,j}}}
\def\DDHij{\delta^2{\mathbf{H}_{i,j}}}

\def\GG{\mathbf{G}}
\def\DG{\delta{\mathbf{G}}}
\def\DGi{\delta{\mathbf{G}_i}}
\def\DGj{\delta{\mathbf{G}_j}}
\def\DGij{\delta{\mathbf{G}_{i,j}}}
\def\DDGij{\delta^2{\mathbf{G}_{i,j}}}
\def\PP{\mathbf{P}}
\def\Gs{{\GG^{\prime}}}
\def\DF{\delta{F}}
\def\DFi{\delta{F_{i}}}
\def\DFj{\delta{F_{j}}}
\def\DFk{\delta{F_{k}}}
\def\DFij{\delta{F_{i,j}}}
\def\DFjk{\delta{F_{j,k}}}
\def\DFik{\delta{F_{i,k}}}
\def\DFijk{\delta{F_{i,j,k}}}

\def\ui{{p}}
\def\uj{{q}}

\def\aH{{\rm H}}
\def\aP{{\rm P}}

\def\Tr{\textrm{Tr}}

\begin{document}

\title[Physical Aspects of Proteins]{Proteins: the physics of amorphous
evolving matter}

\author{Jean-Pierre Eckmann $^{1,2}$, Jacques Rougemont$^{1}$, Tsvi Tlusty $^{3,4}$} 
\affiliation{ $^1$ D\'{e}partement de Physique Th\'{e}orique,Universit\'{e} de Gen\`{e}ve, CH-1211, Geneva 4, Switzerland}
\affiliation{$^2$ Section de Math\'{e}matiques, Universit\'{e} de Gen\`{e}ve, CH-1211, Geneva 4, Switzerland}
\affiliation{$^3$ Center for Soft and Living Matter, Institute for Basic Science (IBS), Ulsan 44919, Korea}
\affiliation{$^4$ Department of Physics, Ulsan National Institute of Science and Technology (UNIST), Ulsan 44919, Korea}

%\ead{tsvitlusty@gmail.com}
\begin{abstract}
Proteins are a matter of dual nature. As a physical object, a protein
molecule is a folded chain of amino acids with multifarious
biochemistry. But it is also an instantiation along an evolutionary
trajectory determined by the function performed by the protein within a hierarchy of
interwoven interaction networks of the cell, the organism and the
population. A physical theory of proteins therefore needs to unify both
aspects, the biophysical and the evolutionary. Specifically, it should
provide a model of how the DNA gene is mapped into the functional
phenotype of the protein.

We review several physical approaches to the
protein problem, focusing on a mechanical framework which treats
proteins as evolvable condensed matter: Mutations introduce localized
perturbations in the gene, which are translated to localized
perturbations in the protein matter. A natural tool to examine how
mutations shape the phenotype are Green's functions. They map the
evolutionary linkage among mutations in the gene (termed epistasis) to
cooperative physical interactions among the amino acids in the
protein. We discuss how the mechanistic view can be applied to examine
basic questions of protein evolution and design.  
\end{abstract}
%\newpage
%\section{Contents}

\maketitle
Sections marked with $^*$ contain more technical material and can be
omitted at first reading.

\tableofcontents
%\newpage

\section{The protein problem: a theoretical physics perspective}
\label{sec:proteinproblem}

The macromolecules that make living matter -- lipids, hydrocarbons, nucleic acids, 
and in particular proteins -- are among the most studied objects of Nature. 
Proteins comprise the central nano-machinery of the cell, whose numerous functions include 
the formation of structural elements, catalyzing metabolic reactions and conveying biochemical signals \cite{alberts1998cell,Fersht1999,Howard2001,Goodsell2009,Whitford2013a}. 
For their significance in life, proteins and the genes that encode them 
have been extensively investigated using various experimental  methods, 
such as crystallography, biochemical assays, mass spectrometry, fluorescence imaging, 
electron microscopy, directed evolution and deep sequencing \cite{Rambo2013,Cohen2001,Collins2011,Mandala2018,Barrera2011,Mehmood2015,Ha2012,Mardis2013,Chapman2011,Fernandez-Leiro2016}.
In parallel, sophisticated computational models, such as molecular dynamics, 
have been developed to predict the structure, function and folding of proteins \cite{Karplus2002,Karplus2005,Adcock2006,Dror2018,Scheraga2018,Isralewitz2001}. 

These experiments and simulations provide valuable data on protein 
structure, dynamics and genetics. However, there remain two inherent challenges: 
(i) Sparsity of data -- the protein is the outcome of long evolutionary search 
in a high-dimensional space of gene sequences, which is impossible to sample, 
even by high-throughput experiments.
(ii) Complexity of interactions -- The function of a protein arises from collective
many-body interactions in the heterogeneous amino acid matter, which
are hard to probe and model.  

In light of these challenges, we focus in this colloquium on a complementary 
theoretical approach that links the protein problem to the realm of condensed matter physics.
Rather than using realistic simulations predicting the dynamics and function
of concrete proteins,  
we shall discuss minimal models that allow, under several simplifying assumptions, 
to examine basic questions of protein evolution, especially how the collective 
physical interactions within the protein direct its
evolution.\footnote{The main body of this colloquium is based on, and 
  expands, ideas from papers
  \cite{Tlusty2007,Eckmann2008,Tlusty2008a,Tlusty2010,Tlusty2016,Tlusty2017,Dutta2018}.} 

The structure of many proteins is known at a resolution of a few angstroms and there are
detailed computational models of the forces between the amino acids. 
Here, however, the protein will be examined at a coarse grained level, 
in the spirit of lattice \cite{Lau1989,Shakhnovich1991} and network \cite{Chennubhotla2005}
models. The protein will be described as a connected network whose nodes represent 
the amino acids. Furthermore, from this conceptual point of view, it suffices to assume 
that there are only two types of amino acids instead of the usual twenty.\footnote{21, when counting the rare pyrrolysine \cite{Hao2002,Srinivasan2002}.} 
(for example the classical HP model \cite{Lau1989} used in
\sref{sec:pnasmodel}, in which amino acids can only be either hydrophobic ($\aH$)
or polar ($\aP$)).
In a real protein, forces between the amino acids are a complicated combination
depending for example on their polarity, hydrophobicity, charge and shape. 
At the coarse grained level, it again suffices to consider instead 
just simple springs between pairs of neighboring sites. This is akin
to using harmonic approximations in mechanics, which provide a generic
understanding and a good physical insight.

With this kind of simplifications, one can translate certain questions of biology
to analogous questions in the physics of amorphous networks. Among the rich set of methods
in this classical subject of physics, some tools seem particularly well adapted 
to the protein problem. The approach is based on the dual nature of the protein; 
it is a physical object whose formation and physical interactions are also represented
in the `dual' gene, a sequence of symbols from a four-letter alphabet of the DNA bases, 
`A', `C', `G', `T'. Evolution progresses by introducing mutations, 
that is, permanent modifications of this sequence. 
There are local mutations (nucleotide substitutions, short insertions and deletions) 
besides larger scale modifications (\eg translocations, inversions, duplications). 
A natural approach to study protein evolution is to model 
the effect of mutations on the physical properties of the amino acids network.

Local mutations amount to short jumps between neighboring sequences in
the genotype space, differing by one letter only, while large-scale
mutations are equivalent to longer jumps.  
Both classes of mutations can be described in terms of alterations of
the mechanical properties of the amino acid network. However, we shall
focus on the class of local mutations.  
Practically, local mutations are easy to treat with classical techniques of condensed matter,
for instance via Green's functions, since they induce localized perturbations 
in the spring network. More importantly, it is possible to statistically sample
the genotype space with continuous trajectories progressing by consecutive local mutations. 
This will be the main axis of this colloquium. 
Along the evolutionary trajectory,  mutations come in three flavors: 
The ones leading to some sort of functional catastrophe or significant disadvantage, 
and therefore get eliminated by selection; others which improve the
properties of a protein and finally, the large `neutral' majority
which do not induce any significant change in the function of the protein  \cite{Kimura1983,Neher2011}.
In this manner, the `learning' evolutionary process reduces 
the problem of improving a protein from an exhaustive combinatorial search approach 
into a biased random walk. This drastically reduces the dimension of the space which one needs to explore. 

The condensed-matter approach to the protein problem may be viewed 
as an example of a potentially general framework that may be used 
to examine other strongly-coupled biological systems. For example, 
one may analyze metabolic and genetic networks in terms 
of localized perturbations and Green's functions. 
Such analysis may suggest common underlying principles.
It might as well turn out that biology is more contingent and depends on the history of the evolutionary process, but at least the few examples we describe give us hope that a rational approach, based on the laws of physics, may be useful in some cases.  

Biological molecules are far from being the spring networks we use as a model.
Still, similar abstractions proved successful in many areas of physics. 
For example, the dynamical systems of the so-called Axiom A class \cite{Eckmann1985}
are systems with a very special, yet simple, structure. 
And although most systems do not belong to the Axiom A class,
it proved very useful to consider that they behave `as if'. 

There is a long history of studies in similar spirit of abstraction 
and simplification, starting with the conformal maps of D'Arcy Thompson \cite{Thompson1942}, 
through the morphogenetic studies based on the theory of catastrophes by Ren\'e Thom \cite{Thom2018}.
In the 21st century  researchers have much more 
data available on biological systems. This allows to test hypotheses 
against measurements, infer from the data other questions to investigate, 
and suggest possible experiments to confirm or refute the theory. 
We close the introduction by two citations which reflect the general outlook of this colloquium:

\textbf{Misha Gromov}, in \cite[Abstract]{Gromov2013}  
\begin{quotation}
  When you read a textbook on molecular/cellular biology you are enchanted 
  by the logical beauty of biological structures. You want to share 
  your excitement with your colleagues, but\dots you find out you are
  unable to do it: there is no language in the 21st century mathematics
  that can express this beauty. You feel there must be a new world of
  mathematical structures shadowing what we see in Life, a new language
  we do not know yet, something in the spirit the `language' of calculus
  we use when describing physical systems.
\end{quotation}

\textbf{Giovanni Jona-Lasinio}, in \cite{Jonalasinio2012}: 
% Original:
% \begin{quotation}
%   La fisica teorica e' stata
%   riconosciuta come disciplina autonoma solo alla fine del
%   diciannovesimo secolo poco prima delle grandi svolte concettuali della
%   relativita' e della meccanica quantistica e le sue forme oggi sono
%   molteplici. Penso che i tempi siano maturi per una piu' precisa
%   caratterizzazione della ricerca teorica in biologia e dei suoi scopi.
% \end{quotation}

\begin{quotation}
  Theoretical physics was recognized as an independent field of research 
  only at the end of the 19th century, shortly before the great conceptual 
  revolutions of relativity and quantum mechanics. 
  Today theoretical physics has multiple facets. I think that
  the time has come for a more precise characterization of the research
  field of theoretical biology, and for an assessment of its scope.
  [Translated from Italian]
  \end{quotation}
%%%%%%%%%%%%%%%%%%%%%%%%%%%%

We are convinced that such outlooks are important and our work
should be viewed as an attempt in this general direction, in the hope
that readers will be encouraged to proceed along this path.

\section{Biology as a challenge to theorists}
\label{sec:challenge}

Biological research has been extremely active in the past decades and 
experimental results have flourished to vastly improve our understanding
of living matter. 
The challenge for theorists is to find subtopics which are at a stage
where theoretical abstraction can be fruitful.

Here we focus on the relation between genes and the functions of proteins:
genes (in DNA) code for amino acid chains that fold into the
three-dimensional configurations of functional proteins. 
This sequence-to-function map is hard to decrypt 
since it links the collective physical interactions inside 
the protein to the corresponding evolutionary forces acting on the genome
\cite{Koonin2002,Xia2004,Dill2012,Zeldovich2008,Liberles2012}. 
Furthermore, evolution selects the tiny fraction of functional sequences 
in an enormous, high-dimensional space
\cite{Povolotskaya2010,Keefe2001,Koehl2002}, which implies that
proteins form non-generic, information-rich matter, outside the
scope of standard statistical methods. Therefore, although the
structure and physical forces within a protein have been extensively
studied, the fundamental question of how a functional 
protein originates from a linear DNA sequence still provides research
challenges, in particular how functionality constrains the accessible DNA sequences. 

To examine the geometry of the sequence-to-function map, we devise
below a mechanical model of proteins as amorphous evolving
matter.\footnote{In his book ``What is Life?'' \cite{schrodinger1944},
  Schr\"odinger uses the term `aperiodic crystal' to describe material which
  contains genetic information.  This  is of course a very
  interesting forethought, but since the advent of quasiperiodic
  crystals, the term `amorphous' leads to a more precise classification.}
Rather than simulating concrete proteins, we construct models which 
describe the hallmarks of the genotype-to-phenotype map 
(the translation of the gene to the protein). 
These models are sufficiently simple so that large-scale simulations 
can be performed, which allow to average over
stochastic noise inherent to evolutionary dynamics.
Furthermore, we restrict our approach to models in which the function of  
a protein arises from large-scale conformational changes, where big chunks of the
protein move with respect to each other.
These motions are central to certain functions 
\cite{Koshland1958,Henzler-Wildman2007,Savir2007,Schmeing2009,Savir2010a,Huse2002,Savir2013},
For example, allosteric proteins are a type of `mechanical transducers' 
that transmit regulatory signals between distant sites
\cite{Perutz1970,Goodey2008,Lockless1999,Ferreon2013}.

We end this section by mentioning a few papers which have dealt with
similar issues, and which highlight the increasing interest in
connecting biological questions with methods from solid state physics.

Common to these studies is a mechanical perspective on protein function. 
The motivation originates from many observations of proteins 
whose functions involve  collective patterns of forces and coordinated
displacements of their amino acids
\cite{Daniel2003,Bustamante2004,Hammes-Schiffer2006,Boehr2006,
Karplus2002,Henzler-Wildman2007,Huse2002,Eisenmesser2005,Goodey2008,
Savir2010a}.  
In particular, the mechanisms of allostery
\cite{Monod1965,Perutz1970,Cui2008,Daily2008,Motlagh2014,Thirumalai2018,Koshland1966},
 induced fit \cite{Koshland1958},
 and conformational selection \cite{Grant2010}
 often involve global conformational changes
by hinge-like rotations, twists or shear-like sliding of protein
subdomains \cite{Gerstein1994,Mitchell2016,Mitchell2017}. 

A now-standard approach to examine the link between function and motion
is to model proteins as elastic networks of amino acids connected 
by spring-like bonds. Early studies that apply this class of models
are from the 1980s and 90s \cite{Levitt1985,Tirion1996}, and in the last 
two decades the methods have been further developed and applied 
to many proteins \cite{Chennubhotla2005,Bahar2010,Lopez-Blanco2016}. 
Decomposing the dynamics of the network into normal modes revealed that
low-frequency `soft' modes capture functionally relevant large-scale
motion \cite{Tama2001,Bahar2010a,Haliloglu2015}, especially
in allosteric proteins \cite{Ming2005,Zheng2006,Hawkins2006,Arora2007,Tehver2009,Wrabl2011,Greener2015}. 

Recent work associates the soft modes of protein conformations with the emergence
of weakly connected regions as described above, but also `cracks'
\cite{Miyashita2003}, `shear bands' or `channels'
\cite{Mitchell2016,Mitchell2017,Tlusty2016,Tlusty2017,Dutta2018,Rocks2019} 
that enable low-energy viscoelastic motion \cite{Qu2013,Joseph2014}. 
Such contiguous domains evolve in models of allosteric proteins
\cite{Hemery2015,Flechsig2017,Tlusty2017}.

A source of inspiration for linking proteins to the physics of amorphous matter are the papers by the late Shlomo Alexander, 
especially \cite{Alexander1998,Alexander1982}.
In these works, Alexander highlighted the essential role 
of `floppy modes' in the mechanical spectrum of amorphous solids.
Also relevant are studies by Thorpe and Phillips on constraint theory 
and rigidity percolation in glasses, such as \cite{Thorpe1985,Phillips1985}.
Those works highlighted the ability to control the rigidity and 
accessible zero-energy modes of mechanical networks by balancing the
number of degrees of freedom and the number of constraints,
as was noted by Maxwell in 1864 \cite{Maxwell1864}.

The link between the dynamical spectra of proteins and amorphous matter has
been further explored in a recent series of works on mechanical metamaterials.
The emergence of long-range allosteric response was used in \cite{Rocks2017}
as a design principle for `programmable' metamaterial made of 
amorphous spring networks \cite{Rocks2018}. 
A similar random network approach was applied in \cite{Yan2017}
to design elastic materials with tailored mechanical response. 
These works suggest that tunable amorphous materials 
have the flexibility required to produce elaborate designs, 
as recently demonstrated by mimicking the cyclical conformational
motion of protein motors \cite{Flechsig2018}. 
These promising approaches to metamaterial design are discussed elsewhere,
for example in \cite{Ronellenfitsch2018,Kim2018,Rocklin2017,Baardink2018}.

The present Colloquium focuses on a different aspect: understanding
fundamental properties of the protein evolution --
in particular the genotype-to-phenotype map --  
within the framework of condensed matter theory.

%%%%%%%%%%%%%%%%%%%%%%%%
\section{Proteins as information machines}\label{sec:informationmachines}

The building plan of a protein is determined by its corresponding
gene, via the genetic code. The gene is a 1-dimensional string in an
alphabet of 4 letters: the nucleotides `A' (adenine), `C' (cytosine), `G' (guanine), and `T' (thymine) (see \cite{Alberts2017}, Ch. 6). The protein is a (folded) chain of
amino acids (AA) which is translated from the gene according to the
genetic code: each three successive letters (each non-overlapping triplet,
called a codon) maps to a single AA.
In principle, this would allow for $4^3$ possibilities, but in general
there are only 20 different AA's, making the code redundant, 
as we shall discuss in \sref{sec:readingerrors}.

We view the gene, \ie the 1-dimensional string of letters as 
the tape of a Turing machine \cite{Turing1936,Herken1992,Condon2018}.
Since any alphabet can be recoded in binary (for example, each of the 4
nucleotides can be recoded as a 2-bits number), one can always think of it as
a string of `0's and `1's. The proteins (and the transcription-translation 
machinery, which is itself made of proteins) would be the computer,
which is able to read and interpret the string.

This particular machine is an example of a self-reproducing Turing
machine \cite{Neumann1966}, since the replication of the genome can be
achieved by genome-encoded proteins. In addition, these machines
are evolving when the genes are mutated. In other words, the machine 
can modify its own tape (see also \cite{Tlusty2016}).
A further study in this direction is \cite{Dyson1970}, but there are
many more, see \eg \cite{Freitag2004}.

\subsection{Handling reading errors}\label{sec:readingerrors}

Translation of the gene into its corresponding string of amino
acids requires a specialized machinery, which includes the ribosome \cite{Alberts2017}.\footnote{In addition to the ribosome, the machinery includes two sets
  of molecules, 
tRNAs, which carry the amino acids, and aminoacyl-tRNA synthetases, 
which charge the tRNAs with amino acids. The translation is preceded
by a transcription step in which the DNA gene (a segment of the
genome) is copied into a mRNA (a single molecule).}
The translation machinery `reads' the code through chemical affinity,
and might therefore mis-read the tape.  
Most amino acids are encoded by more than one codon, and this
hard-coded redundancy of the genetic code helps to reduce the impact of 
such misreadings (see \cite{Tlusty2007,Tlusty2008,Tlusty2008a,Tlusty2008b,Tlusty2010} for a theoretical study 
and \cite{Eckmann2008} for an illustration).

\begin{figure}
  %%\centering
 \includegraphics[width=1.0\columnwidth]{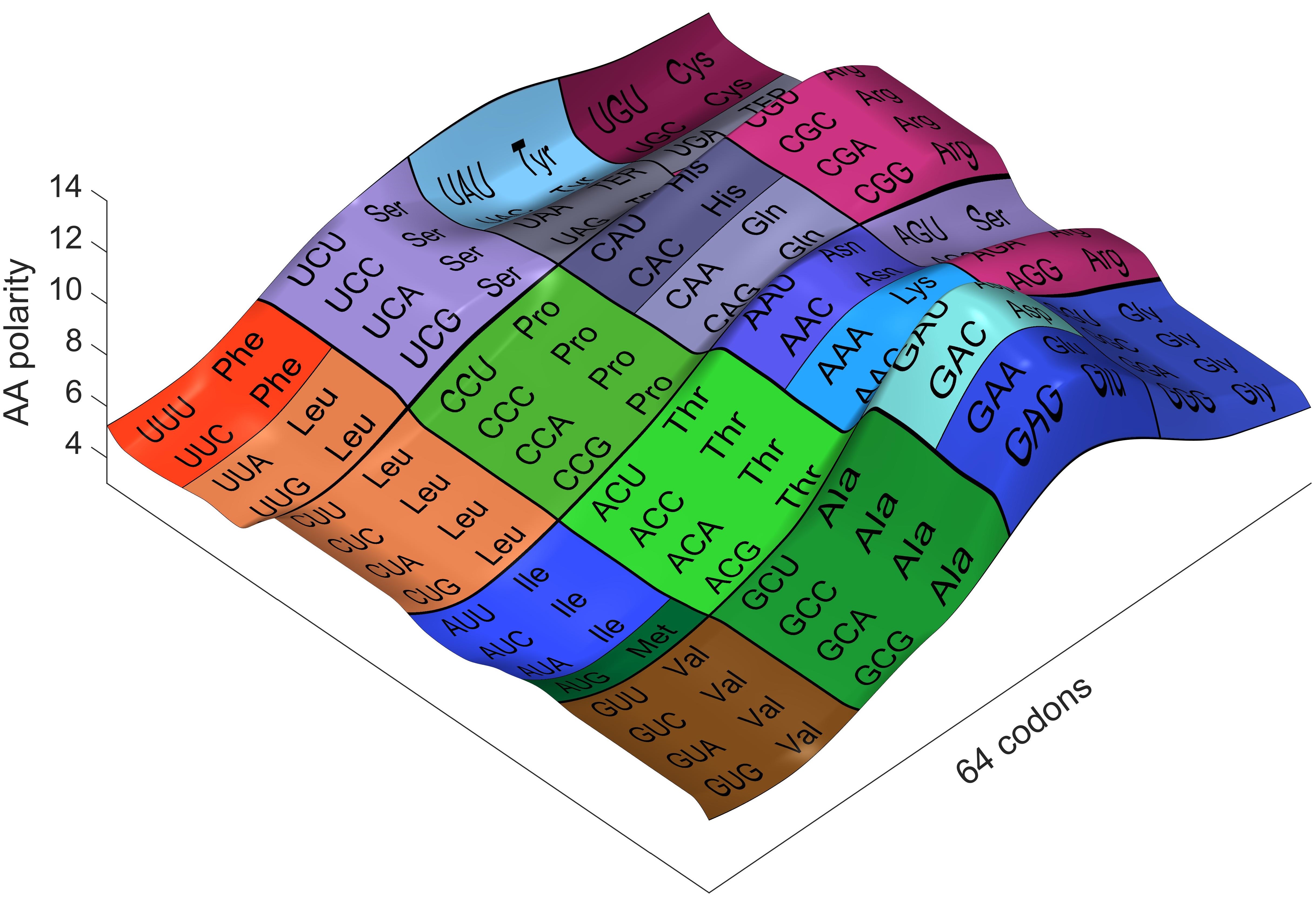}
 \caption{A representation of the genetic code, as function of
   the measured polarity of each codon (values from \cite{Haig1991,Woese1965}). 
   The smoothness of the landscape shows that moving from one AA 
   to its neighbor does not change the polarity too abruptly.}
   \label{fig:landscape}
\end{figure}

As noted above this system allows for $64=4^3$ different codons
(number of triplets from an alphabet of 4 nucleotides),
but they generate only 21 different symbols.\footnote{Some terminology: The individual  
  symbols (A, C, G, T) refer to nucleotides. The triplets of 3
  nucleotides form the 64 codons. The 64 codons code for 20 AA and the
  stop symbol (which does not generate an AA). 
  One of the AA is Methionine (codon ATG) which marks the start of a
  protein.}
The geometric aspects of this arrangement 
of 21 among 64 possibilities can be understood in graph-theoretical terms:
One presents the 64 codons as the nodes of a codon-graph, 
and two nodes are connected by a link if the corresponding codons 
differ in only one symbol. Note that swapping  `C' and `T' 
in the codon's third position always results in the same AA (\fref{fig:landscape})
and we can therefore reduce the graph to $48=4^2\cdot 3$ nodes.\footnote{This
  graph is difficult to draw, as each node has $8=3+3+2$ neighbors 
  which differ in exactly one position. So a representation would have to be
  in 8 dimensions. Recall that in a cube in 3-dimensions, every corner
  has 3 neighbors.} 
In the codon-graph, each amino acid is coded as a simply connected region, 
as shown in \fref{fig:landscape}, with the exception of Serine (ser)
(Arginin (arg) is disconnected in the 2D table, but not in the graph). 
Such an arrangement minimizes the ratio of surface by area for each region. 
This reduces the probability of coding the wrong AA,
under the assumption that most reading errors involve only one-letter
differences. 

Additionally, amino acids with similar chemical properties (for example
polarity) tend to be neighbors in this graph. 
This can be visualized by plotting the measured polarity as a 
function of the codon, which produces a relatively smooth landscape. 
The smoothness manifests the chemical similarity between 
neighboring amino acids, and implies that most misreadings change
the polarity of AA only moderately. 
We note that, unlike the 2D landscape of \fref{fig:landscape},
an ideal representation should wrap the surface so that each
AA would have 8 neighbors (and can therefore be embedded only in high dimension). 

For the connection between the numbers 21 and 48, 
an inequality can be given in terms of the genus of the codon-graph 
\cite{Tlusty2007,Tlusty2007a,Tlusty2010}
(this uses results from \cite{Colin1993,Banchoff1965}). 
Without going into further detail we conclude: 
\emph{The optimal code must balance contradicting needs for tolerance to errors 
     (with the smoothness of the mapping between codons and chemical space) 
     and chemical diversity, which is essential for the versatility of protein function}.
 
\subsection{Folding}

\begin{figure}
  \includegraphics[width=.9\columnwidth]{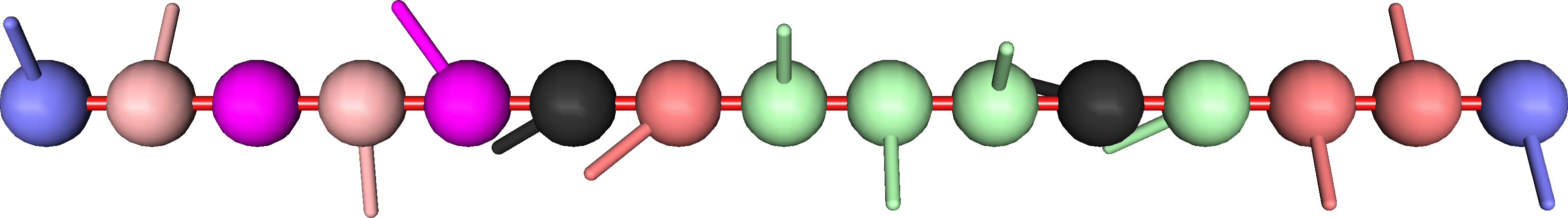}
  \includegraphics[width=0.5\columnwidth]{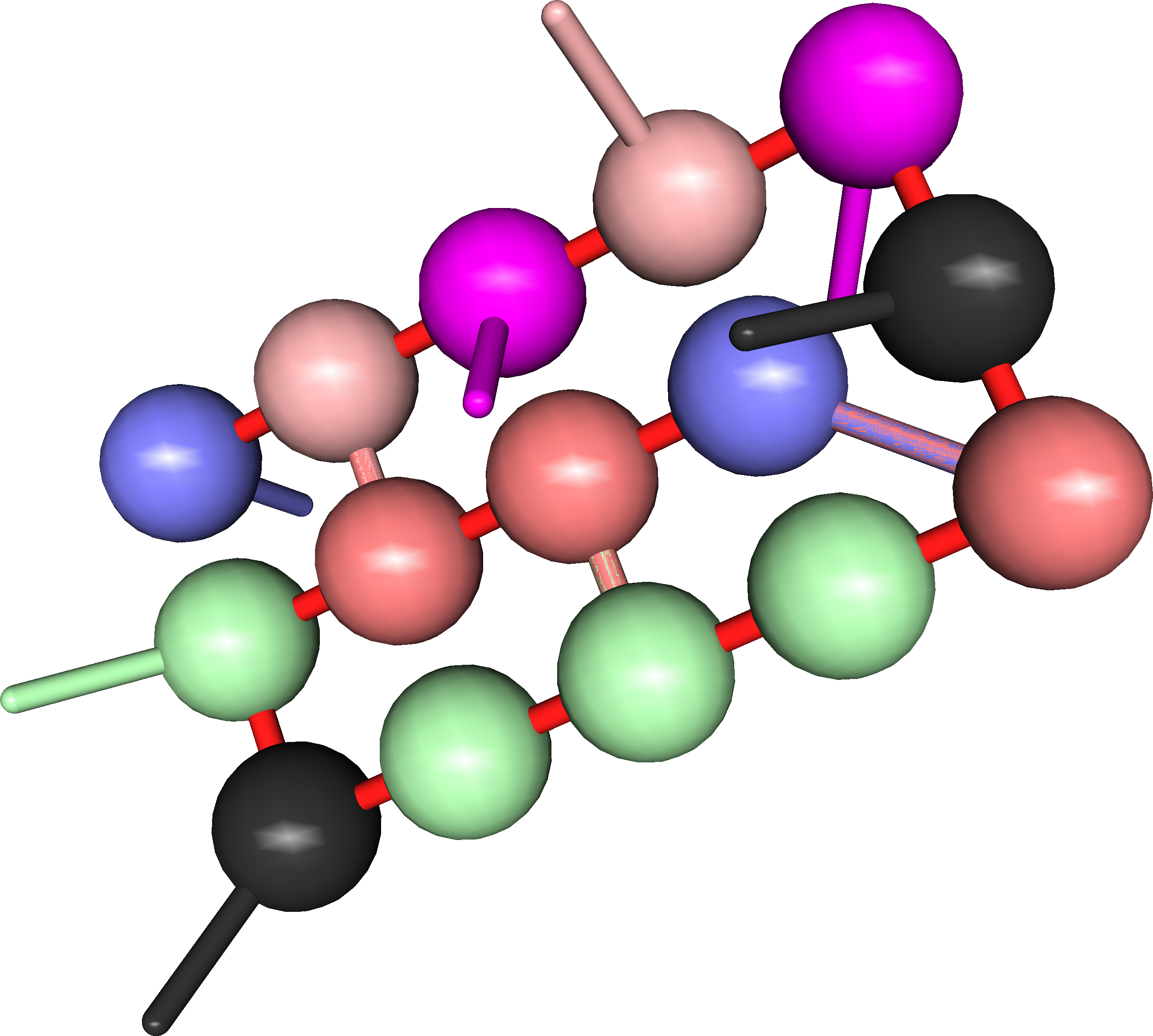}
  \caption{Schematic illustration of protein folding: 
  Proteins are polypeptides, linear heteropolymers of AAs (colored spheres), 
  linked by covalent peptide bonds (red sticks), which form the protein backbone.  
These peptide bonds are much stronger than the non-covalent interactions 
among the AAs (side chains) and do not change when the protein mutates.}
\label{fig:backbone}
\end{figure}

Having translated the gene into a linear chain of amino acids 
(the backbone, see \fref{fig:backbone}) via the genetic code 
(and modulo translation errors), this chain will spontaneously fold 
into a 3-dimensional shape which gives rise to its function.
How this folding proceeds is an important and difficult question,
which we shall not address here. Instead we will assume that a certain 
folding pattern is preserved (see \cite{Petsko2004} for a discussion
of these issues). This assumption is practical, as we shall be 
mostly interested in how the function of the protein changes under
point mutations of the gene, \ie bit flips of the code in the tape.
Such mutations often do not seriously affect the overall shape of the
protein (see also \cite{Bussemaker1997}).

We can next model the function of this folded amino acid chain and we
will show that there is yet another level of redundancy besides the
redundancy of the genetic code and the robustness of the folding.
we shall see that there are many mutations which have no effect on performance. 
\emph{Namely, there is high redundancy in the AA sequences that are mapped to 
the same or similar enough protein function}. 
we shall quantify this property in terms of dimension 
\cite{Grassberger1983,Eckmann1992}.

%%%%%%%%%%%%%%%%%%%%%%%%
\section{Mechanical views on protein evolution}
\label{sec:views}
Consider a protein interacting with a small molecule. 
Presence of the latter often induces a conformational change 
at some distance from the interaction site. 
One important example is the class of allosteric proteins 
for which an active site is regulated by binding at another site, 
resulting in a reconfiguration of the active site.
More specifically, we shall examine the role of large-scale, 
functionally-relevant dynamical modes, 
and their link to long-range genetic correlations.

\begin{figure}[t]
%%\centering
\includegraphics[width=0.8\columnwidth]{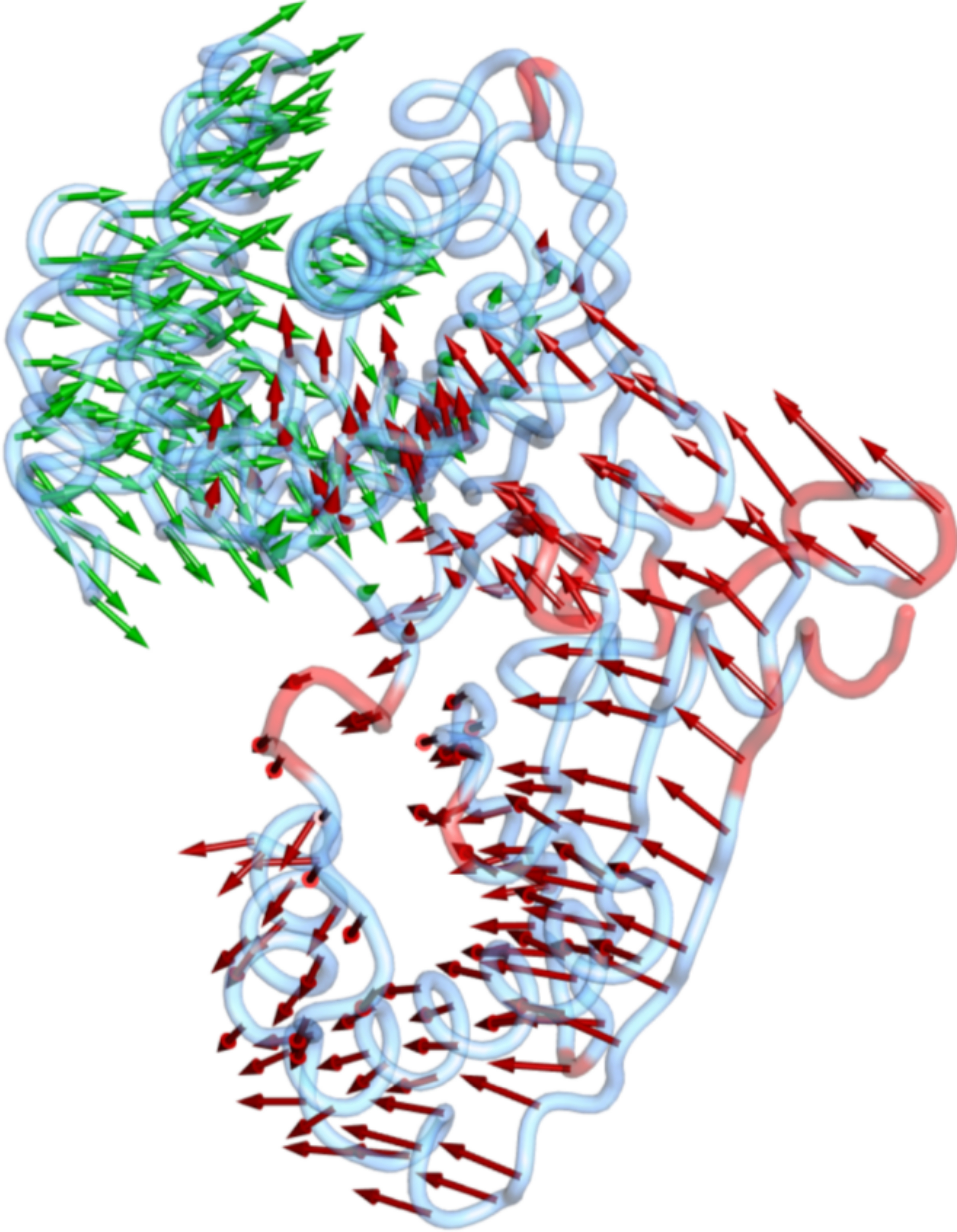}
\caption{The motion and deformation between two states of glucokinase
  in state 1v4s and 1v4t. The arrows are scaled up for better
  visibility. They are colored green, resp.~red depending on whether
  they move down or up relative to a plane passing horizontally
  through the center of the protein. Galilean motions have been
  eliminated. The red coloring of the tube corresponds to concentration
  of shear and is the same as in the leftmost panel of \fref{fig:3gluco}. 
  See \sref{sec:shear}.}
\label{fig:gluco}
\end{figure}

Before reviewing the literature on this issue, we illustrate 
such a mechanical effect on a particular example: human glucokinase 
(which is involved in sugar metabolism), see \fref{fig:gluco}.
The data were obtained from crystallographic structure of
two conformations of that protein: the first 
(PDB\footnote{PDB = protein data bank, https://www.rcsb.org.} accession 1v4s) 
corresponds to the binding of glucose to its active site and is compared to the
conformation in the absence of glucose (PDB 1v4t) \citep{Kamata2004}.

The backbone, see \sref{sec:backbone}, is shown as a light blue
curled tube, and the arrows indicate the displacement from one
shape to the other (any Galilean motion between the two is
eliminated). The color of the arrows indicates up/down motion relative
to a horizontal plane. The red coloring in the twisted tube 
shows the high shear region separating two low-shear domains 
that move as rigid bodies (shear calculated by the method of
\cite{Mitchell2016,Rougemont2099}).

\begin{figure}[t]
%%\centering
\includegraphics[width=0.8\columnwidth]{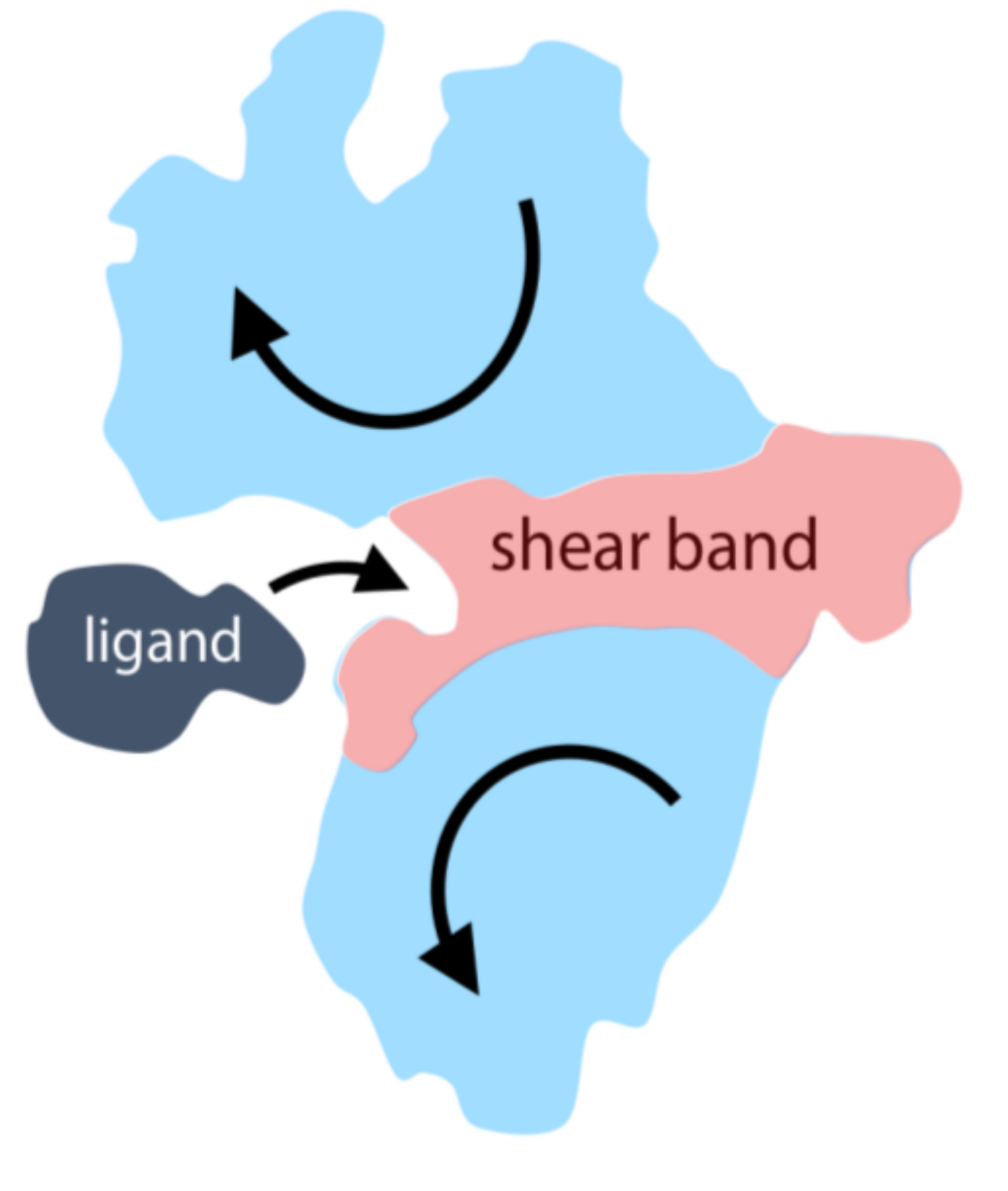}
\caption{A schematic interpretation of \fref{fig:gluco}. 
  Emphasis is given to the two moving pieces, 
  with a hinge between them. This kind of hinge 
  will be called the `fluid channel' or `shear band'.}
\label{fig:glucosymbolic}
\end{figure}

On a conceptual level, one can simplify the figure as shown in
\fref{fig:glucosymbolic}. The protein seems to have a central shear
band and two external flaps which perform a rotating motion when a ligand
attaches to the protein. This kind of mechanical phenomenology is
accessible to the language of physics.

Large-scale motions take part in several basic biological functions
and mechanisms. For example, in the induced fit \cite{Koshland1958}
and conformational selection \cite{Bahar2007,Grant2010} mechanisms,
the presence of a substrate induces reshaping of the enzyme to
properly align the catalytic groups in the active site. 
Such reshaping is a dynamic mechanism of
\textit{specific recognition} that allows the selection of a target
ligand among similar competing molecules \cite{Savir2007,Savir2013}.
In \textit{allostery}, reconfiguration of the active site is  regulated by binding at a secondary, allosteric site, often via long-range mechanical
interactions \cite{Motlagh2014,Thirumalai2018}.
%Allostery may have evolved from specific recognition, by adding a
%second binding site to that is connected to the active site by
%pre-existing dynamical interaction pathway.    
In this Colloquium, we describe simple physical models for the emergence of these mechanisms via evolutionary tuning of the protein's mechanical response.

\begin{figure*}
%%\centering
\def\x{0.6}
\includegraphics[width=\x\columnwidth]{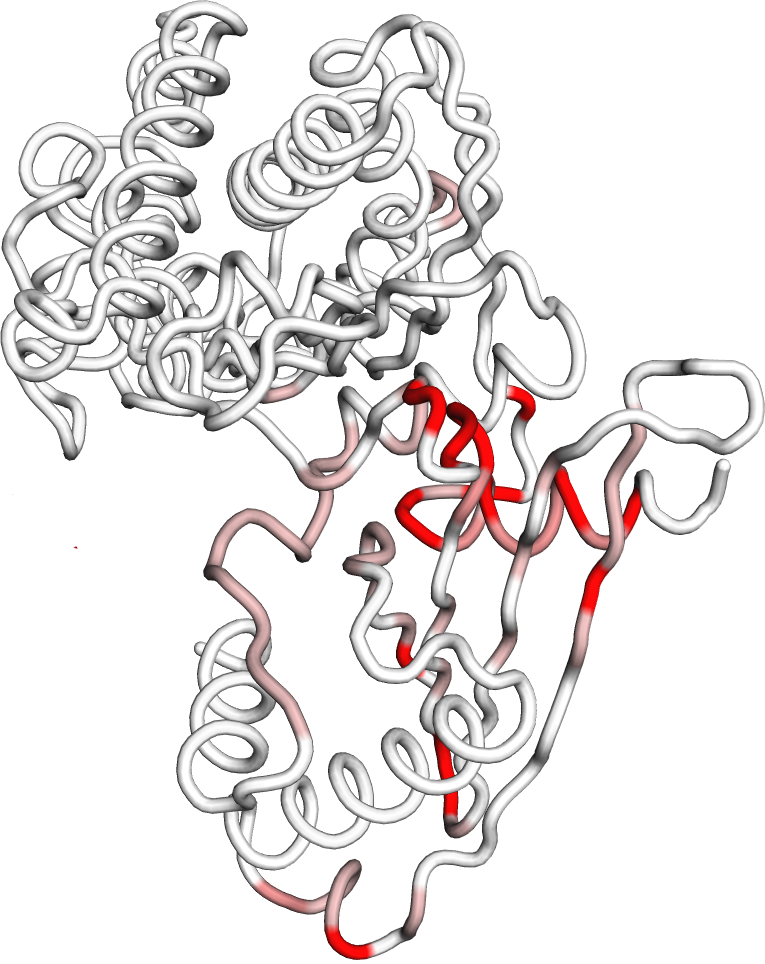}\includegraphics[width=\x\columnwidth]{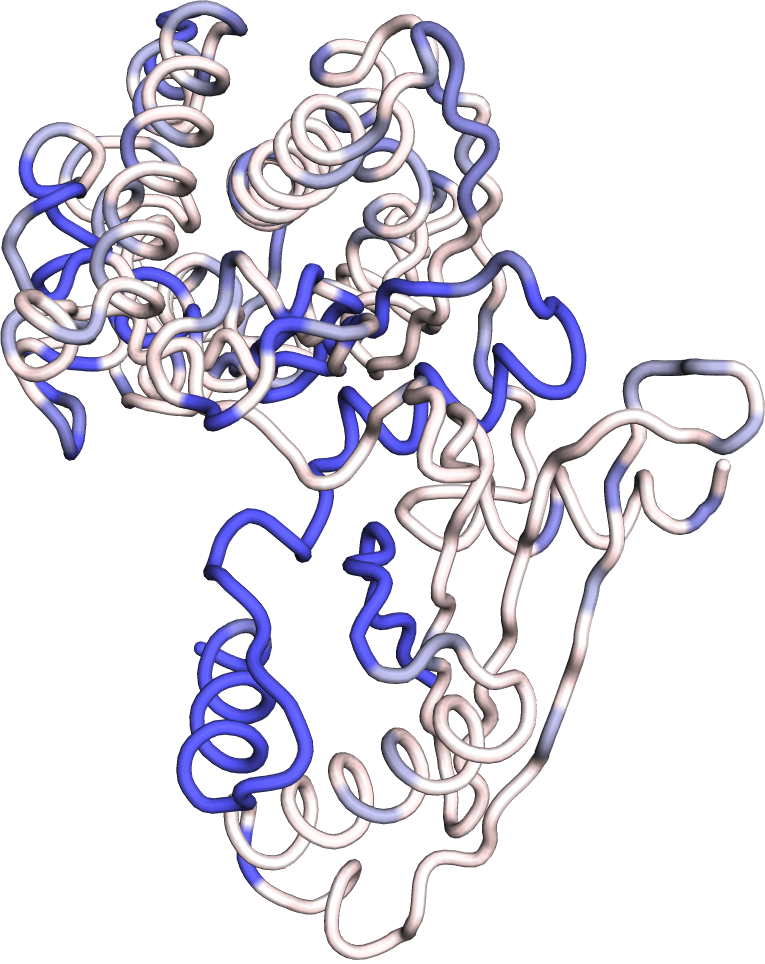}\includegraphics[width=\x\columnwidth]{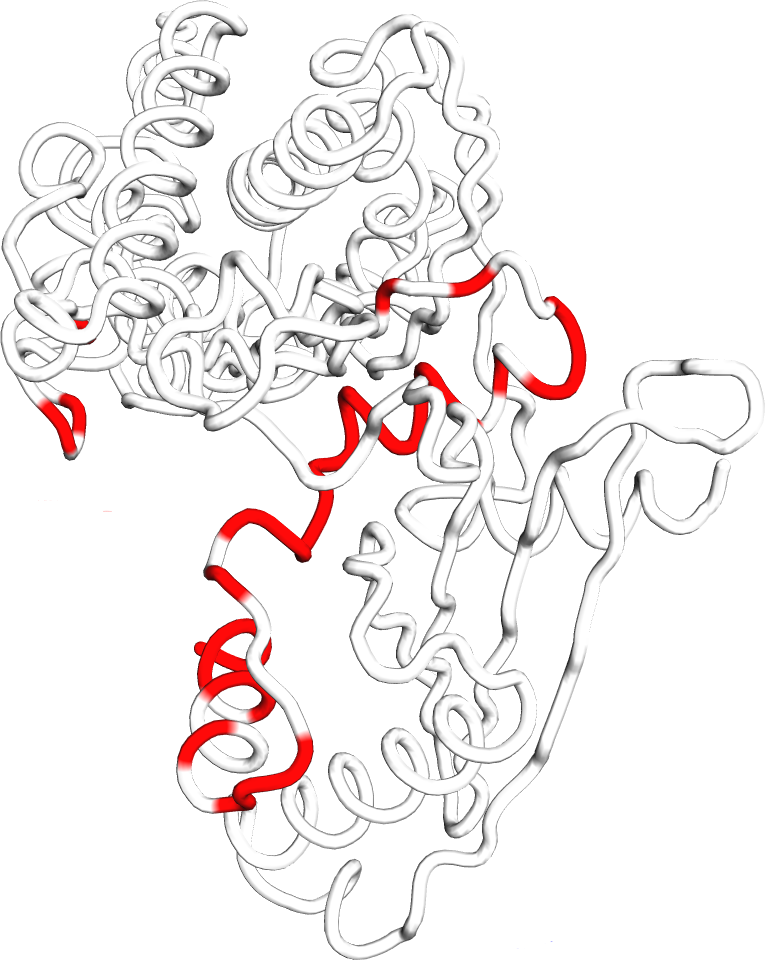}
\caption{\textbf{Evolutionary and mechanical properties of glucokinase}. 
  Left: Shear (darker is more shear), discussed in \sref{sec:shear}.
  Center: Conservation (darker = fewer mutations), discussed in \sref{sec:conservation}. 
  Right: Correlation (dark red = mutations correlate), discussed in \sref{sec:epistasis}.
  }\label{fig:3gluco}
\end{figure*}

Like their dynamic phenotypes, proteins' genotypes (their gene sequences),
as explained in \sref{sec:informationmachines}, are remarkably collective. 
The history of protein evolution can be traced by gathering 
evolutionary related proteins in different species (homologous proteins)
and aligning their sequences. Genes of these proteins sometimes display long-range
correlations \cite{Goebel1994,Marks2011,Jones2012,Lockless1999,
Suel2003,Hopf2017,Poelwijk2017,Halabi2009,Tesileanu2015,Juan2013}.
The correlations indicate epistasis, the compensatory mutations
that take place among residues linked by physical forces or common function. 
As an example \cite{Rougemont2099}, consider again glucokinase. 
We aligned about 120 variants of this molecule and asked 
where along the gene have mutations preferentially occurred (\fref{fig:3gluco}).

Still, the relationship between sequence correlation, 
epistasis and selection pressure are not fully understood. 
As discussed in \sref{sec:proteinproblem}, the two main challenges
are the intricacy of the physical forces among the amino acids,
and the high dimensionality of the of the genotype-to-phenotype map
\cite{Koonin2002,Povolotskaya2010,Liberles2012}. 
These inherent difficulties motivated the development 
of complementary approaches which utilized 
simplified coarse-grained models, such as lattice proteins
\cite{Lau1989,Shakhnovich1991} or elastic networks
\cite{Chennubhotla2005}.
Network and lattice models have been recently used to study 
the evolution of allostery in proteins 
and in biologically-inspired allosteric matter
\cite{Tlusty2016,Tlusty2017,Hemery2015,Flechsig2017,Rocks2017,Yan2017}.
Our aim here is different: 
to construct a simplified condensed-matter model 
in terms of how the mechanical interactions within the protein shape its evolution.

%%%%%%%%%%%%%%%%%%%%%%%%
\section{Condensed-matter theory of proteins}

This section will review a theory of proteins in terms of evolvable
condensed matter. we shall discuss the conceptual roots of this
approach in the physics of amorphous matter (mainly glasses) 
and spectral theory. 
We will introduce the basic setting of modeling proteins as
evolving amino acid networks. The emergence of function is associated 
with the evolution of a weakly connected region, which enables 
a low-energy `floppy' mode to appear. This minimal network approach 
allows one to examine basic questions of protein evolution. 

%===========================================================

we shall discuss two different models in this review. One will be
called the `cylinder-model' and the other the `HP-model'. 
The first model is simpler, but the second comes somewhat closer the
biological reality. Before distinguishing the two models, we describe
their common features.

\subsection{Lattice models}\label{sec:lattice}

Our protein is modeled by a finite (regular) lattice in 2 (or 3)
dimensions. We assume that the lattice forms a cylinder (periodic
boundary conditions) or an open rectangle (open boundary conditions) 
of width $w$ and height $h$ (see the examples in
\sref{sec:cylinder}--\ref{sec:pnasmodel}).

It is important to note that $w$ and $h$ are \emph{finite} 
while otherwise quite arbitrary. This is so because the protein 
should not be viewed as a problem of thermodynamic limits, 
but rather in the context of small amorphous objects. 
This being said, other aspects of the geometry seem less important. 
The points may also be chosen as lying on small perturbations 
from the regular lattice to avoid effects 
of lattice symmetry on the spectrum.
The number of AAs should typically be in the range
200--2000, corresponding to the typical size of the protein.

Amino acids interact via electrostatic forces, 
van der Waals forces, hydrogen bonds, disulfide bonds and hydrophobicity
\cite{Petsko2004,Fersht1999}. All these are short range interactions, 
which amount to \emph{local} coupling between lattice points.
we shall therefore assume that each AA interacts with its
nearest and next nearest neighbors. For example on a hexagonal lattice,
with nearest and next-nearest neighbors linked, 
the number of connections (the node's degree) is at most 12;
all nodes in the protein interior have 12 links (\ie bonds) 
while those at the boundary have fewer (but at least 3), 
see \fref{fig:1}.

Finally, the coupling itself is modeled by harmonic springs
carried by each graph link
\cite{Born1954,Alexander1998,Chennubhotla2005,Tirion1996}.
Its strength is determined only by the types of AAs 
at each end of the link.

\subsection{The lattice Laplacian}\label{subsec:lattice}

The lattice and its links may be viewed as an abstract graph. 
This means that one can define gradients and Laplacians 
\cite{Biggs1993,Chung1997}.\footnote{The first book is more
combinatorial, and the second introduces more spectral concepts.} 
In the graph, there are $n_a=w\times h$ amino acid nodes, 
indexed by Roman letters, and $n_b$ bonds, indexed by Greek
letters.\footnote{If a bond $\alpha$  
connects nodes $i$ and $j$ ($i\ne j)$, we also write $\alpha =(i,j)$.}

First, one endows every bond in the graph with an arbitrary but fixed
orientation, and then the incidence matrix of a graph is the
$n_b\times n_a$ matrix defined by
$$
\nabla_{\alpha i} =\left\{
\begin{array}{@{}rl@{}}
  -1~, &  \text{if $i$ is the initial vertex of edge $\alpha$}~,\\
   1~, &  \text{if $i$ is the final vertex of edge $\alpha$}~,\\
   0~, &  \text{if $i$ is not in $\alpha$}~.\\
\end{array}\right .
$$
Remark that for any function $f$ on the vertices, 
the map $f \mapsto \nabla f$ is the co-boundary 
mapping of the graph, namely
$$
\nabla f (\alpha )= f(j)-f(i)~,
$$
where $\alpha$ is the link connecting $i$ to $j$.

As in the continuum case, the Laplace operator $\Delta$ is the product
$\Delta = \nabla^\T\nabla$, where $^\T$ denotes the adjoint. 
The non-diagonal elements $\Delta_{ij}$ are $-1$ if $i$ and
$j$ are connected and $0$ otherwise. The diagonal part of $\Delta$ is
the degree $\Delta_{ii}= z_i$, \ie the number of nodes
connected to $i$. Note that this is a discrete graph Laplacian, 
and no coordinates are involved so far.

We next embed the graph in a Euclidean space $\mathbb{R}^d$ ($d=2$), by
assigning positions $r_i \in \mathbb{R}^d$ to each AA, \ie to each
lattice point $i=1,\dots,n_a$. This is coded as a $n_a\times d$ real matrix $\rb$.
Finally, to each bond $\alpha $ we assign a spring with constant
$k_{\alpha}$ which we view as the diagonal elements of an $n_b
{\times} n_b$ matrix $\Kb$:
\begin{equation}
  \Kb_{\alpha \beta}\,=\, k_\alpha\delta_{\alpha\beta}~.
\end{equation}
This defines a deformable spring network which has an internal energy, an
equilibrium configuration. 

To account for the energy cost of deformations in the lattice protein, 
one considers the elastic tensor $\HH$ (or Hamiltonian) which we now 
describe in detail \cite[pp.~618--619]{Chung1992}. 
The quantity $\HH$ is a tensor because the deformations 
are not scalars, but vectors in $\mathbb{R}^d$. 
We first denote by $\nb_{\alpha}$ the (normalized) 
direction vectors for each bond $\alpha=(i,j)$:
$\nb_{\alpha}=\left(\rb_i - \rb_j\right)/\left\vert\rb_i -\rb_j\right\vert$. 
Then, we define the `embedded' gradient tensor $\Db$ 
(of size $n_b\times n_a\times d$) which is obtained 
by multiplying each element of the graph gradient $\nabla$ 
by the corresponding vector $\nb$:
\begin{equ}
\Db_{\alpha i}\,= \,\nb_{\alpha}^\T\nabla_{\alpha i}~,
\label{eq:mba}
\end{equ}
namely each projection of $\Db$ on a bond $\alpha$, $\Db_{\alpha, :}$,
is a $n_a\times d$ matrix containing only $2d$ non-zero entries in
rows $i$ and $j$, which
correspond to the components of the unit vector along the bond $\alpha=(i,j)$.

Let $\ub_i$ be the displacement vectors of
each vertex from $\rb_i$ to $\rb_i+\ub_i$, therefore $\ub$ is a $n_a\times d$ matrix.
The elastic energy of such a perturbation is 
\begin{equ}\label{eq:Hamiltonian}
  \elastic\,=\, \HALF\ub^\T \HH \ub
  \,=\, \HALF\sum_{ij}\sum_{k\ell}\ub_{ik} (\HH_{ij})_{k\ell} \ub_{j\ell}~,
\end{equ}
where the Hamiltonian tensor is defined as
\begin{equa}
\HH\,&=\,\Db^\T\Kb\Db\\
\,&=\,\Bigl(
  \sum_{\alpha\beta}\Db_{\alpha ik}\Kb_{\alpha\beta}\Db_{\beta j\ell}
  \Bigr)_{i,j=1,\dots,n_a;~k,\ell=1,\dots,d}~ .
\end{equa}
The $d\times d$ off-diagonal components are:
\begin{equa}
\HH_{ij}  =& \sum_\alpha \nabla _{\alpha i}\nb_{\alpha}
\Kb_{\alpha\alpha} \nb_{\alpha}^\T\nabla _{\alpha j}\\
=&\Delta_{ij} k_{(i,j)}\nb_{(i,j)}\nb_{(i,j)}^\T~,
\end{equa}
which we complete with the diagonal blocks ($i=j$) so that 
rows and columns sum to zero: $\HH_{ii}=-\sum_{j \ne i}{\HH_{ij}}$. 

In this construction, we have assumed that the
equilibrium configuration of the network (described by the vectors $\rb_i$)
is such that all springs are at their
equilibrium length, disregarding the possibility of `internal
stress' \cite{Alexander1998}, hence the initial elastic energy is
$0$. The extension of the theory to networks that are
initially frustrated, \ie where some springs are stretched or squeezed
is a difficult subject. A paper which studies the conjectures by
Alexander on internally stressed networks is \cite{Kustanovich2003}.
The spring constant is therefore the derivative of the interaction
at the equilibrium length of the spring.

We next consider only small deviations of the AAs from their
equilibrium, \ie the linear mechanical response of the protein
to an applied force. While this approximation 
cannot account for plastic and non-affine deformations that often occur 
in real proteins, it certainly simplifies the analysis, 
in contrast to the inherent difficulties 
of studying fully nonlinear systems.

Given a prescribed `protein fold' (the lattice positions $\rb_i$, $i=1,\dots,n_a$), a
gene first determines the spring constants via a
`genetic code' which maps codons to AAs on the lattice and thereby determines
the interaction strength between neighbors on the lattice.
This in turn defines a phenotype of the
protein, namely its mechanical response under deformations.
Each choice of the gene, \ie the set of codons $\cb=\{c_i\}$, 
defines a Hamiltonian $\HH=\HH(\cb)$. 
Component-wise, each $d {\times} d$ block $\HH_{ij}$ ($i \neq j$) 
depends only on the codons $c_i$ and $c_j$: 
$\HH_{ij}(\cb)=\HH_{ij}(c_i, c_j)$.

In summary, note that $\HH$  depends on three things:
\begin{myenum}
\item The position $\rb_i$ of each amino acid, $i=1,\dots,n_a$,
\item The type $c_i$ of each amino acid, $i=1,\dots, n_a$,
\item The spring constants $k(c,c')$ representing the interaction strengths between amino-acids types $c$ and $c'$.
\end{myenum}
This definition is clearly versatile enough to be generalized to other systems,
such as proteins made of the standard 20 AAs with 
specific interaction constants for each of the possible AA pair. 

\subsection{Hooke's law}

We have now a map from genes $\cb$ to Hamiltonians $\HH(\cb)$, and we
want to study the deformability of the network as a function of $\cb$.
In the linear regime of relatively moderate deformations, 
one can use Hooke's law (see \eg \cite{Alexander1998}) to relate a (small)
deformation $\ub$ to the force by 
\begin{equ}
  \fb= \HH\,\ub~,
\end{equ}
where $\fb$ is a force vector field. We are interested in the inverse
relation, since we want to know the deformation of the network
(protein) as a function of the applied force $\fb$. 
This inverse will be described by Green's function $\GG$:
\begin{equ}\label{eq:gf}
  \ub= \GG\,\fb~.
\end{equ}

\section{Simulating evolution}\label{sec:simulating}

Next, we consider modeling evolution, for a general `genetic code'.  
As described above, to each gene $\cb$ there is a natural
Hamiltonian $\HH(\cb)$ associated with it.  
This is the mechanical genotype-to-phenotype map.
We assume that a fitness function $F$ is given,  
mapping every $\HH$ to its fitness score $F(\HH)\in\real$. 
The observable we take later as $F$ will be an expectation
value for some components of the force field $\fb$.
The evolutionary process alters this fitness, by mutating
individual random positions in the gene $\cb$ 
(the collection of $c_i$).
This is realized by a Metropolis algorithm 
\citep{Metropolis1953}:

In an evolution simulation, one exchanges a randomly 
selected codon with another one (at the same position), 
while demanding that the fitness change $\delta F$ 
is positive or non-negative. We call $\delta F >0$ 
a beneficial mutation, whereas $\delta F=0$ corresponds
to a neutral one. Deleterious mutations, $\delta F<0$, 
are generally rejected.

As in statistical physics, variants of this algorithm 
can be envisaged, for example, by asking for an increase 
of $F$ by a minimal factor $|F|\to |F|\cdot (1+\epsilon )$ 
with $\epsilon >0$, for a step to be accepted. 
Other possibilities include the introduction of `temperature', 
\ie accepting or rejecting even deleterious mutations, $\delta F<0$, 
with some probability. 
The rationale behind using these variants of Metropolis algorithms 
lies in the nature of natural mutations. For a review of the role of
deleterious mutations, see \citep{Kondrashov2017}.
Details of $F$ will be given when we discuss 
various models in \sref{sec:models}.

%%%%%%%%%%%%%%%%%%%%%%%%
\section{Green's function as a link between the theory of
 amorphous solids and living matter}\label{sec:green}

In the previous section, the ground was prepared for studying the
connection between the genes and the mechanical properties of the proteins
they code. we shall use the mapping from the gene $\cb$ to 
the Hamiltonian $\HH(\cb)$ introduced above. 
One of the questions to be examined is how the protein reacts 
to forces applied to it, and how this response is encoded in the gene. 
Such forces occur when a small ligand molecule
attaches to a binding site on the protein's surface, 
inducing a mechanical response in other regions of the protein.

Intuitively, this means that we are looking for a relatively
strong reaction to a weak signal. Such phenomena are captured by soft modes. 
Such modes are given by zero eigenvalues of the Hamiltonian $\HH$,
and the corresponding deformations are described by the eigenvector $\ub$ 
of displacements of $\rb$ (corresponding to the zero eigenvalue).

Among the many approaches to the zero eigenvalue problem, 
we use the methods of Green's functions, which are well adapted 
to the emergence of soft modes in protein evolution. 
Green's function (also called the resolvent, matrix inverse) is useful here 
because of the following observation:
Consider a mutation that alters just one $c_i$.
Given the short-range nature of $\HH$, this implies that only
a small number of terms in \eref{eq:Hamiltonian} will change, 
independently of the size of the system. 
For example, for the hexagonal lattice in \fref{fig:fig1}, no
more than 12 terms change with each mutation.

Since, by Hooke's law, $\fb=\HH(\cb)\ub$, the response of the system to an
external force is given by the inverse relation $\ub=\GG(\cb)\fb$, where
$\GG$ is Green's function, \ie the inverse of $\HH$. So, $\GG$
maps the genotype $\cb$ to the reaction of the protein 
to an external force $\fb$. A typical example of such a stimulus 
appears when $\fb$ `pinches' 2 neighboring AAs
towards each other;  we would like to measure the effect of the pinch
on another AA pair (usually on the opposite side of the protein).

In dimension $d{=}2$, the Hamiltonian $\HH$ has always $d(d-1)/2=3$ zero eigenvalues, owing to the rigid Galilean transformations (2 translations and 1 rotation) of the lattice as a whole. Therefore, since $\HH$ is bound to be singular, it lacks a proper inverse. instead, one may compute the inverse 
on the subspace of $\real^{n_a}\times\real^{d}$
in the complement of the 3 Galilean directions. 
This is called the pseudo-inverse \cite{Penrose1955} and is usually denoted by $\GG(\cb)=\HH(\cb)^{\pseudo}$.

Let $\PP$ be the projection on the subspace spanned by the generators
of the Galilean transformations, then
\begin{equ}
  \GG(\cb)= \bigl((\id-\PP) \HH (\cb) (\id-\PP)\bigr)^{-1}\equiv \HH(\cb)^{\pseudo}~.
\end{equ}
It is easy to verify that if $\ub$ is orthogonal to the zero modes,
$\ub=(1-\PP)\ub$, then $\ub = \GG \HH\ub$. 
The pseudo-inverse obeys the four requirements:
(i) $\HH \GG \HH = \HH$~, (ii) $\GG \HH \GG = \GG$~, 
(iii) $(\HH \GG)^\T = \HH \GG$~, and (iv) $(\GG \HH)^\T = \GG \HH$~.

The projection onto the complement of the 0-space
commutes with the action of mutations, since changing the AA at a
site does not change the Galilean invariance of the lattice. 
Therefore, the pseudo-inverse can be used for our purposes 
just like the standard inverse.

\subsection{Woodbury's formula}

When one changes a gene $\cb$ to some $\cb'$, then the change in the Hamiltonian is $\DH=\HH(\cb')-\HH(\cb)$ 
and correspondingly the changes in Green's function from 
$\GG(\cb)$ to $\GG(\cb')$.
The Woodbury formula \cite{Woodbury1950,Deng2011}
relates $\DG=\GG(\cb')-\GG(\cb)$ to $\DH$ as follows:
First, one notes that the rank of the change tensor $\DH$ is
equal to the number $r$ of bonds altered by the mutation.
$\DH$ can therefore be written as
$$
\DH = \MM \Bb \MM^\T~,
$$
where the $r {\times} r$ diagonal matrix $\Bb$ records the strength change  
of the bonds (\eg from weak to strong or vice versa). 
$\MM$ is a $r\times n_a\times d$ tensor which is the restriction of $\Db$
(see \eref{eq:mba}) to bonds which were changed.
This allows one to easily calculate changes in Green's function:
%(with pseudo-inverses):
\begin{equ} \label{eq:Woodbury}
  \DG= -\GG\MM \left(\Bb^{-1} + \MM^{\T}\GG\MM \right)^{\pseudo}\MM^\T\GG~.
\end{equ}
% Note that $\GG$ is a square matrix of size $d\cdot n_a$, while $\Bb$
% is of a much smaller size $r$ (the number of affected links), $r \ll d\cdot n_a$.
The reader who is not familiar with \eref{eq:Woodbury} 
can compare it to the resolvent formula (in the commutative, scalar  case):
$$
\frac{1}{x+y} -\frac{1}{x}=
- \frac{1}{x}\left(\frac{1}{\frac{1}{x}+\frac{1}{y}}\right)\frac{1}{x}~,
$$
with $x^{-1}$ corresponding to $\GG$, and $y$ to $\Bb$ (and $\DH$) . 

The Woodbury formula is especially useful since one has to invert only 
square matrices of size $r$ ($\Bb$ and the term in brackets 
in \eref{eq:Woodbury}), instead of inverting the larger tensor $\HH$ 
of size $d\times n_a$ \cite{Henderson1981}.
For point mutations, this difference is dramatic, since the rank 
$r$ can be at most $z$, the number of neighbors of the mutated AA, 
implying that $r \ll n_a \times d$. For example, in the hexagonal model
\fref{fig:fig1}, $r \le z = 12$ while $n_a\times d = 1080$.

\subsection{Dyson's formula}
Another useful (and more common) identity is Dyson's formula
\cite{Dyson1949,Dyson1949a,Abrikosov1963}. It can be obtained by applying the 
resolvent identity to $\GG'=\GG+\DG$, leading to
\begin{equ}\label{eq:Dyson}
 \GG' = \GG -\GG \,\DH\, \GG'~.
\end{equ}
Since $\GG'$ appears on the r.h.s., one may successively iterate 
this identity to get the Dyson series,
\begin{equ}\label{eq:dysonseries}
 \DG =\GG'-\GG= -\GG\,\DH\, \GG+ \GG\,
 \DH\,\GG\,\DH \, \GG - \cdots ~.
\end{equ}
The series is widely used in potential scattering,
and is interpreted there as expansion in multiple scattering. 
The first term is usually called the Born term. 
We will interpret this identity in terms of multiple mutations 
and this will be another contact of methods known from 
the physics literature with questions in evolution.

\section{Models: Protein as an evolving machine}\label{sec:models}

After introducing the basic principles of our approach, 
we now discuss how to apply them in specific models. 
As in any simplifying model, there is an intrinsic conflict: 
On the one hand, one would like to keep the model as simple as possible, 
because the goal is to tests basic principles, not specific proteins. 
On the other hand, there should still be some connection to real proteins. 
As mentioned before, one cannot apply the thermodynamic limits 
of standard statistical mechanics (infinite number of particles, 
long range potentials, and the like), since the protein boundary
plays an important role. So the protein is treated 
as a finite, amorphous system.

\subsection{A model with very simple structure (Cylinder-model)}
\label{sec:cylinder}

This model, introduced in \cite{Tlusty2017}, assumes that
the coupling between nodes will only depend on one of the two AAs 
linked by a bond.
Although we reformulate the problem differently, it is in fact
equivalent to a lattice model as described above (\sref{sec:lattice}),
in the limit of infinite spring constants (bonds are solid rods). 
%%%In this limit, the deformations are restricted to 
%%%a constant energy subspace. 

\begin{figure}
  \includegraphics[width=\columnwidth]{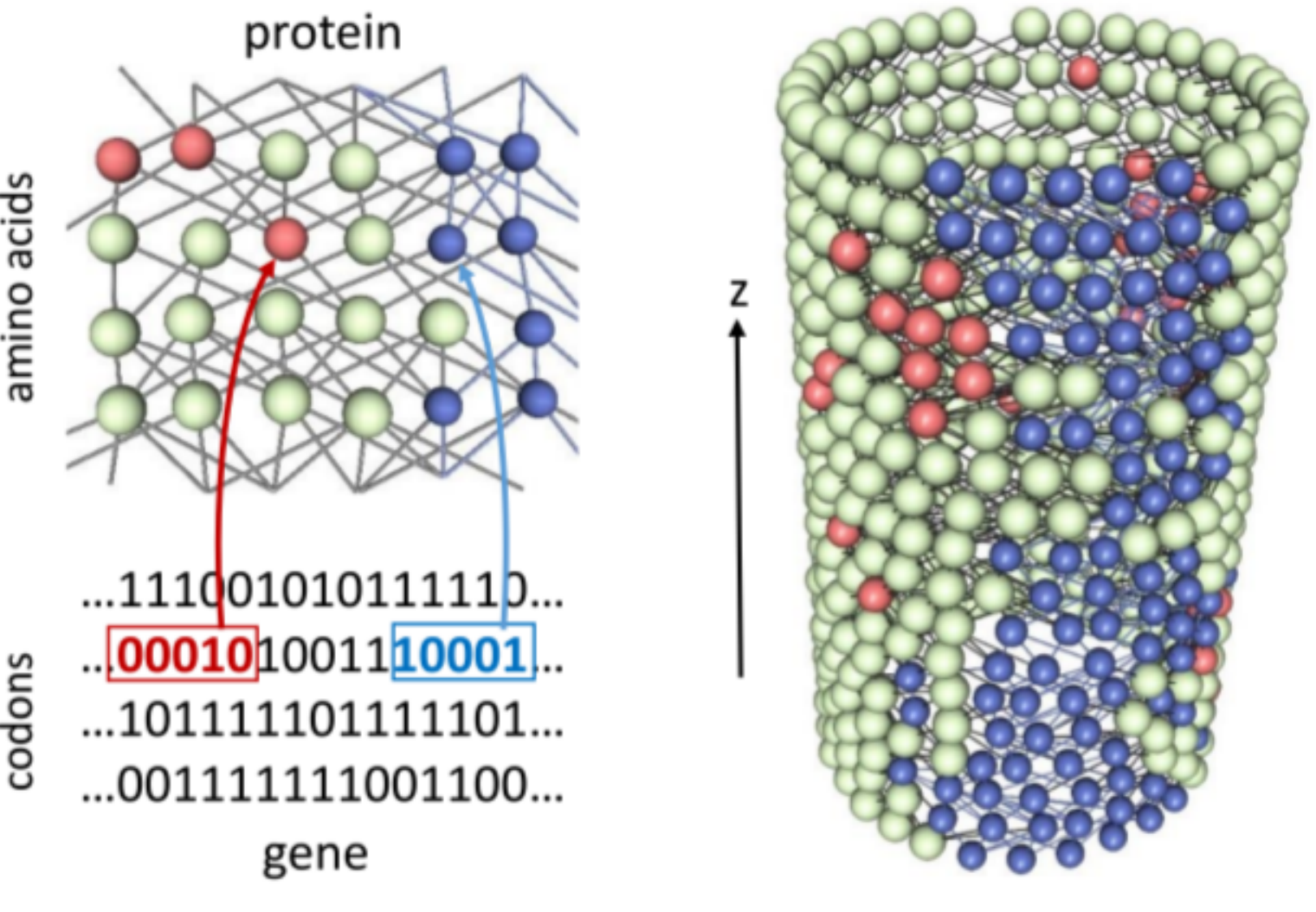}
  \caption{
    \textbf{The main features of the cylinder model}:
    Left: the mapping from the binary gene to the connectivity of the
    amino acid (AA) network that makes a functional protein. 
    The color of the AAs represents their rigidity state 
    as determined by the connectivity according to the algorithm of
    \sref{sec:rigidity}. Each AA can be in one of three states:  
    rigid (gray) or fluid (\ie non-rigid), which are divided between
    shearable (blue) and non-shearable (red). \\ 
    Right: the AAs in the model protein are arranged in the shape of a
    cylinder, in this case with a fluid channel (blue region). 
    Such a configuration can transduce a mechanical signal 
    of shear or hinge-like motion along the fluid
    channel.}\label{fig:fig1}
\end{figure}

To get somewhat closer to the standard genetic code with its 20 AAs, 
we introduce $2^5=32$ species of AAs; each AA is coded by a 5-bit 
codon written in a binary alphabet of 0s and 1s.\footnote{So there are 2 nucleotides and 32 non-redundant codons.}
The geometry of the model is a square lattice with periodic 
boundary conditions in the horizontal direction, forming a cylinder. 
One realization is shown in \fref{fig:fig1} (right) 
where the blue region corresponds to the shear band. 
This should be compared to \fref{fig:gluco} where the shear
band is between the red and green arrows 
(in \fref{fig:glucosymbolic} the shear band is shown in red). 
we shall see later that the motion around shear bands 
in the models is similar in nature to the one of \fref{fig:gluco}.

\subsubsection{The cylindrical amino acid network}

We now define the model in further detail:
We consider a geometry with height $h=18$, \ie the number of layers 
in the $z$ direction, and width $w=30$, \ie the circumference 
of the cylinder. The row and column coordinates of an AA are $(r,q)$, 
with $r$ for the row $(1,\dots,h)$ and $q$ for the column $(1,\dots,w)$.
The cylindrical periodicity is realized by taking the horizontal
coordinate $q$ modulo $w$, $q \rightarrow \textrm{ mod}_w(q-1) + 1$.

Each AA in row $r$ can connect to any of its five nearest neighbors
in the next row below it. This defines $2^5 = 32$ effective
species of amino acids that differ by their `chemistry', 
\ie by the pattern of their bonds. An AA at $(r,q)$ is
encoded in the gene as a $5$-letter binary codon $\ell_{\rck}$, $k=-2,\dots,2$, where the $k$-th letter denotes the 
existence ($= 1$)\ or absence ($ = 0$) of the $k$-th bond. 
The full gene is therefore a binary sequence of length 
$2700 = 5\cdot w\cdot h$. Each of its $w\cdot h=540$ codons 
specifies which of the 5 bonds are present or absent.
The effective size of the problem is only 
$n_s=2550 = w\cdot(h-1)\cdot 5$ because the bonds of the bottom row 
are never used and do not affect the configuration of the protein
and the resulting dynamical modes.

\subsubsection{Evolution searches for a mechanical function }

We next define the evolutionary fitness of the cylindrical protein 
as following: To become functional, the protein has to evolve 
a configuration of AAs and bonds that can transduce a mechanical signal
from a prescribed input at the bottom of the cylinder to
a prescribed output at its top. This signal is a large-scale,
low-energy deformation where one domain moves rigidly
with respect to another in a shear or hinge-like motion, 
which is facilitated by the presence of a fluidized, `floppy' channel
separating the rigid domains \cite{Alexander1983,Phillips1985,Alexander1998}.

The definition of the fluid channel is described in detail in
\sref{sec:rigidity}, but can be summarized by
two features of amino acids in the channel (\fref{fig:fig1}):
(A) Fluidity -- these AAs are not part of rigid sub-networks
in the protein. Locally, this means that 
fluid AAs cannot be linked to too many rigid neighbors.  
(B) Shearability -- the AAs in the channel should have 
enough fluid AAs around them to sustain low-energy shear motion.

The fluidity/rigidity and shearability propagate in a manner   
reminiscent of percolation. Note that, while the system
`learns', through mutations, to form a fluid channel, this learning
is not by presenting it with many inputs, but by only checking the
quality of the output under random mutations.

In \cite{Tlusty2016}, Figure 8, the author imagined a feedback of the
following type: a protein can evolve the ability to activate its own
transcription in response to a stimulus, which is the first step towards
cellular regulatory networks, see \cite{Lee2002} 
for how this appears in the biological context, and \cite[Fig.~8]{Djordjevic2003,Laessig2007,Tlusty2016}, for
theoretical studies.

%%%%%%%%%

\subsubsection{Rigidity propagation algorithm$^*$ }\label{sec:rigidity}

The aim of this subsection is to define a model in which some local
rigidity 
rules, spelled out below, are able to transmit deformability from the
bottom of the cylinder to the top.
There are many ways in which this can be realized, and the
rules we give are a compromise between simplicity and the ability to
fulfil this aim. We have tested other variants with similar outcome.

\begin{figure}
  %%\centering
 \includegraphics[width=1.0\columnwidth]{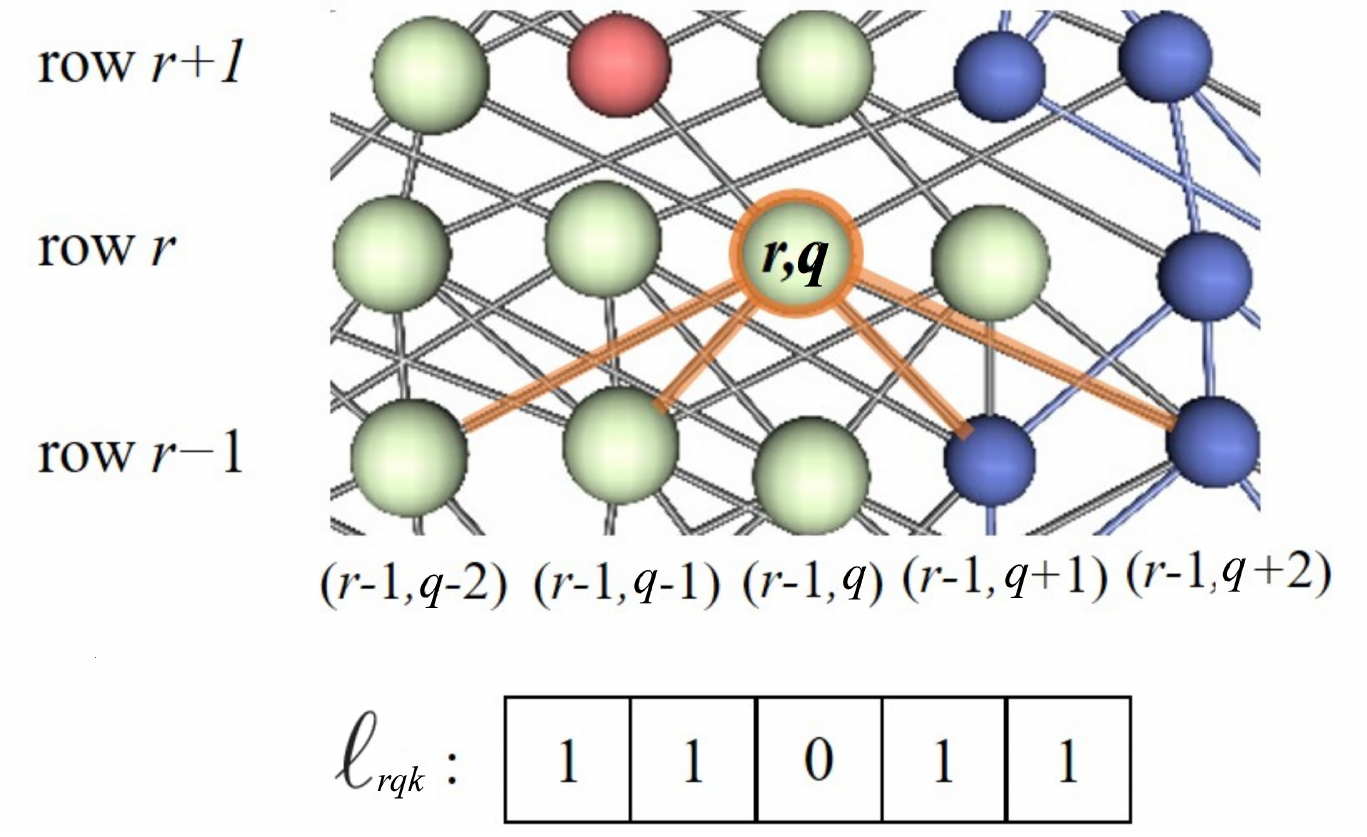}
 \caption{
Illustration of the percolation rules for
  shearability and rigidity states. Note that site $(r,q)$ was
  turned solid because it is attached to 2 solid sites below it. 
  Also note that the red site above it is fluid, 
  because it is attached to less than 2 solid sites below it. 
  But it is not shearable because it does not connect 
  to a shearable site below it. On the other hand, the top right site 
  is shearable and fluid, since it is attached to only one solid site
  (namely $(r,q)$) and no others on the invisible part of the structure 
  (as seen by its blue connections), and it is also  connected to the blue site at $(r,q+2)$.}\label{fig:st}
\end{figure}

The large-scale deformations are governed by the rigidity pattern of
the protein, which is determined by the connectivity of the AA
network via a simple majority rule (\fref{fig:fig1}, \ref{fig:st}), 
as follows. These large scale deformations could in principle change
the ability of the protein to bind a target, and in this way implement the response
trigger in the feedback
loop mentioned above. Each AA position will have two binary properties 
which define its state:
\emph{rigidity} $\sigma$ (an AA is either \emph{solid}, 
$\sigma=1$, or \emph{fluid}, $\sigma=0$) and 
\emph{shearability} $s$ (an AA is either \emph{shearable}, 
$s=1$, or \emph{non-shearable}, $s=0$). 
Only 3 of the 4 possible combinations are allowed:
\begin{myenum}
 \item non-shearable and solid (yellow): $\sigma = 1; s = 0$,
 \item non-shearable and fluid (red): $\sigma = 0; s = 0$,
 \item shearable and fluid (blue): $\sigma = 0; s = 1$.
\end{myenum}
Non-shearable protein domains tend to move as rigid bodies 
(\ie via translation or rotation), whereas shearable regions 
are easy to deform. The non-shearable domains are mostly rigid, 
but can still have pockets of fluid AAs.

Given a fixed sequence and an input state in the bottom row of the
cylinder, $\{\sigma_{1,q}$, $s_{1,q} \}$ the state of the cylinder is
completely determined  by `percolation' of the two properties,
rigidity/fluidity and shearability, through the network, as follows.

In a first sweep through the rows, we establish the rigidity property
$\sigma$. The rigidity of AAs in row $r = 1$ are prescribed initially.
In all other rows ($r = 2,\dots,h$) the bonds determine the value of 
the rigidity of $(r,q)$ through a majority rule:
\begin{equ}\label{eq:1}
\sigma_{r,q}= \theta \left( \sum_{k=-2}^2 
\ell_{\rck} \sigma_{r-1,q+k} - \sigma_0 \right),
\end{equ}
where $\theta$ is the step function ($\theta(x \ge 0) =1$,
$\theta(x<0)=0$)). The parameter $\sigma_0 = 2$ is the minimum number
of rigid AAs from the $r-1$ row required to rigidly support an AA:
In 2D each AA has two coordinates which are constrained if it is
connected to two or more static AAs. 
In this way, the rigidity property of being pinned in place propagates
through the lattice as a function of the initial row and of
the bonds as encoded in the gene. 

We next address the shearability property $s$ which is
determined by the rigidity as follows: We assume
that all fluid AAs in row $r=1$ are also shearable 
(blue: ($\sigma = 0; s = 1$)). A fluid node $(r,q)$ in row $r$ 
will be shearable if any of its neighbors at $(r-1,q)$ or
$(r-1,q\pm1)$ is shearable:
\begin{equ}\label{eq:2}
s_{r,q}= \left(1-\sigma_{r,q}\right)
\cdot\theta \left( \sum_{k=-1}^1 {s_{r-1,q+k}} - s_0 \right),
\end{equ}
where $s_0 = 1$. The first factor on the r.h.s.\ 
ensures that a solid AA is never shearable. 

\subsubsection{Fitness and mutations}

\begin{figure}
  \includegraphics[width=\columnwidth]{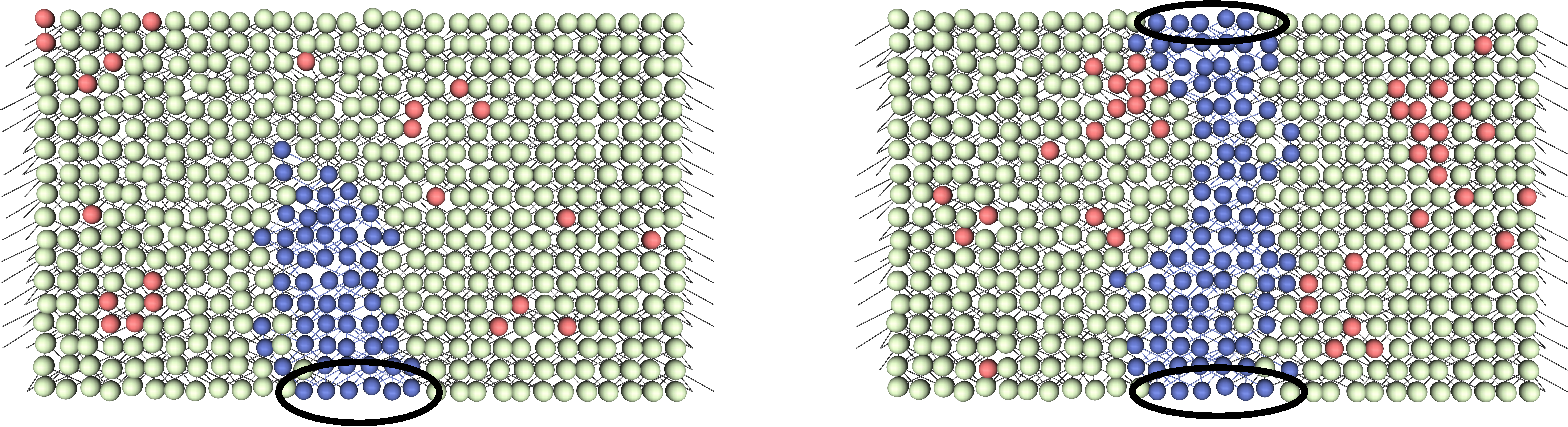}
\caption{
\textbf{Evolution of a mechanical function:}\\
A configuration (left) with a prescribed input (black
ellipse at bottom) and random connectivity pattern
eventually evolved to form a fluid channel (right). 
The initial state has 6 fluid points (in black ellipse), 
our fitness requires 5 fluid points at the top (black ellipse).}
\label{fig:fig1b} 
\end{figure}

As explained before, evolution searches for a functional protein which
can transfer forces. The simulation of this search starts from a random
sequence (of 2550 codons), and from an initial state (input) in the bottom
row of the cylinder. For most simulations, this initial state
consisted of only solid beads except a stretch 5 consecutive shearable
beads, as shown in \fref{fig:fig1b}.

We next define a fitness function which will direct the
evolutionary process. The state with maximal fitness (\ie the `target')
is a chain of $w$ values, fluid and shearable ($\sigma = 0; s = 1$)
or solid ($\sigma = 1; s = 0$), in the top row, which the protein
should yield as an output: we call it $x^*\equiv\{\sigma^*_{q}$,
$s^*_{q} \}_{q=1,\dots,w}$. Given
\begin{myenum}
\item a gene sequence $\cb$, which determines the connectivity $\ell_{\rck}$,
\item the input state, $\{\sigma_{1,q}$,$s_{1,q} \}_{q=1,\dots,w}$,
\end{myenum}
the algorithm described above uniquely defines the
output state in the top row, $\{\sigma_{h,q}$, $s_{h,q} \}_{c=1,\dots,w}$.
At each step of evolution, the output state is compared to 
the fixed target by measuring the Hamming distance\footnote{The Hamming distance between two sequences is the
number of indices where they differ.} to the target $x^*$:
\begin{equ}\label{eq:3}
F = -w+\sum_{q=1}^w\left(1-\lvert s_{h,q} - s_{q}^* \rvert\right)\cdot
\left(1-\lvert \sigma_{h,q} - \sigma_{q}^* \rvert\right)~.
\end{equ}
This is the fitness function $F$ of \sref{sec:simulating}.

\textit{Remark}: It is an important feature of this model
that the fitness of the network is only measured 
at the target line. This corresponds to the biological fact
that the protein can only interact with the outside world
through its surface (in our case, the ends of the cylinder).
One of the major outcomes of the model is that this fitness
still has a strong influence on the connectivity deep inside the
interior of the protein. 
While similar, the propagation of fluidity should not be confused
with learning in neural networks: In the learning case, the system is
presented with several inputs and learns to recognize others,
while here there is a fixed task, and the connections 
are only driven by the target function $F$.

\subsubsection{Simulation of evolutionary dynamics}

Thanks to the simplicity of the model, one can easily perform $10^6$
simulations in a short time, and gain much better statistical
insight than is possible with typical bioinformatic data 
(of course, at the price of disregarding many biochemical details).
We present results for one specific fitness:
the input at the bottom is a fluid region of length 6 and output target
at the top is a fluid region of length 5. For other variants of this
model, \textit{cf}.~\cite{Tlusty2017}.

We study 200 independent initial states (genes), starting from a
random sequence with about $90\%$ of the bonds present at the start. 
Given a sequence, we sweep according to the rules
of \eref{eq:1}--(\ref{eq:2}) through the net, 
and measure the Hamming distance $F$ (\eref{eq:3}) 
between the last row and the desired target. 

Solutions are then searched by successive mutations, 
with a Metropolis algorithm, \cite{Metropolis1953}. 
At each iteration, a randomly drawn digit in the gene 
is flipped, that is, the values of 0 and 1 are exchanged. 
This corresponds to erasing or creating a randomly
chosen link of a randomly chosen AA.  After each flip, 
a sweep is performed, and the new output at the top row
is again compared to the target. 
If $F$ (which is negative) decreases, we backtrack 
and flip another randomly chosen bond. This procedure 
is repeated until optimal fitness is reached $(F=0)$. 
This will happen with probability 1 if such a state exists,
and typically requires $10^3$-$10^5$ mutations. 

\begin{figure}
 %%\centering
  \includegraphics[width=0.7\columnwidth]{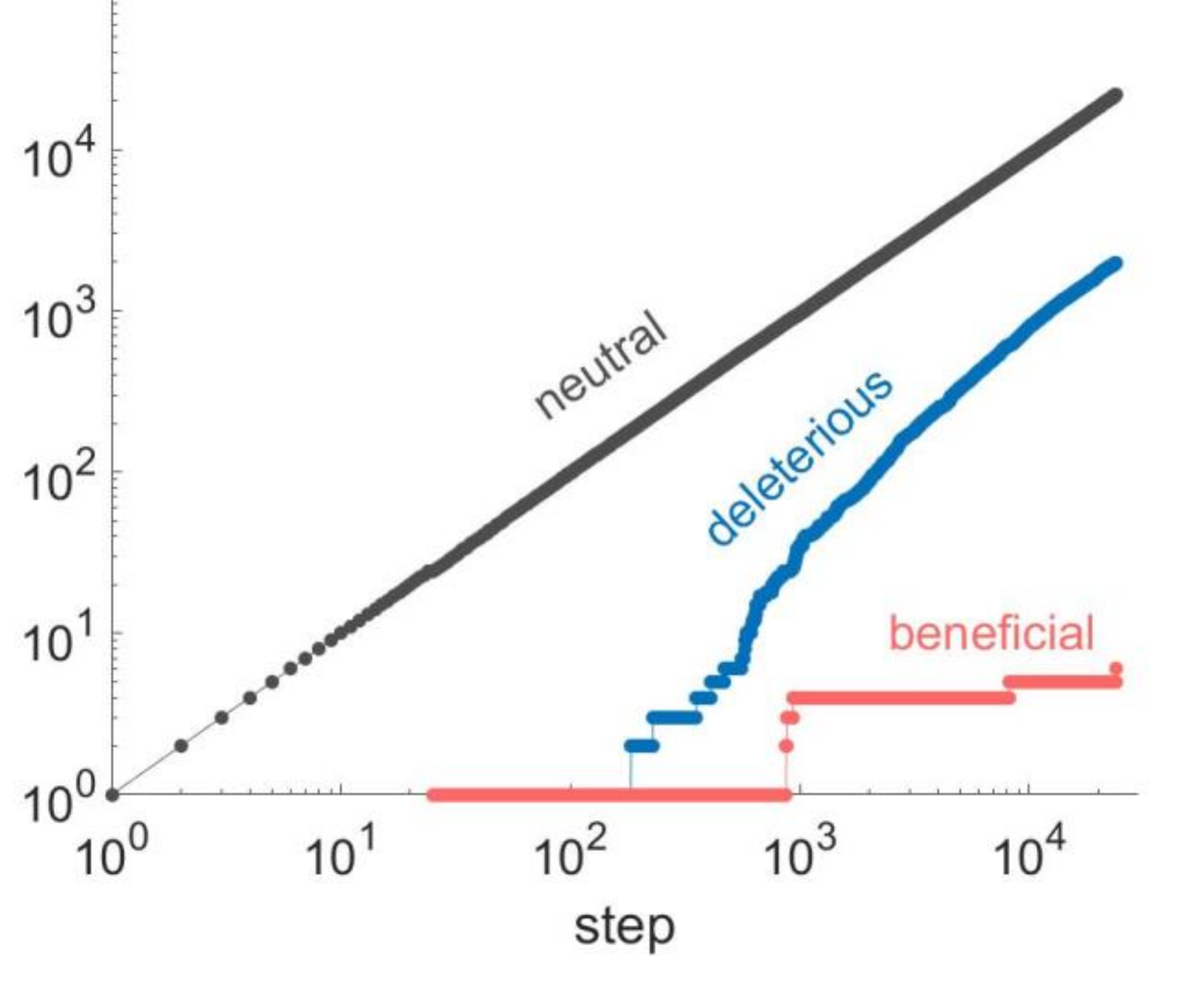}
 \caption{Following the progress of evolution during a typical run.
  It is a sequence of mostly neutral steps, a fraction of deleterious
  ones, and rare beneficial steps. The vertical axis is the 
  accumulated number of steps of each type.}\label{fig:deleterious} 
\end{figure}%

Although the functional sequences are extremely sparse among the
$32^{510}=2^{2550}$ possible sequences, the small bias 
for getting closer to the target in configuration space 
directs the search rather quickly.

Once a maximal $F$ is reached, we move away from it  
by further mutations and then look again for a new optimum.
Reaching a state with $F = 0$ takes around 11.2 beneficial mutations 
on average. Getting from an initial sequence to a maximizer 
is called a `generation'. For each of the 200 initial random genes, 
we followed 5000 generations, finding a total of $10^6$ optima.
The typical length of a generation between two maxima 
is about 1500 mutations (most of them neutral, see
\fref{fig:deleterious}),
similar to the time it takes 
starting from a totally random gene.
We also simulated $1$-generation paths starting from $10^6$ random genes.
The two cases are very similar, but the destruction-reconstruction
simulations show some correlations between consecutive generations,
which disappear after about 4 generations.

\subsection{A model with more realistic interactions (HP-model)}\label{sec:pnasmodel}

This model differs from the cylinder-model in several respects:
\begin{myenum}
\item Geometry -- The lattice is hexagonal with open boundary conditions,
\item Two-body interactions -- the bonds depend on the nature of the AA at \emph{both} ends
  of the link,
\item Amino acids species -- There are only 2 species, $\aH$ (hydrophobic) or $\aP$ (polar).
\end{myenum}
The two amino acids species are encoded in a binary genetic alphabet
and a codon size of 1: each AA chain is encoded in a gene $\cb$
of the same length $n_a$, where $c_i=1$ encodes $\aH$ and $c_i=0$
encodes $\aP$, $i=1,\dots, n_a$ 
(in other words, the genetic code is the identity map).

\begin{figure}
%%\centering
\includegraphics[width=0.6\columnwidth]{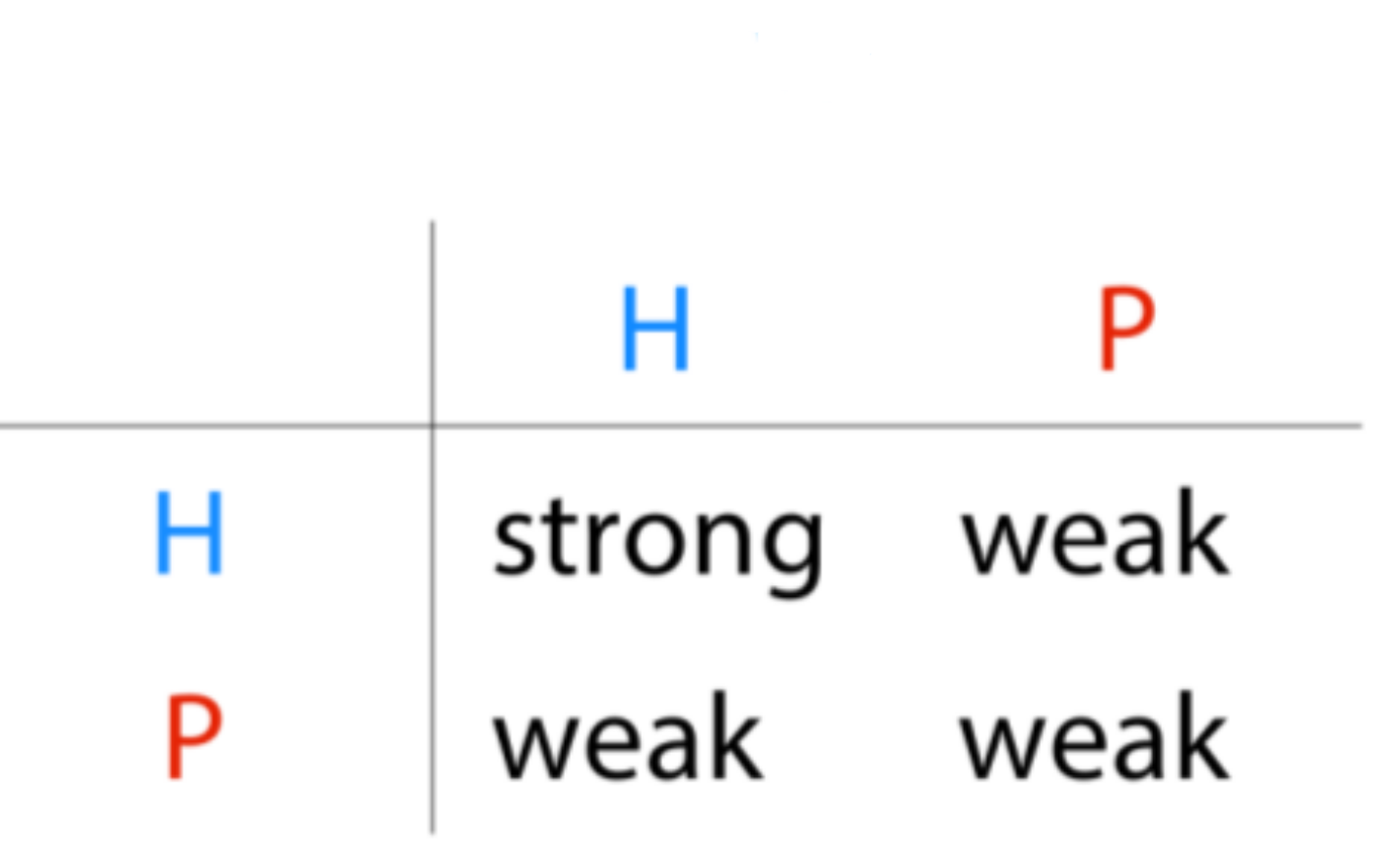}
\includegraphics[width=0.33\columnwidth]{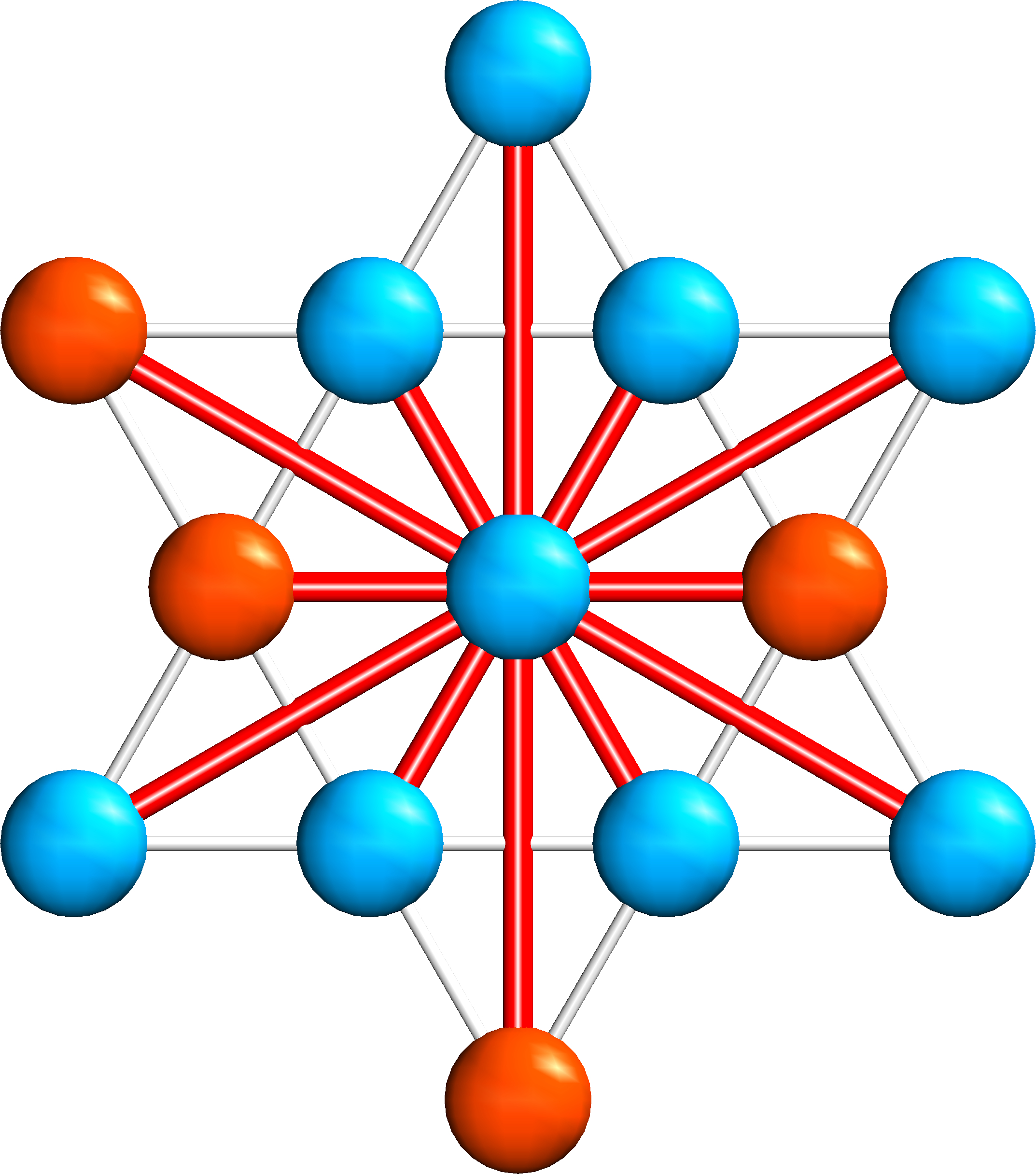}
\caption{
  The protein is made of two species of AAs, polar ($\aP$, red) 
  and hydrophobic ($\aH$, blue) whose sequence is encoded in a gene. 
  Each AA forms weak or strong bonds with its 
  $12$ nearest neighbors on the hexagonal lattice (right)
  according to the interaction rule in the table (left).}
\label{fig:1}\label{fig:lattice}
\end{figure}

We give next details of how the model is constructed.
The lattice has width $h=20$ and height $w=10$, and therefore
$n_a=200$ AAs (see \sref{sec:lattice}). The bonds stretch 
over the 12 nearest and next-nearest neighbors of an AA 
(see \fref{fig:lattice}, right panel, for the connectivity 
and any of the panels in \fref{fig:pinch} for the
global arrangement of the lattice).

The strength of the springs is given by the HP model \cite{Lau1989}
according to the rule:
\begin{equ}
  k_{(i,j)} = \kw + (\ks -\kw ) c_i c_j~,
\end{equ}
where $c_i$ and $c_j$ are the (binary) codons of the AA 
connected by bond $\alpha=(i,j)$, with $c_i=1$ corresponding 
to $\aH$ and $c_i=0$ to $\aP$. This implies that 
a strong $\aH{-}\aH$ bond has $k_\alpha=\ks $, 
whereas the other bonds $\aP{-}\aP$, $\aH{-}\aP$, and
$\aP{-}\aH$ are weak with $k_\alpha=\kw $.
In our simulations below we used $\ks =1$ and $\kw =0.01$.

\subsubsection{Pinching the network}\label{sec:pinch}

The network is subjected to an external `pinch' which is 
a localized force $\fb$ applied at the boundary of the network. 
This force acts in the complement of the subspace 
of Galilean invariance, \ie $\PP\fb=0$ (see \sref{sec:green}).
The pinch acts on a pair of neighboring boundary vertices, 
$p'$ and $q'$, in the direction $\nb_{p'q'}=\rb_{p'}-\rb_{q'}$
which is parallel to the boundary (the L face of the network).
Thus $\fb$ is a force dipole, \ie  two opposing forces,
$\fb_{q'}=-\fb_{p'}=f\cdot \nb_{p'q'}$, 
which can be related to the deformation $\ub$ of
the network by Green's function \eref{eq:gf}.
The pinch stimulus may represent localized interactions, 
for example a ligand biding at a specific binding site
(\fref{fig:glucosymbolic}).

The biological fitness is specified by how well the response
of the network fits to a prescribed deformation vector $\vb$.
This vector is zero except at $p$ and $q$ which are
%on the same columns as $p'$, $q'$ but 
on the opposite side of the network (R face).
The fitness function $F$ is therefore
\begin{equ}\label{eq:fitness}
  F=\vbT\ub  =\vbT \GG \fb=\vb_p\ub_p+\vb_q\ub_q~. 
\end{equ}

Note that \eref{eq:fitness} is a specific way of defining $F$, 
adapted to the phenomenology of building a fluid channel that can transmit
allosteric interaction between two specific sites. 
Other choices of $\fb$ and $\ub$ could be treated similarly, 
such as multi-site force patterns 
and multi-domain dynamical modes. For example, if the response can occur at any $(p,q)$ site at the R face, the model may describe the emergence of
induced fit or conformational selection mechanisms (\sref{sec:views}). 
To model the emergence of specific recognition, one sets as target strong response to a stimulus $\fb$, but hardly any response to a similar `competitor' stimulus $\tilde{\fb}$.

\begin{figure}
  %%\centering
  \includegraphics[width=1.0\columnwidth]{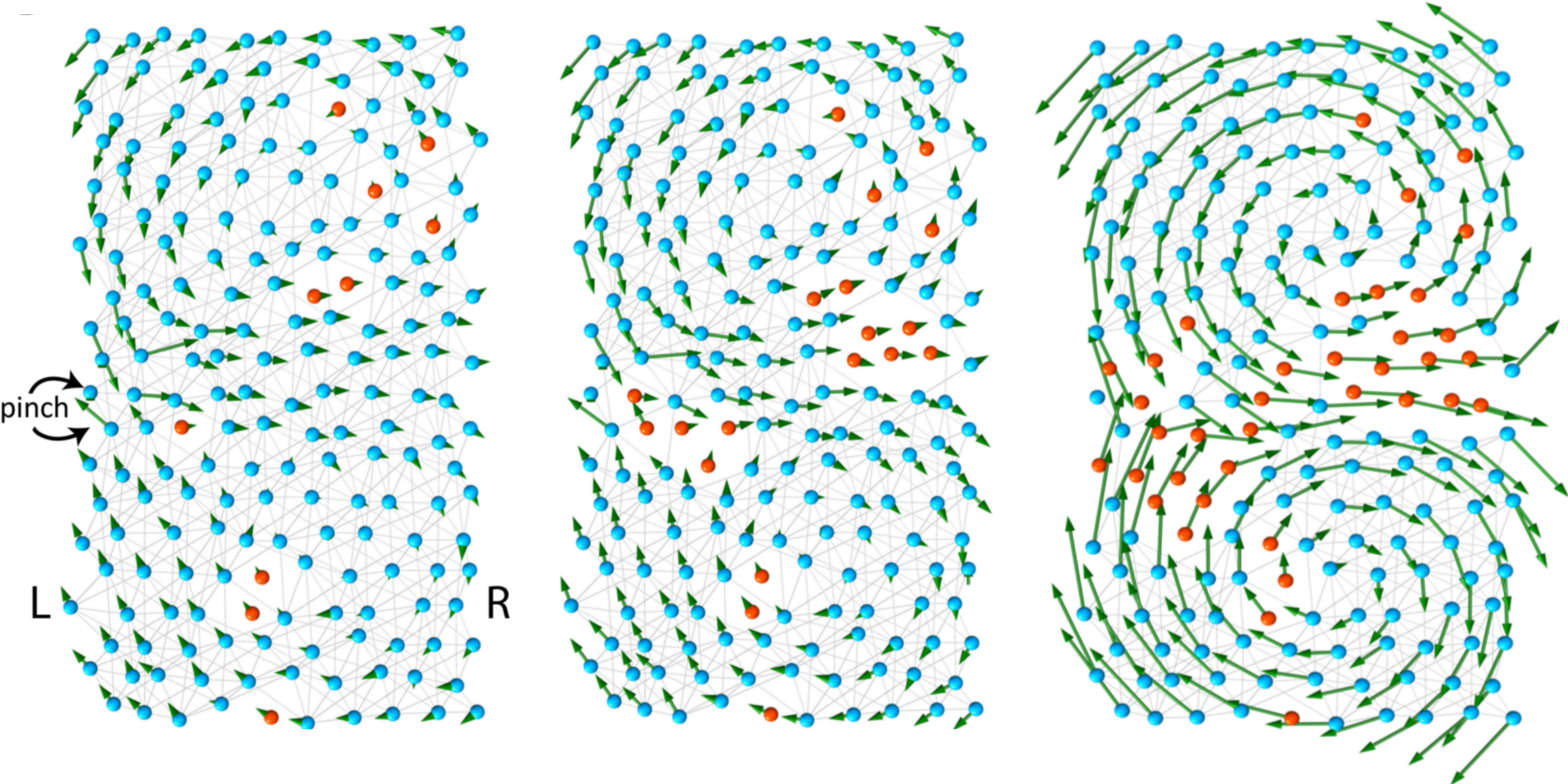}
  \caption{Illustration of the deformation field,
    $\ub(\cb)=\GG(\cb)\fb$, \eref{eq:gf} 
    for three choices of $\cb$. The force
    $\fb=f\cdot\nb_{p'q'}$ is applied on the left of the lattice. 
    The three panels show, from left to right, how the response
    $\ub$ (shown in small arrows) evolve as the network fitness  
    $F=\vb^\T\,\ub(\cb)$, increases. 
    The choice of $\vb$ corresponds to the vertical separation
    of the two central points on the right.}
  \label{fig:pinch}

\end{figure}%

In the spirit of the Metropolis algorithm used in the cylinder model,
one exchanges randomly AAs between $\aH$ and $\aP$ while looking
for changes in the fitness $F$. A gene $\cbs$ is considered 
a solution if $F$ exceeds a certain large value.\footnote{It is not reasonable to ask for $F=\infty$, but it
  suffices to look for $F>F_\textrm{crit}$. In our case, 
  $F_\textrm{crit}=5$ is a good choice since in general, the channel 
  will have already formed, and increasing $F_\textrm{ crit}$ 
  will only enlarge the channel somewhat.}
\fref{fig:pinch} illustrates the vector field $\ub$ of the 
deformation for three genes $\cb$, along an evolutionary 
trajectory, improving the fitness value $F$ from left to right.

\subsubsection{The protein backbone}\label{sec:backbone}

One hallmark of proteins is that they are made from a long 
chain of amino acids connected by strong covalent bonds, 
called a backbone (see \fref{fig:backbone}). 
This backbone is then folded in an intricate way 
to form the protein, but the chain is not broken.
Here, we assume that the folding process is just given 
and that the mutations we consider are moderate enough
so that they do not change the general folding.
Given this restriction, one still can ask whether 
the existence of the backbone affects such studies. 
From a conceptual point of view, having a backbone 
just means that some springs in the lattice are much stronger
than the others, and therefore, it is not surprising 
that adding a backbone does not change the general picture.

\begin{figure}[htb]
%%\centering
  \includegraphics[width=1\columnwidth]{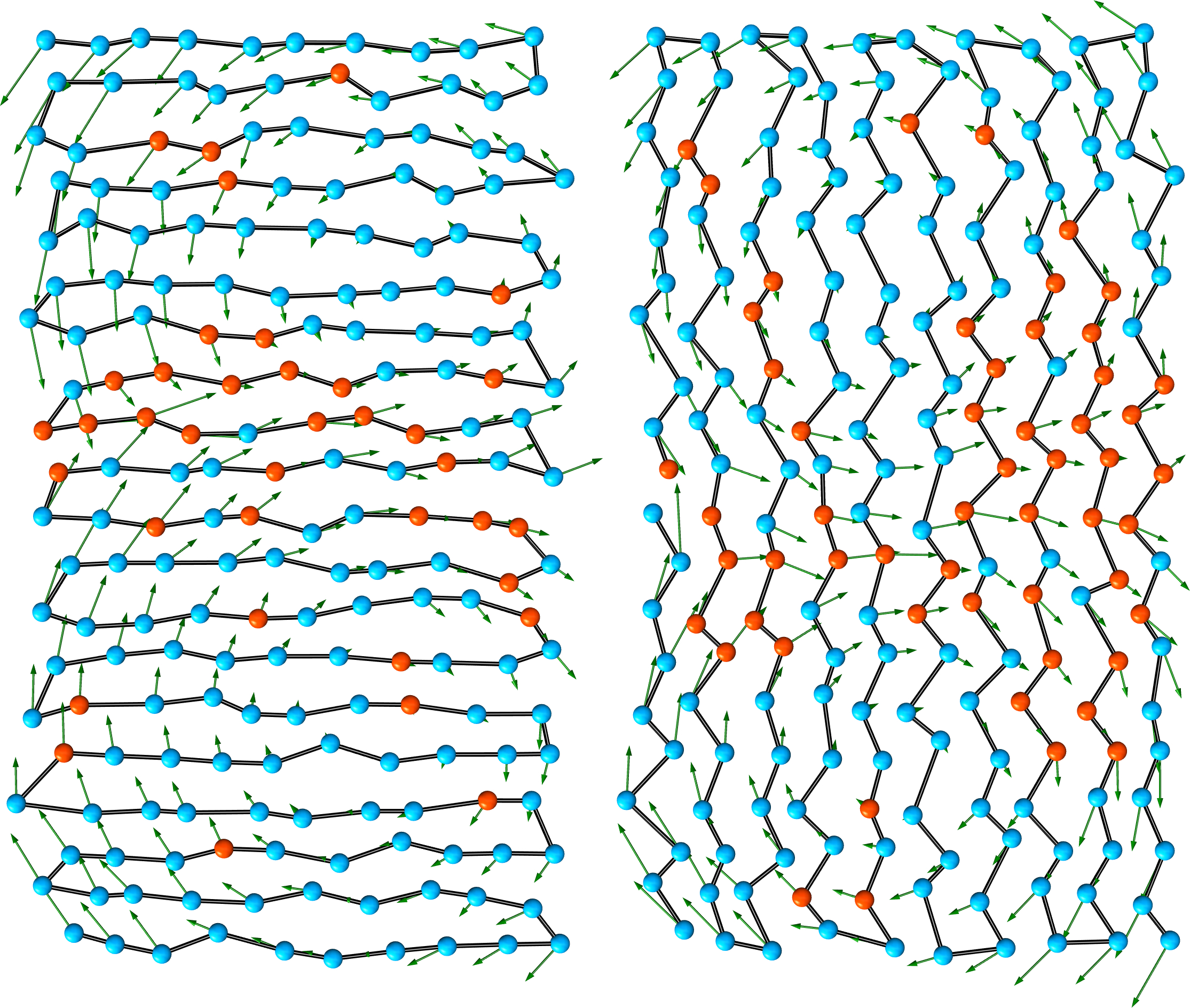}
\caption{\textbf{Illustration of the backbone.}
The backbone is shown as solid black serpentine curve. 
AAs in neighboring sites along the backbone tend to move together. 
We show two configurations: parallel to the channel (left)
and perpendicular to the channel (right).
Parallel: The backbone favors the formation of 
a narrow channel along the fold (compared to \fref{fig:pinch}).
Perpendicular: The formation of the channel 
is `dispersed' by the backbone.}
\label{fig:6}
\end{figure}%

In \fref{fig:6} we show two extreme cases, a serpentine backbone
either parallel to the shear band or perpendicular to it.
The presence of the backbone does not interfere with the emergence
of a low-energy mode of the protein whose flow pattern 
(\ie displacement field) is similar to the backbone-less 
case with two eddies moving in a hinge-like fashion. 
In the parallel configuration, the backbone constrains the channel
formation to progress along the fold (\fref{fig:6}, left).  
The resulting channel is narrower than in the model without backbone
(\fref{fig:pinch}).
In the perpendicular configuration, the evolutionary progression of
the channel is much less oriented (\fref{fig:6}, right).

We expect that, in a realistic 3D geometry, the backbone will have a
weaker effect than what we observed in 2D networks, since the extra
dimension adds more options to avoid the backbone constraint.

\subsubsection{Pathologies and broken networks$^*$}

As our criterion for evolution is the floppiness (large eigenvalue of
Green's function), there is of course the trivial case 
where the network is just broken in two disjoint pieces 
or into pieces with dangling ends.
Such broken networks exhibit floppy modes owing to the 
low energies of the relative motion of the disjoint domains 
with respect to each other. Any evolutionary search 
might end up in such non-functional unintended modes. 
The common pathologies one observes are:
\begin{myenum}
  \item isolated nodes at the boundary that become 
  weakly connected via $\aH {\to }\aP$ mutations,
  \item `sideways' channels that terminate outside 
  the target region (which typically include around $8{-}10$ sites),
  \item channels that start and end at the target
  region without connecting to the binding site.
\end{myenum}
All these are floppy modes that can vibrate independently of
the pinch and cause the response to diverge (${>}F_\textrm{crit}$)
without producing a functional mode. To avoid such
pathologies, we apply the pinch force symmetrically:
pinch the binding site on face L and look at responses on face R
and vice versa. Thereby we not only look for the transmission 
of the pinch from the left to right but also from right to left. 
The basic algorithm is modified to accept a mutation only
if it does not weaken the two-way response and enables hinge motion
of the protein. This prevents the vibrations from being localized at
isolated sites or unwanted channels.
Of course, the presence of a backbone (see \sref{sec:backbone} 
and \fref{fig:6}) will make disconnection of the network 
more difficult. This is also a more realistic model.

One may also impose a stricter minimum condition, 
$\DF \ge \varepsilon\ F$ with a small positive $\varepsilon$, 
say $1\%$. An alternative, stricter criterion would be
the demand that each of the terms in $F$, $\vb_{\ui}\ub_{\ui}$ 
and $\vb_{\uj}\ub_{\uj} $, increases separately.

%%%%%%%%%%%%%%%%%%%%%%%%%%%%%%%%%%%%%%
% SECTION

\section{Connecting the models to biological concepts}

The theoretical methods introduced earlier lead to 
a family of easily implementable numerical simulations. 
We discuss these simulations and show that they 
can explain several basic observations from the biology literature. 
They also suggest new connections to be explored.
The main idea is that mechanical properties of the protein
constrain the genetics in multiple ways.

Each subsection introduces a technique of analysis, 
application to one of the models described earlier, 
and an interpretation in terms of biological questions.

%%%%%%%%%%%%%%%%
\subsection{Dimension of the solution set in the genotype and phenotype spaces }\label{sec:dimension}

We describe the set of solutions for the cylinder-model of
\sref{sec:cylinder}. The (genotype) sequence space is $\{0,1\}^{2550}$
(see \fref{fig:fig1}, bottom left). One can view 
this space as a 2550-dimensional hypercube, with $2^{2550}$ corners, 
and any flip of a digit will move along an axis from one corner 
to another (the dimension of a hypercube equals 
the number of directions in which one can move from a corner). 
The space of configurations (phenotypes) is the arrangements of colors 
(see \fref{fig:fig1}, top left), which is $\{0,1,2\}^{540}$, 
since there are 3 colors (red, blue, yellow).

\begin{figure}
  %%\centering
  \includegraphics[width=1.0\columnwidth]{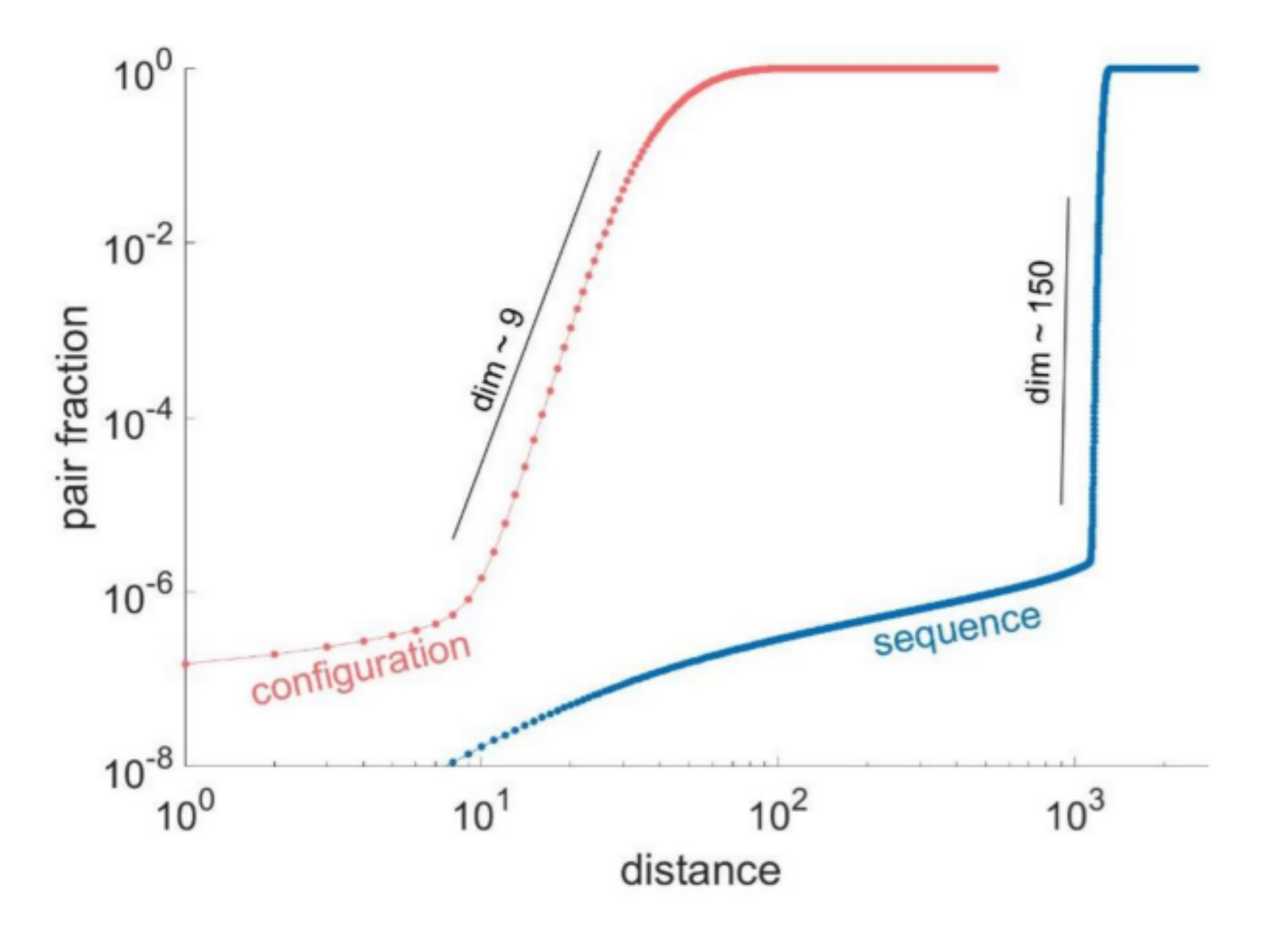}
\caption{
\textbf{Dimensional reduction of the genotype-to-phenotype map:}\\
Dimension measurement from $10^6$ independent configurations 
(phenotypes) for the cylinder-model. The dimension of 
the configuration space is about 9 (red curve), 
while in sequence space it is basically infinite (blue curve). 
All pairs seem to have the same distance, namely
$1275=2550/2$, which is the typical distance between two random sequences.
}\label{fig:fig2} 
\end{figure}%

The set $\mathcal S$ of solutions to the mutation problem
is a subset of this hypercube (\fref{fig:fig2}). 
To determine the dimension of experimental data, 
with large sample size, it is convenient to use
the box-counting algorithm \cite{Grassberger1983}. 
First, one counts the number $N(\rho)$ of pairs of points
in $\mathcal S$ at Hamming distances $\le \rho$, \ie with not more
than $\rho $ changes.
One then plots $\log  N(\rho)$ vs $\log \rho$ and the dimension
is the slope in this log-log plot, as indicated by black lines
in \fref{fig:fig2}. We see that the dimension in the
space of configurations (phenotypes) is about 8-9, while, 
in the space of sequences (genotypes), the dimension 
is basically `infinite', namely just limited by 
the maximal slope one can obtain \cite{Procaccia1988} 
from the $10^6$ simulations.\footnote{For explanation of the flat
  pieces of the graph, see \cite[p.~647]{Eckmann1985}.} 
It would be interesting to discuss this problem 
as a special case of the problem of hitting times 
of small sets in hypercubes (these hitting times are
usually exponentially distributed). The novelty 
in the current context is the use of a very small drift, 
namely, we do not allow steps which increase the distance
to the set $\mathcal S$.\footnote{We thank
G.~Ben Arous for helpful discussions on this point.}

The dramatic dimensional reduction in mapping genotypes to phenotypes
stems from the different constraints that shape them
\cite{Savir2010,Savir2013,Kaneko2015,Friedlander2015}. 
In the phenotype space, most of the protein is rigid, 
and only a small number of shear motions are low-energy modes, 
which can be described by a few degrees-of-freedom. 
In the genotype space, in contrast, there are many
neutral mutations which do not affect the motion of the protein.

The biological interpretation is that
\textbf{the gene is much more random than the phenotype of  the protein it forms}. 
However, we shall see below, in particular in \sref{sec:similarity}, 
that the gene still needs to be quite precise in certain 
well-defined positions. In the case of allosteric proteins
these critical positions are the hinges and other locations 
of strong stress.

\subsection{Expansion of the protein universe}\label{sec:expansion}

\begin{figure}
  %%\centering
  \includegraphics[width=1.0\columnwidth]{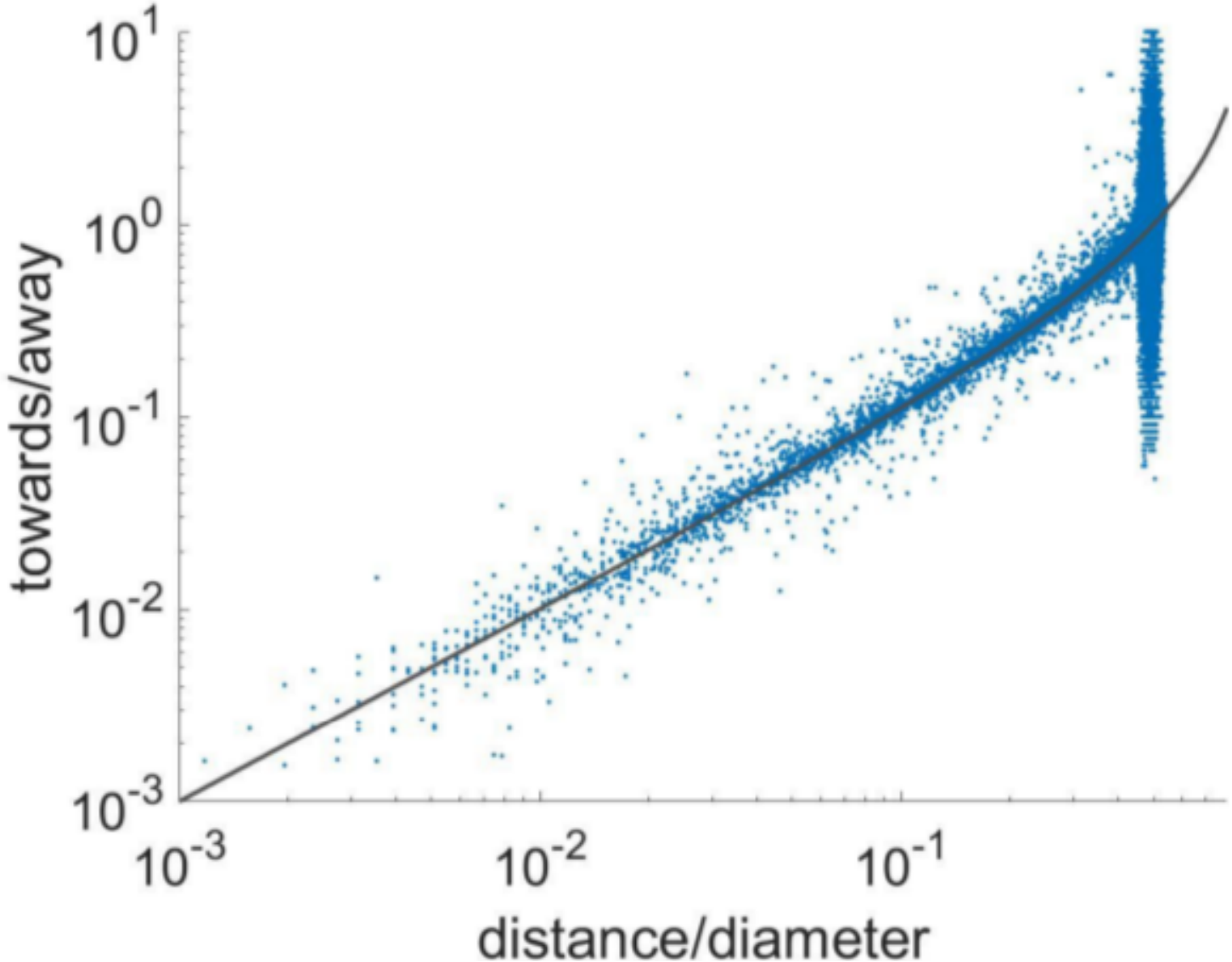}
  %%  /disk/home/eckmann/tsvi/sheargenedistances
  \caption{
    \textbf{Distribution of solutions in the sequence universe: }\\
    A measure for the expansion of functional genes in the sequences
    universe is the backward/forward ratio, the fraction 
    of point mutations that make two sequences closer 
    vs.~the ones that increase the distance \cite{Povolotskaya2010}. 
    The Hamming distances $D$ (normalized by the universe diameter
    $d_{\max}= 2550$) show that most sequences reach the edge 
    of the universe, where no further expansion is possible. 
    The black curve, $D/(1-D)$, is the backward/forward ratio 
    of purely random mutations. Given the overwhelming number
    of samples near the maximal $D$, the Gaussian distribution
    is well visible (in the vertical direction).
  }\label{fig:fig2b} 
\end{figure}

Here, we test the cylinder-model against the ideas of
\cite{Povolotskaya2010}. Our results will give some insight about 
the nature of the set of solutions, \ie genes of functional proteins. 
In \cite{Povolotskaya2010}, the authors consider any two solutions
with gene sequences $\textbf{s}_1$ and $\textbf{s}_2$.
They ask how much the solution $\textbf{ s}_3$, 
one generation after $\textbf{ s}_2$, differs from $\textbf{ s}_1$, 
and define the following observable:

%Let $w_i=1$ if $s_{1,i}=1$ and $w_i=-1$ if
%$s_{1,i}=0$, for all $i$.
Let $x_i=(2s_{1,i}-1)\cdot (s_{3,i}-s_{2,i})$ (since $s_{1,i}\in\{0,1\}$,
$x_i>0$ if the change between $s_{3,i}$ and $s_{2,i}$  
is towards $\textbf{ s}_{1}$ and $x_i<0$ otherwise).
%In \fref{fig:fig2b}, we plot the analogous result for the
%cylinder-model, normalizing the Hamming distances by dividing by 2550.
Finally, $N_{\textrm{ away}}=\#\{i\,:\,x_i<0\}$ and
$N_{\textrm{ towards}}=\#\{i\,:\,x_i>0\}$.
In \fref{fig:fig2b}, the ratio $N_{\textrm{ towards}}/N_{\textrm{ away}}$
is plotted as a function of the distance $D$ between $\textbf{s}_1$ 
and $\textbf{s}_2$, normalized by the diameter $d_{\max}=2550$. 
The interested reader will notice the similarity to Fig.~3 in
\cite{Povolotskaya2010}: In their case, because of the small number of
experimental samples, they only see the low-$D$ region
of \fref{fig:fig2b}, far from the diameter of the `protein universe'.

The set of solutions is a very dilute, but complex, subset $\mathcal S$ 
of the hypercube. The search for a good gene corresponds 
to a slightly biased random walk along path of monotonically increasing
fitness ($\delta F\ge0$). While we do not have a good mathematical
description of such intricate walks, we can compare them 
to the null model of purely random walks. In this case,
one gets a simple expression for the towards/away ratio, 
as a function of $D$, the normalized Hamming distance: 
$D/(1-D)$, which is shown as black curve \fref{fig:fig2b}
(\ie $D$ is the proportion of sites which differ between 
the pair of solutions $\textbf{ s}_1$ and $\textbf{ s}_2$).
The good fit shows that the fitness-constrained evolutionary paths
expand \emph{as if one performed a random walk on the full cube}.

It is interesting to note: First, that this result 
must be intimately connected to the high dimension of the problem, 
since for low dimensional hypercubes it does not hold. 
Second, most samples are near the edge $D=1$ of the universe,
where the Hamming distances among the sequences are close to
the typical distance between any two random sequences. 
To conclude: {\bf While maintaining functionality, 
the divergence of acceptable gene sequences has 
all aspects of a random walk (on a hypercube)}.
This conclusion is close to the `expansion of the protein universe' 
(in honor of E. Hubble), described in \cite{Povolotskaya2010}.

\subsection{Spectrum in phenotype and genotype spaces}\label{sec:spectrum}

\begin{figure*}
%%\centering
\includegraphics[width=0.7\textwidth]{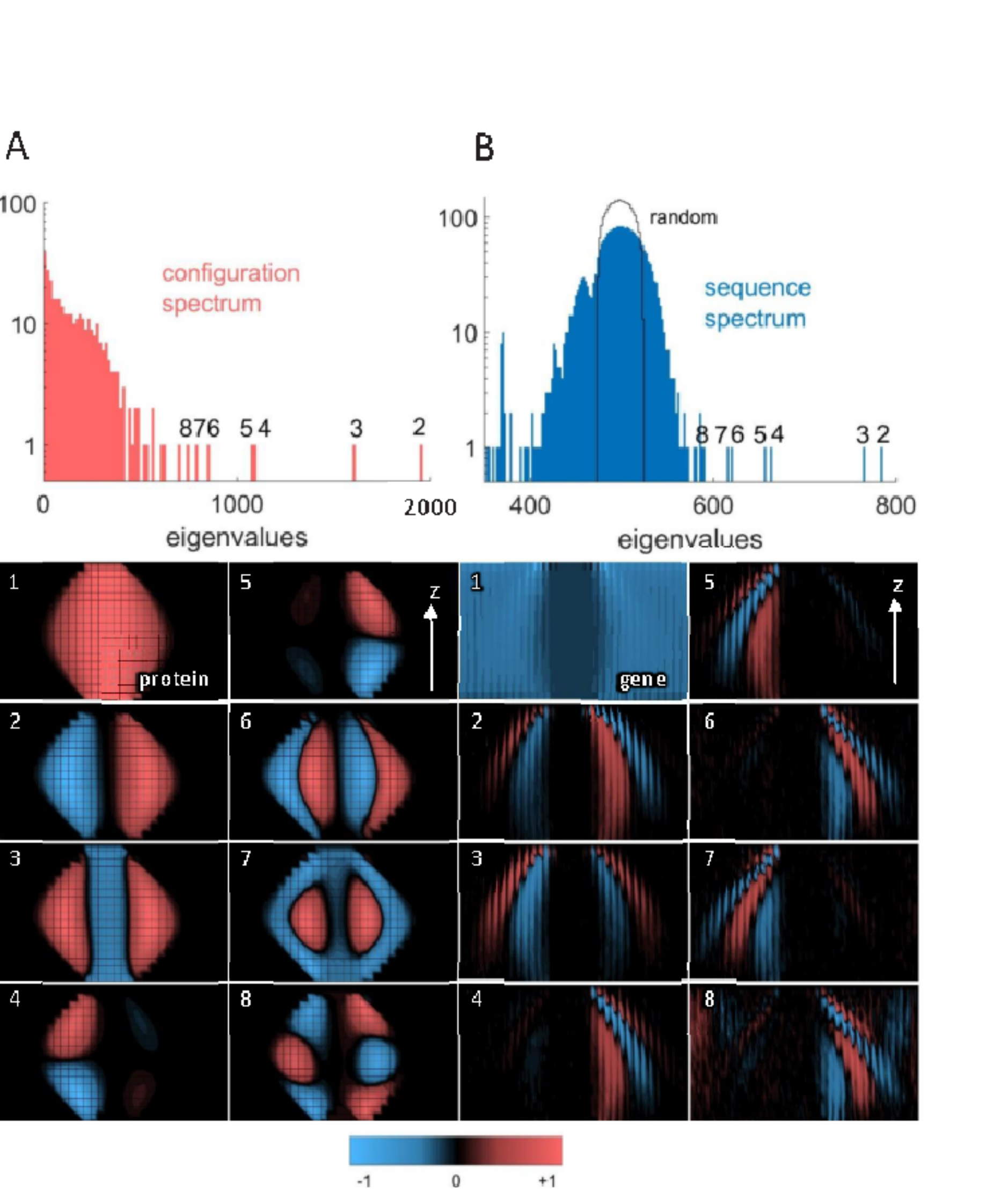}
\caption{\textbf{Correspondence of modes in sequence (genotype) 
and configuration (phenotype) spaces for the cylinder-model:}\\
 We produced the spectra by singular value decomposition of the
 $10^6$ solutions.\\
(A) Top: the spectrum in configuration space exhibits about 8-10
 eigenvalues outside the continuum (large $1^{\textrm{ st}}$ eigenvalue
 not shown).\\ 
 Bottom: the corresponding eigenvectors describe the
 basic modes of the fluid channel, such as side-to-side shift
 ($2^{\textrm{ nd}}$) or expansion ($3^{\textrm{ rd}}$).\\  
(B) Top: The spectrum of the solutions in sequence space is similar to
 that of random sequences (black line), except for about 8 to 9 high
 eigenvalues that are outside the continuous spectrum. 
 (Note that the $x$-axis does not start at 0.)\\ 
 Bottom: the first 8 eigenvectors exhibit patterns
 of alternating +\textbackslash- stripes \---  
 which we term correlation `ripples' \--- around the fluid channel region.
 Seeing these ripples through the random evolutionary noise required 
 at least $10^5$ independent solutions \cite{Tesileanu2015}.
}\label{fig:fig3}
\end{figure*}

Another useful method to analyze large sets of solutions is 
by spectral analysis in terms of Singular Value Decompositions (SVD).
For the cylinder model, we have $10^6$ binary vectors with $n=2550$
components each. To find the typical correlation spectrum of the 
solution, one forms a matrix $W$ of size $m\times n=10^6\times 2550$.
The SVD of this matrix is a generalization of 
the spectral decomposition of positive (semi-definite) square matrices:
$W$ is decomposed as $U\cdot D\cdot V^\T$, where $U$ is $m\times m$,
$V$ is $n\times n$ and $D$ is an $m\times n$ diagonal matrix 
(only the elements $D_{ii}$ with $i=1,\dots,n$ are nonzero). 
In our case, $m \gg n$, which is required to obtain 
good statistics of the random process.
The singular values $\lambda^{G}_i = D_{ii}$ are 
in general positive and in this case the decomposition is unique. 
The columns of $V$ are the (generalized) eigenvectors of $W$, 
the first few of which are shown in \fref{fig:fig3}.

The singular values $\lambda^{G}_i$ are the square roots of
the spectrum of the covariance matrix $W^\textrm{ T}W$,\footnote{The
  definition of the covariance requires to subtract the mean. 
Instead we project out the first eigenvalue.} 
which has the same eigenvectors as $W$. Therefore, the high values 
correspond to the principal covariance components, 
the directions with maximal variation in the solution set.
Mutatis mutandis, we perform the same SVD for the configurations,
using the 540 $s$-values (that is, of the shearability \eref{eq:2}) 
of vectors of the configurations. 
% This is reasonable, because, 
% in general, there are very few non-shearable and fluid AAs.

Figure \ref{fig:fig3} illustrates the difference between the
configuration space (phenotype) and the sequence space (genotype):

\textit{Configuration space} (The eight figures on the bottom left): 
The first mode is proportional to the average configuration. 
The next modes reflect the basic deviations of
the solution around this average. For example, the second mode is
left-to-right shift, the third mode is expansion-contraction
etc. Since, the shearable/non-shearable interface can
move at most one AA sideways between consecutive rows, the modes are
constrained to diamond-shaped areas in the center of the
protein. This is the overlap of the influence zones 
of the input and output rows. 

\textit{Sequence space} (The eight figures on the bottom right): 
The first eigenvector is the average bond occupancy 
in the $10^6$ solutions. The higher eigenvalues reflect 
the structure in the many-body correlations among the bonds. 
The typical pattern is that of `diffraction' or `oscillations' 
around the fluid channel. This pattern mirrors the
biophysical constraint of constructing a rigid shell around the
shearable region. Higher modes exhibit more stripes, until they become
noisy, after about the tenth eigenvalue. 

The sequence-spectrum, top right in \fref{fig:fig3} 
has some outliers, which correspond to the localized modes 
shown in the eight panels below. Apart from that, the
majority of the eigenvalues seem to obey the Mar\v cenko-Pastur
formula, see \cite{Marcenko1967}. 
If the matrix is $m\times n$, $m>n$, then
the support of the spectrum is $\frac{1}{2}(\sqrt{m}\pm\sqrt{n})$. 
In our case, since we have a $10^6\times 2550$ matrix, 
one expects (if the matrices were really random) to find the
spectrum at $\frac{1}{2}(\sqrt{10^6}\pm \sqrt{2550})$, 
which is close to the simulations, and confirms that most 
of the bonds are just randomly present or absent. 
The slight enlargement of the spectrum is attributed 
to memory effects between generations in the same branch. 
This corresponds to phylogenetic correlations among
descendants in the same tree \cite{Felsenstein1985}.\footnote{The
  continuous
  part of the sequence spectrum, 
which is not quite of the standard form, could in principle be studied
by taking into account the known correlations. However, even the techniques of \cite{Guhr1998} seem difficult to implement.}
We conclude: \textbf{The small number of discrete eigenvalues shows
 that a small number of parameters characterizes both the
 phenotype and the non-random part of genotype of proteins}.

\subsubsection{Geometry of the genotype and phenotype solution spaces$^*$}

\begin{figure}
  %%\centering
  \includegraphics[width=\columnwidth]{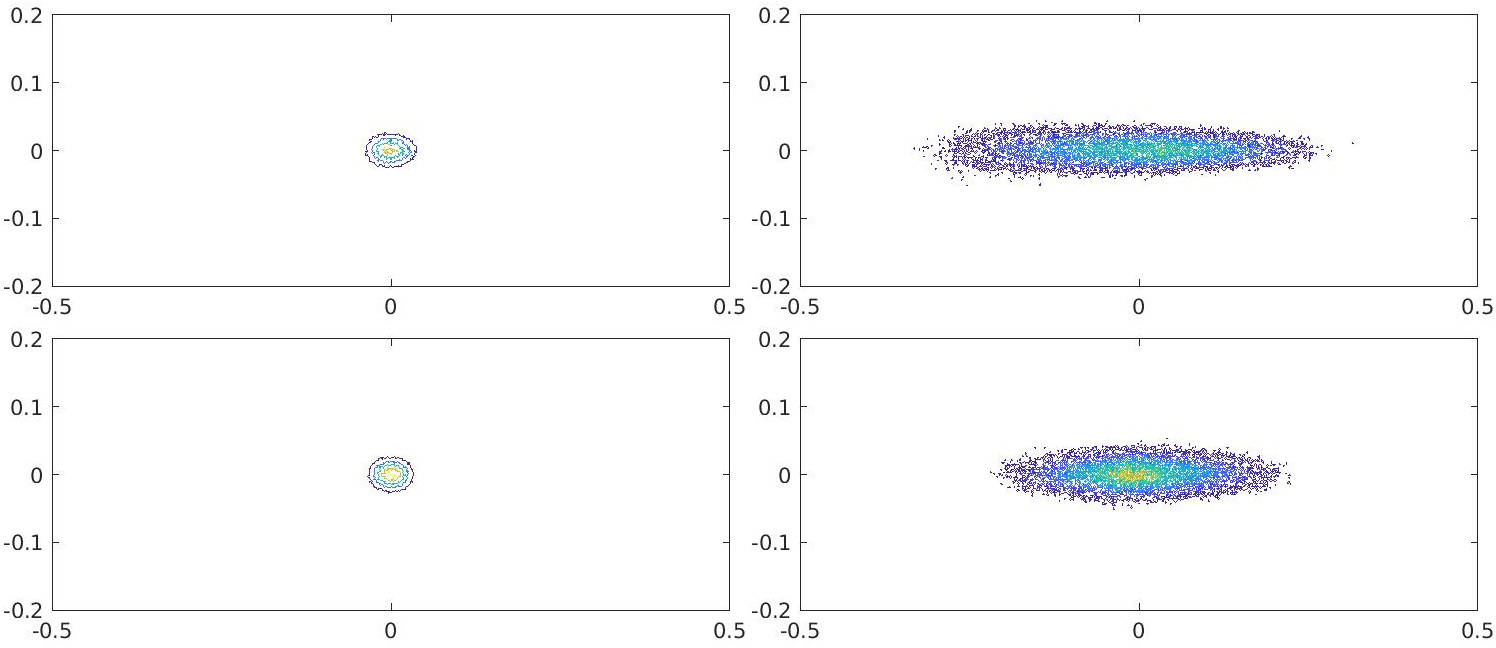}
  \caption{The projections of the set of $10^6$ solutions as
    2550-dimensional gene sequences 
    (left column) and corresponding 540-dimensional phenotypes 
    (right column)
    onto their SVD eigenvectors $\cb$ and $\ub$.
    Top row shows axes 2 (horizontal) and 100 (vertical), bottom row
    shows axes 3 (horizontal) and 100 (vertical).
    Note that the phenotypes have larger variation along their 2 and 3-components 
    than their 100-component, unlike the genotypes which appear evenly
    distributed in all directions.}
    \label{fig:figure5b}
\end{figure}

The $10^6$ genotype vectors form
a `cloud' of points in a 2550-dimensional space.
The geometry of the cloud can be explored 
by plotting projections along the axes defined by the eigenvectors
$\cb_i$, $i=1,2,\dots,2550$, \fref{fig:figure5b}.
Consider for example the projection of the cloud
onto the subspace spanned by $\cb_2$, $\cb_3$ and $\cb_{100}$. 
The variation along the $3$ axes is of comparable size.
However, the equivalent projection of the 540-dimensional phenotypes along their
eigenvectors $\ub_2$, $\ub_3$, and $\ub_{100}$ shows
very small variation along the vertical ($\ub_{100}$) axis, similar to
the projections of a flat ellipsoid.

The projections reflect the differing shapes of the solution clouds:
in the genotype space the cloud is a $2550$-dimensional spheroid object,
while in the phenotype space it is a flat discoid of dimension $\sim 10$.

\def\x{0.49}

\subsection{Stability of the mechanical phenotype under mutations}\label{sec:stability}

Any protein is the outcome of a long evolutionary trajectory
starting from a distant ancestor. It is likely to find 
other descendants of the same ancestor in closely related species.
In practice, one fetches the pairwise most similar proteins from a
collection of species and aligns their sequences to identify 
homologous amino acids (all descendants of a given amino acid 
in the ancestor protein) \cite{Karlin1993}. Once this has been
accomplished, one can study mutation patterns in this
multiple sequence alignment. There are two questions of interest here:
\begin{myenum}
  \item Which positions in the gene are \emph{conserved}, \ie
    they encode the same AA?
  \item Which pairs of positions are \emph{co-varying}: 
    a mutation at one position is frequently compensated
    by a mutation at the other position?
\end{myenum}
\begin{figure}
 %%\centering 
\includegraphics[width=1.0\columnwidth]{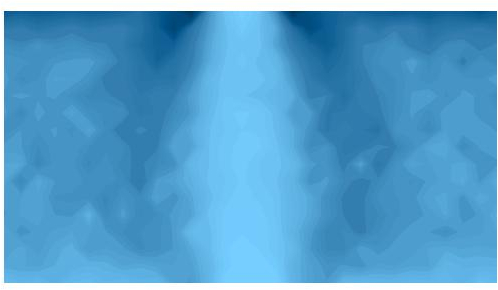}
\caption{The sensitivity of solutions to a single mutation, as a function of the mutation position.
  }\label{fig:mutateone}
\end{figure}%

The second question, about genetic correlations, will be discussed in detail in \sref{sec:epistasis}. In this subsection, we discuss the first
question, regarding AA conservation, in the context of the cylinder-model.
To produce \fref{fig:mutateone} one takes $10^6$ sample solutions and
mutates, for each of them, every possible position in the AA network. 
One then asks which mutations destroy the solution. 
At every position, the intensity of the blue color is proportional
to the probability that a mutation at that position destroys the
solution. The end of the channel is very sensitive, 
but also the boundaries between the channel and the bulk. 
This should be compared to \fref{fig:3gluco} (center) 
where the analogous question was asked for glucokinase
\cite{Rougemont2099}, and answered by aligning 122 
homologs of glucokinase.
We conclude: \textbf{Sensitivity to mutations is localized near
  mechanically critical regions}.

\begin{figure*}
%%\centering
\includegraphics[width=0.7\textwidth]{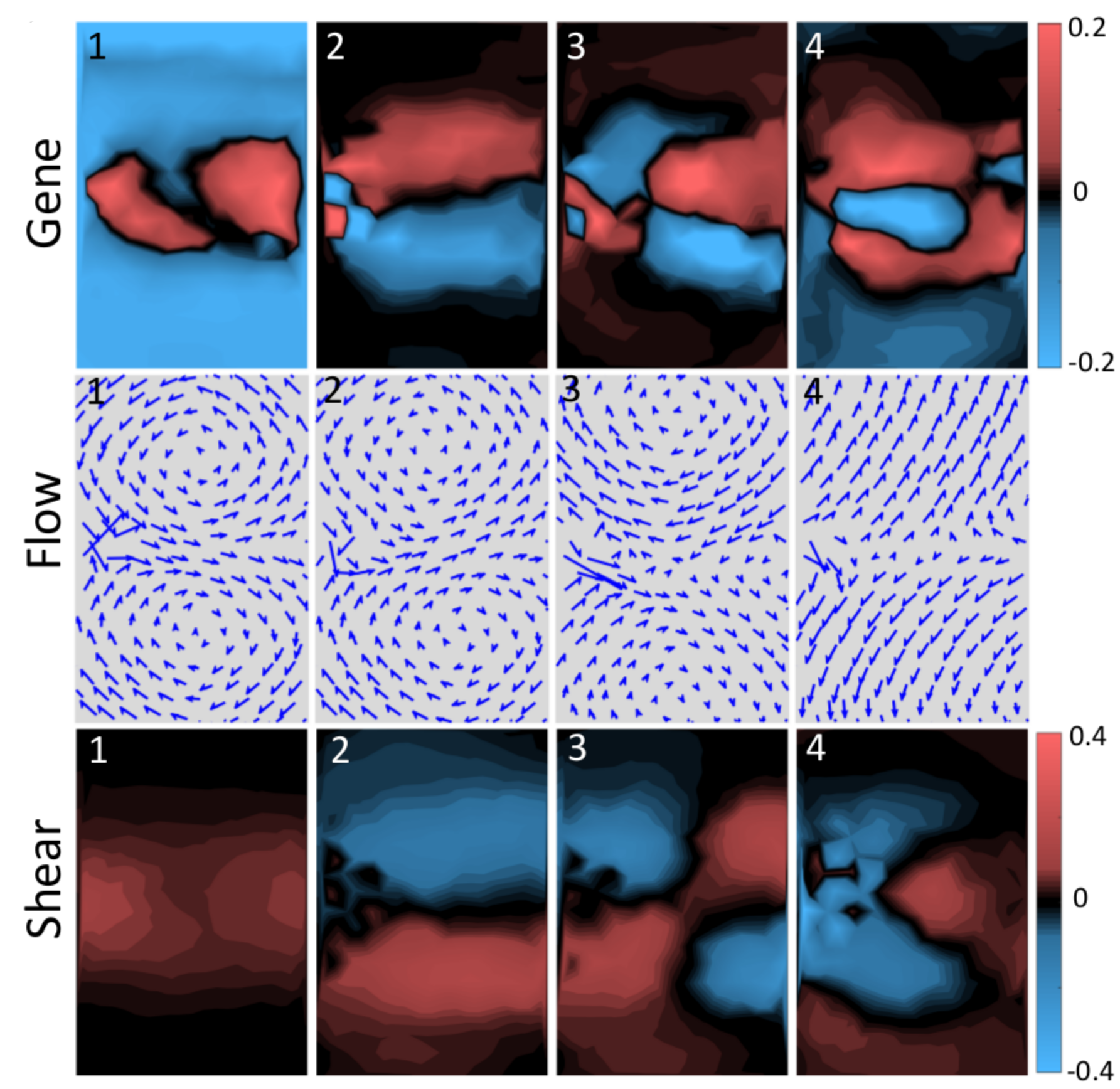}
\caption{\textbf{The vector fields for the HP-model:} The first 4 eigendirections for the three
  vector fields, $k=1,\dots,4$.\\
Top: The first four SVD eigenvectors of the gene $\Ck$,\\
Center: The corresponding displacement flow field $\Uk$,\\
Bottom: The corresponding shear intensity $\Sk$. 
}\label{fig:5}
\end{figure*}

\subsection{Shear modes in the amino acid network}\label{sec:shear}

We focus here on the HP-model, although similar results hold for the
cylinder-model. In \sref{sec:pinch} we have shown a pinch stimulus
$\fb$ leads to a deformation field $\ub=\GG\fb$, see \eref{eq:gf}.
As the fitness $F$ of \eref{eq:fitness} improves, the system forms a fluid
channel and the response field $\ub$ shows a hinge-like rotation, 
as is visible in \fref{fig:pinch}. This should be compared to
\fref{fig:gluco}, showing the experimentally measured deformation 
of  glucokinase. Opening a hinge in a network will not only move 
the two sides of the hinge but also shear the connecting bonds, 
especially at the hinge itself. Furthermore, remaining links near the
opening of the hinge (at the opposite side of the protein) 
will be stretched as well. These observations can be quantified 
by measuring the shear. The shear $s$ at any (lattice) point is 
the symmetrized derivative of the displacement field $\ub$,
which is computed as follows.

% We first recall the definitions for the case of 
% a displacement field on $\mathbb{R}^d$.
First, the displacement vector at the point $\xb\in\real^d$ 
is $\ub(\xb)$ with $u_i(x)=x_i'(x)-x_i$, the difference 
between the `new' points $\xb'(\xb)$ and the `old' points 
$\xb$ in $\real^d$.
The local deformation matrix $\DD(\xb)$ is then given by
\begin{equ}
  D_{ij} =\frac{\partial x_i'(\xb)}{\partial x_j}~,
\end{equ}
so that $\nabla \ub(\xb) = \DD(\xb)-\id$.
(This matrix is also called `compatibility matrix' in the review \cite{Lubensky2015}.)
The shear matrix $\epsilonb$ is
\begin{equ}
  \epsilon _{ij}(\xb)=
  \frac{1}{2}\left(\frac{\partial u_i}{\partial x_j}+\frac{\partial u_j }{\partial
    x_i}+
  \sum_{k=1}^d \frac{\partial u_k}{\partial x_i}\cdot\frac{\partial u_k}{\partial x_j}\right) ~.
\end{equ}
In short notation
\begin{equa}\label{eq:stress2}
 {\epsilonb}&=\frac{1}{2}\biggl((\nabla \ub(\xb))^\T+ \nabla \ub(\xb)
 +(\nabla \ub(\xb))^\T \nabla \ub(\xb)  \biggr) \\
 &= \frac{1}{2}\biggl(\CC(\xb)-{\id}\biggr)~,
\end{equa}
with $\CC(\xb)=\DD(\xb)^\T \DD(\xb)$, 
which is the metric of the coordinate transformation. 

As a measure of the magnitude of the shear one can use
\begin{equa}\label{eq:sx}
  s(\xb) &= \Tr\left({\epsilonb }^2\right) - 
  \frac{1}{d}\left(\Tr ({\epsilonb})\right)^2\\
  &=\Tr\left({\epsilonb }-\frac{1}{d}
  \Tr(\epsilonb )\cdot {\id}\right)^2~,
\end{equa}
which is the square of the Frobenius norm ($L^2$)  of the traceless
part of $\epsilonb $.\footnote{As all norms on 
finite dimensional spaces are equivalent, other norms amount 
basically just to a rescaling.}
The trace of $\epsilonb$ is related to the isotropic dilation, 
which in the protein is a smaller effect than the shear,
and therefore will not be further considered.\footnote{There are many
  variants of the shear calculation see, \eg \cite{continuum}.}

\subsubsection{Implementation in the case of protein structure data$^*$}

As proteins are discrete objects,
we replace the derivatives by difference
operators \cite{Gullett2008,Mitchell2016}. 

We consider the crystallographic data of two conformations
of a given protein (for example the PDB structures 1v4s and 1v4t in \fref{fig:gluco}
\cite{Kamata2004}).
To produce \fref{fig:gluco} from these data \cite{Rougemont2099},
we take a ball of  radius $\rho$ of about 10\si{\angstrom} around
each atom $X$ in the protein.\footnote{It usually
  suffices to look at atoms N, C, O,  
  along the backbone, while one may also include sidechains.}
This will encompass $m =m(X)$ other atoms, at positions $\rb_i\in\real^3$, $i=1,\dots,m$. 
Let $\rb_0$ be the coordinates of $X$, and
let $\AAA(X)$ be the $m\times 3$ matrix of the $m$  
distance vectors $\rb_i-\rb_0$, in the first configuration
(\ie one of the PDB structures). 
Let $\BB(X)$ be the analogous matrix for the second
configuration and compute
\begin{equ}\label{eq:q}
\QQ=\HALF(\AAA^\T \AAA)^{-1} \, \AAA^\T\, (\BB \BB^\T - \AAA \AAA^\T)\, 
\AAA\, (\AAA^\T \AAA)^{-1} ~.
\end{equ}
The matrix $\QQ$ is an approximation of $\epsilonb$ and is obtained by observing that
$\BB=\AAA\DD^\T$:
\begin{equ}
  \CC=\DD^\T\DD= (\AAA^\T\AAA)^{-1} \AAA^\T \BB\, 
  \BB^\T \AAA(\AAA^\T \AAA)^{-1}~.
\end{equ}
By substituting this into \eref{eq:stress2}, we verify 
that \eref{eq:q} holds. With this approximate shear tensor $\epsilonb$
one computes its magnitude $s(\rb_0)$ using \eref{eq:sx}.

The discrete approximation of the shear field 
in glucokinase is shown in the leftmost panel of \fref{fig:3gluco} 
(with $m$ typically around  50). The strain is found to be large
in the hinge of the protein, and also at a somewhat loose outer surface.\footnote{The dilatation (the trace) is much smaller.} 
In \fref{fig:5}, the corresponding shear magnitude fields $s(\xb)$ 
are shown for the HP-model, using a statistical average over
many solutions (see \sref{sec:computation} for more details). 
One again observes strong shear in the hinge. 
We conclude: \textbf{{shear is critical 
in the hinges among moving domains of the protein}}.

\subsubsection{Details of shear computation$^*$}\label{sec:computation}

We describe here in detail the procedure of calculating 
the shear in the HP-model, leading to \fref{fig:5}. 
This figure shows averages over many realizations 
of the random process, in the following sense. 
One starts with $10^6$ solutions.\footnote{The characteristics we are looking for 
do not show cleanly unless there are at least $10^5$ samples.}
Each solution $\cb_*$ together with the (fixed) pinch $\fb$, 
defines 3 vectors
\begin{myenum}
\item the gene of the functional protein, $\cbs$, 
  (a vector of length $n_a=200$ codons),
\item the flow field (displacement),  $\ub(\cbs)=\GG(\cbs) \fb$, 
 (a vector of length $n_d =400$ of the $x$ and $y$ velocity components),
\item the shear field $\sb(\cb_*)$ (a vector of length  $n_a=200$).
\end{myenum}
The $10^6$ solutions are then written as three corresponding matrices 
$W_C$, $W_U$ and $W_S$, of size $200\times 10^6$ resp,~$400\times
10^6$, where each row of these matrices is one 
of $\cbs$, $\ub(\cbs)$, and $\sb(\cbs)$.    

Next, one calculates the singular eigenvalues and corresponding 
eigenvectors of the three matrices (using SVD, 
as in \sref{sec:spectrum}) and isolates the leading eigenvalues.
The central row in \fref{fig:5} shows that the flow field 
can be decomposed into successively weaker motions $\Uk$, 
with the strongest being a rotating hinge motion 
around the fluid channel. One should note that evolution in
this case did not impose this \emph{global} rotation, 
but only the \emph{localized} response to a pinch 
on the left side of the sample.

\subsection{Similarity of gene and shear}\label{sec:similarity}

The results for the HP-model reveal a tight relation between 
the gene fields $\Ck$ and the shear intensities $\Sk$, as shown in \fref{fig:5}. Comparing the top and bottom rows, one observes 
a similar structure of the corresponding eigenfunctions.\footnote{One could in principle measure the distance between the
  corresponding pairs using an $L^2$ norm over the whole area.}
A similar relation is visible in \fref{fig:fig3} for the
cylinder-model.
The functions of many proteins are known to involve 
large-scale motions of the amino acid network, 
such as hinge rotation, shear sliding, or twists \cite{Gerstein1994}.
Recently, the strain that occurs during such conformations changes was computed in several proteins by comparing structures obtained from X-ray and NMR studies \cite{Mitchell2016}. 
However, the tight correspondence between the shear tensor and 
the genetic correlations, that we observe here (\fref{fig:5})
has not yet been measured in real protein. 
In principle, one would need to follow a procedure similar 
to the one presented here: first, to calculate the mechanical shear 
using the methods of \cite{Gullett2008,Mitchell2016},
and then to compare it to the genetic correlations 
from sequence alignment \cite{Rougemont2099}.

%%%%%%%%%%%%%%%%%%%%%%%%%%%%%%%%%%%%%%%%%%%%%%
\subsection{Point mutations are localized mechanical perturbations}\label{sec:scattering}
A mutation in the HP-model may vary the strength 
of no more than $z=12$ bonds around the mutated AA (\fref{fig:lattice}).
The corresponding perturbation of the Hamiltonian $\DH$ is therefore
localized, akin to a defect in a crystal \cite{Tewary1973,Elliott1974}. 
The mechanics of mutations can be further explored 
by examining perturbations of Green's function, $\Gs = \GG +\DG$.
They obey the Dyson equation and the Dyson series,
\eref{eq:Dyson}--(\ref{eq:dysonseries}). 
This series has a straightforward physical interpretation 
as a sum over multiple scatterings (\fref{fig:2}B):
As a result of the mutation, the elastic force field 
is no longer balanced by the imposed force $\fb$, 
leaving a residual force field $\dfb = \DH\, \ub = \DH \, \GG\, \fb$.
The first scattering term in the series balances $\dfb$ by the
deformation $\dub = \GG \,\dfb = \GG\, \DH \,\GG \fb$. Similarly, the
second scattering term accounts for further deformation induced by
$\dub$, and so forth.\footnote{In problems of this local nature,
  calculating a mutated Green's function using the Woodbury formula \eref{eq:Woodbury} accelerates the
  computation by a factor of ${\sim}10^4$ as compared to standard matrix
  inversion.} We conclude: {\bf Standard expansions of Green's functions
  correspond to hierarchical organization of the effects of mutations
  in terms of multiple scattering}.
%\end{showcomment}

\subsection{Mechanical function emerges as a sharp transition}

As the evolution reaches a solution gene $\cbs$, there
emerges a new (almost) zero energy-mode, $\ubs$, in addition to the
Galilean symmetry modes (which we already projected away).
As the other eigenvalues of $\GG(\cbs)$ remain typically distant 
from this small eigenvalue $\lams$, there will be a gap between
$\lambda_*$ and the rest of the spectrum.
While we do not have a proof of that such a gap should appear, 
this is found to be the generic case in the models described here. 
The response to a pinch will be mostly through this soft mode, 
as we show now.

Consider a sequence of mutations $\cb_k$ which converges to
$\cbs$ (in the Hamming distance) as $k\to k_*$.
The corresponding sequence of fitness values is $F_k=F(\cb_k)$. 
For the HP-model with the pinch introduced earlier, 
the fitness is (\eref{eq:fitness}):
\begin{equ}
  F_k =\vbT\ub(\cb_k) =\vbT\GG(\cb_k)\fb ~.
\end{equ}
When $\cb_k$ gets closer to $\cbs$, the almost-zero eigenvalue 
$\lambda _k$ of $\HH(\cb_k)$ will dominate Green's function,
$\GG(\cb_k) = \HH(\cb_k)^\pseudo$
\begin{equ}
  \GG(\cb_k)  \simeq \frac{1}{\lambda_k}
  \ket{\ub(\cb_k)}\bra{\ub(\cb_k)}\sim
  \frac{1}{\lambda _k} \ket{\ubs}\bra{\ubs}~.
\end{equ}
The fitness sequence is therefore
\begin{equ}
\label{eq:divergence}
F_k \simeq \frac{\left( \vbT \ubs\right)\left( \ubsT \fb \right)}{\lambda_k}~. 
\end{equ}
\begin{figure}
%%\centering
\includegraphics[width=1.0\columnwidth]{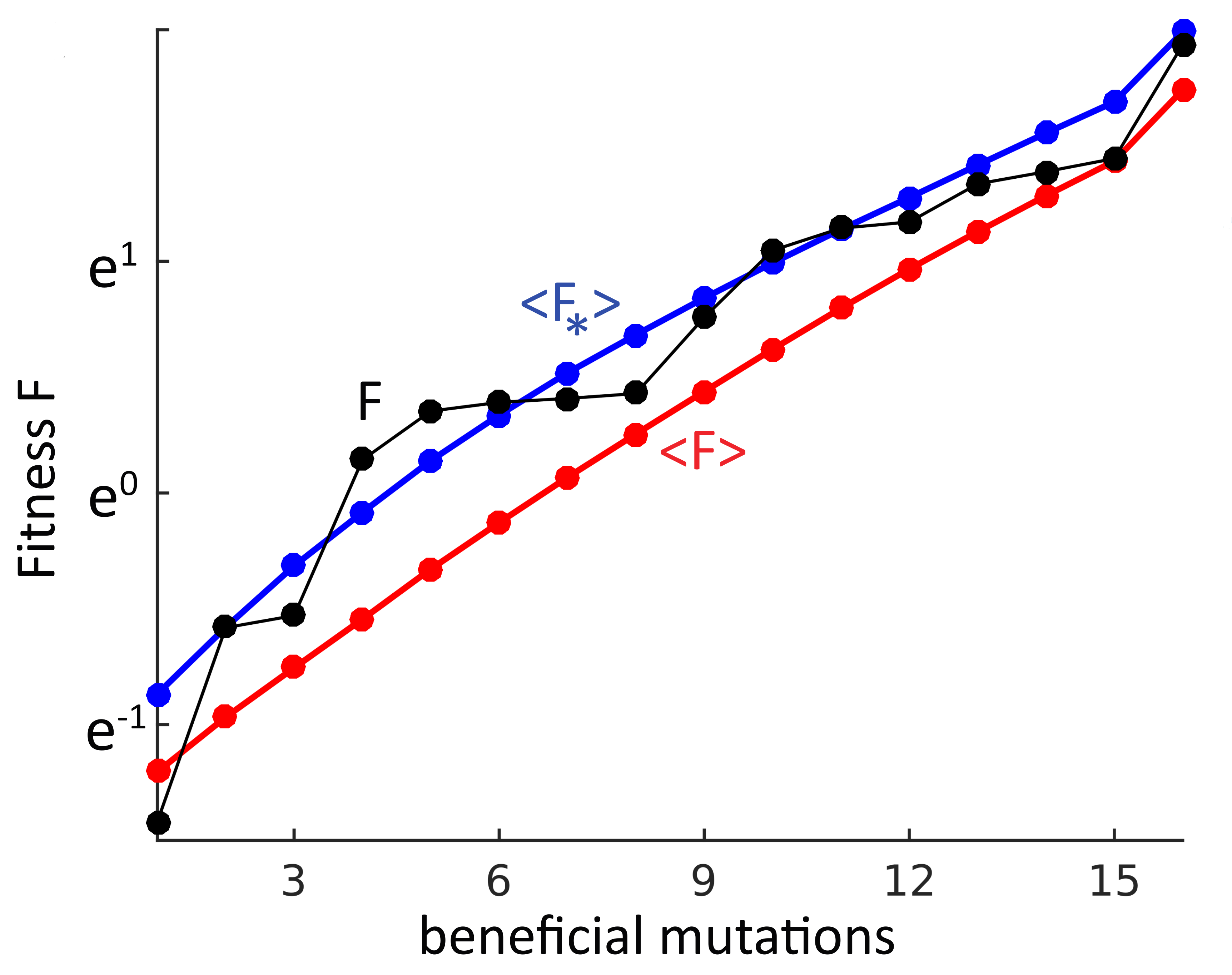}
\caption{Progression of the fitness $F$ corresponding to the evolution
  of \fref{fig:pinch} (black).  The fitness trajectory averaged over
  ${\sim} 10^6$ runs $\langle F \rangle$ is shown in red. Shown are
  the last 16 beneficial mutations towards the formation of the
  channel. The contribution of the emergent low-energy mode $\langle
  F_k \rangle$ alone, shown in blue, dominates the fitness
  (according to \eref{eq:divergence}).  
}\label{fig:3a}
\end{figure}%

On average, the fitness increases exponentially with the number of
beneficial mutations as shown in \fref{fig:3a}. 
The growth of the fitness follows the formation of the channel 
and the narrowing of the remaining rigid `neck', in the middle
of the channel. We lack, however, a quantitative explanation for the 
generic exponential dependence, which is probably related to the 
structure of the Hamiltonian $\HH$.\footnote{Note that the exponential
  increase is much stronger 
  than what could by explained by the choice 
  of the factor $\DF>\epsilon F $ of \sref{sec:simulating}.}
In the particular instance of the two models considered here, 
one can argue that the mutations which improve the channel, 
all act multiplicatively on the fitness $F$. Since this discussion is
`spectral,' we expect it to hold for models with more colors (\ie AAs) than the
HP model.

\begin{figure}
%%\centering
\includegraphics[width=1.0\columnwidth]{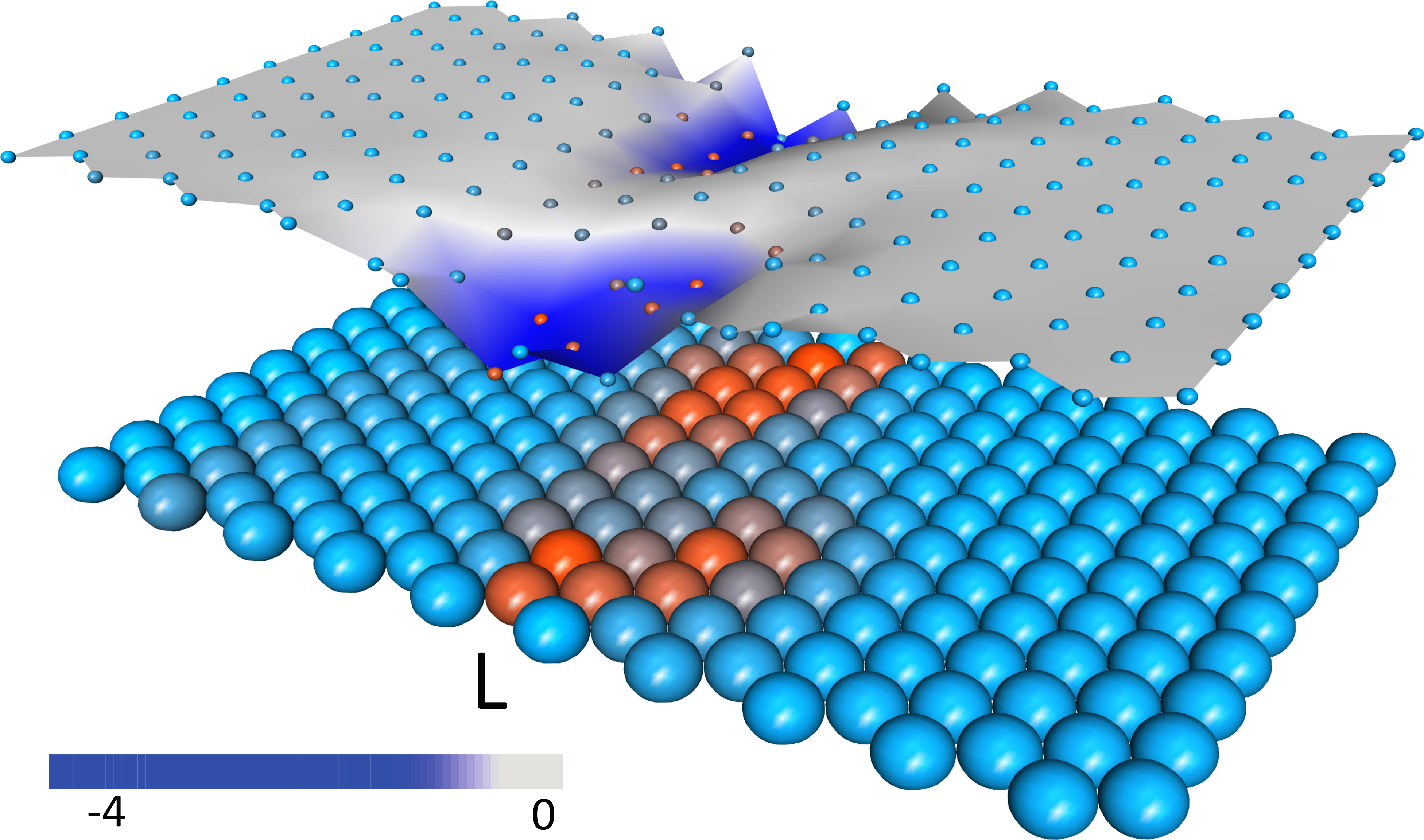}
\caption{Landscape of the fitness change $\DF = \vbT \DG\, \fb~$,
  averaged over $ 10^6$ solutions, for all 200 possible positions of
  point mutations at a solution.  Underneath, the average AA
  configuration of the protein is shown in shades of red ($\aP$) and
  blue ($\aH$). 
In most sites, mutations are neutral, while mutations in the channel
are, on average, deleterious (blue, below the flat surface).
}\label{fig:3b}
\end{figure}

As noted in \fref{fig:deleterious}, beneficial mutations are rare, 
and are separated by long stretches of neutral mutations. 
One may ask where the neutral mutations take place, 
and this is illustrated in \fref{fig:3b}, 
which shows that, in most sites, the effect of mutations 
is practically neutral.
%% 

The vanishing of the spectral gap, $\lambda _k \to 0$, can further be
viewed as a topological transition in the system: the AA network is
being divided into two domains that can move independently of each
other at low energetic cost. The relative motion of the domains
defines the emergent soft mode and the collective degrees-of-freedom,
for example the rotation of a hinge or the shear angle.
\begin{figure}
  %%\centering
\includegraphics[width=1.0\columnwidth]{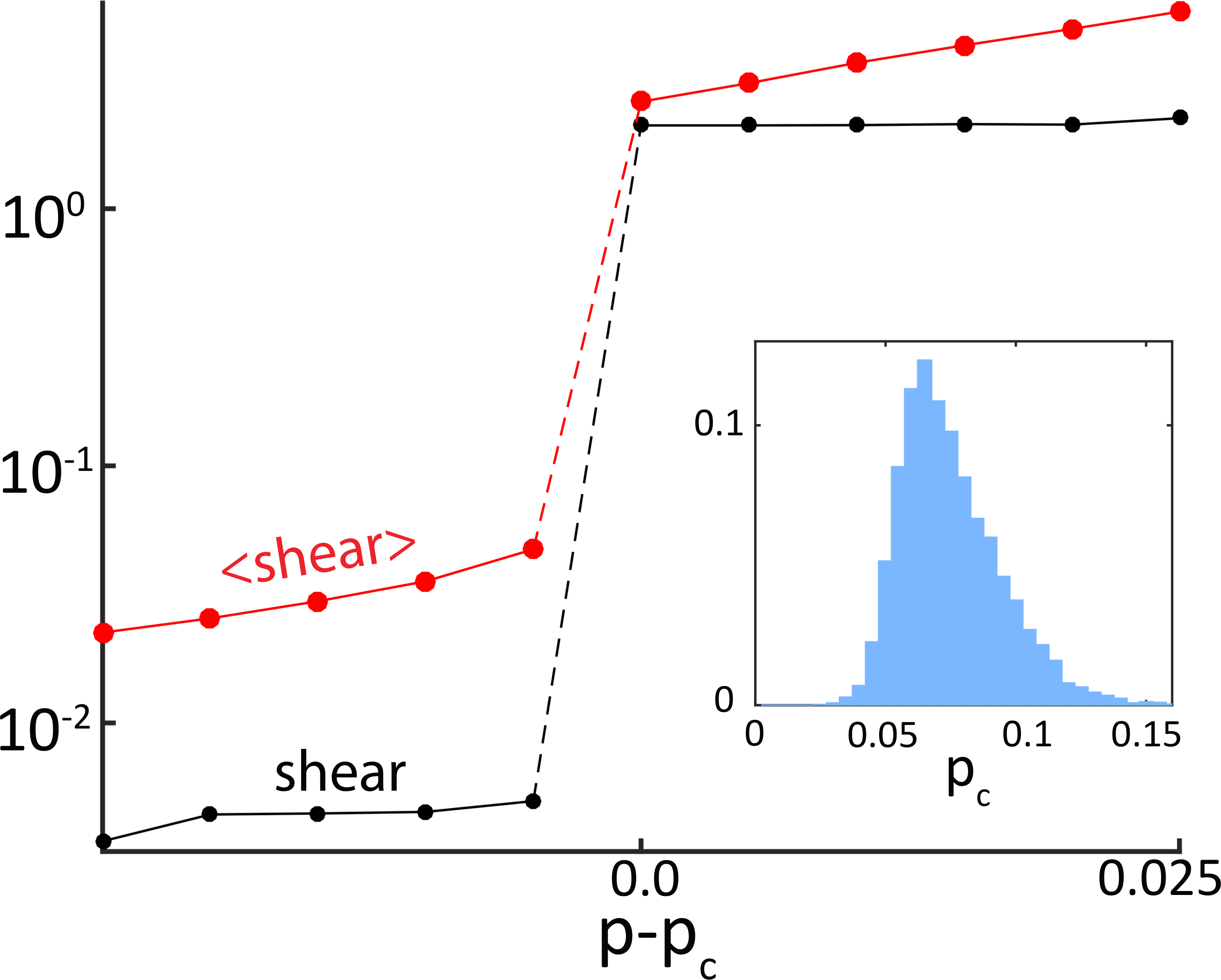}
  \caption{The mean
shear in the protein in a single run (black) and averaged over $10^6$
samples (red) as a function of the fraction $p$ of $\aP$-amino acids. The
values of $p$ are 
shifted by the position of the jump, $p_\textrm{ c}$. Inset:
Distribution of $p_c$.}
\label{fig:figure3d} 
\end{figure}%%

The soft mode appears at a dynamical transition, where the
average shear in the protein jumps abruptly as the channel is formed
and the protein can easily deform in response to the force probe
(\fref{fig:figure3d}). The trajectories are plotted as a function of
$p$, the fraction of AAs of type $\aP$. The distribution of the
critical values $p_c$ is rather wide owing to the random initial
conditions and finite-size effects.

Another connection is provided by the Kirchhoff matrix-tree theorem
(see \eg \cite{Tutte1948,Chaiken1982}). 
Let $\MM$ be an $n$-by-$n$ graph Laplacian, with links
$i\leftrightarrow j$ given by $\MM_{ij}=-1$ and $\sum_j \MM_{ij}=0$ 
for all $i$. The matrix $\MM$ has an eigenvalue 0, with eigenvector
$(1,1,\dots,1)$. 
Take the submatrix $\MM_0$ 
where one row and one column are omitted. In analogy with the
weighted links of the HP-model, assume now that $\det \MM_0=0$, 
\ie that a second eigenvalue vanishes. Then, by the Kirchhoff theorem, 
the number of spanning trees of the full graph is equal to 0. 
In other words, the graph is disconnected, in analogy 
to the formation of the fluid channel.

\subsection{Correlation and alignment}\label{sec:correlation}

\begin{figure}
%%\centering
\includegraphics[width=1.0\columnwidth]{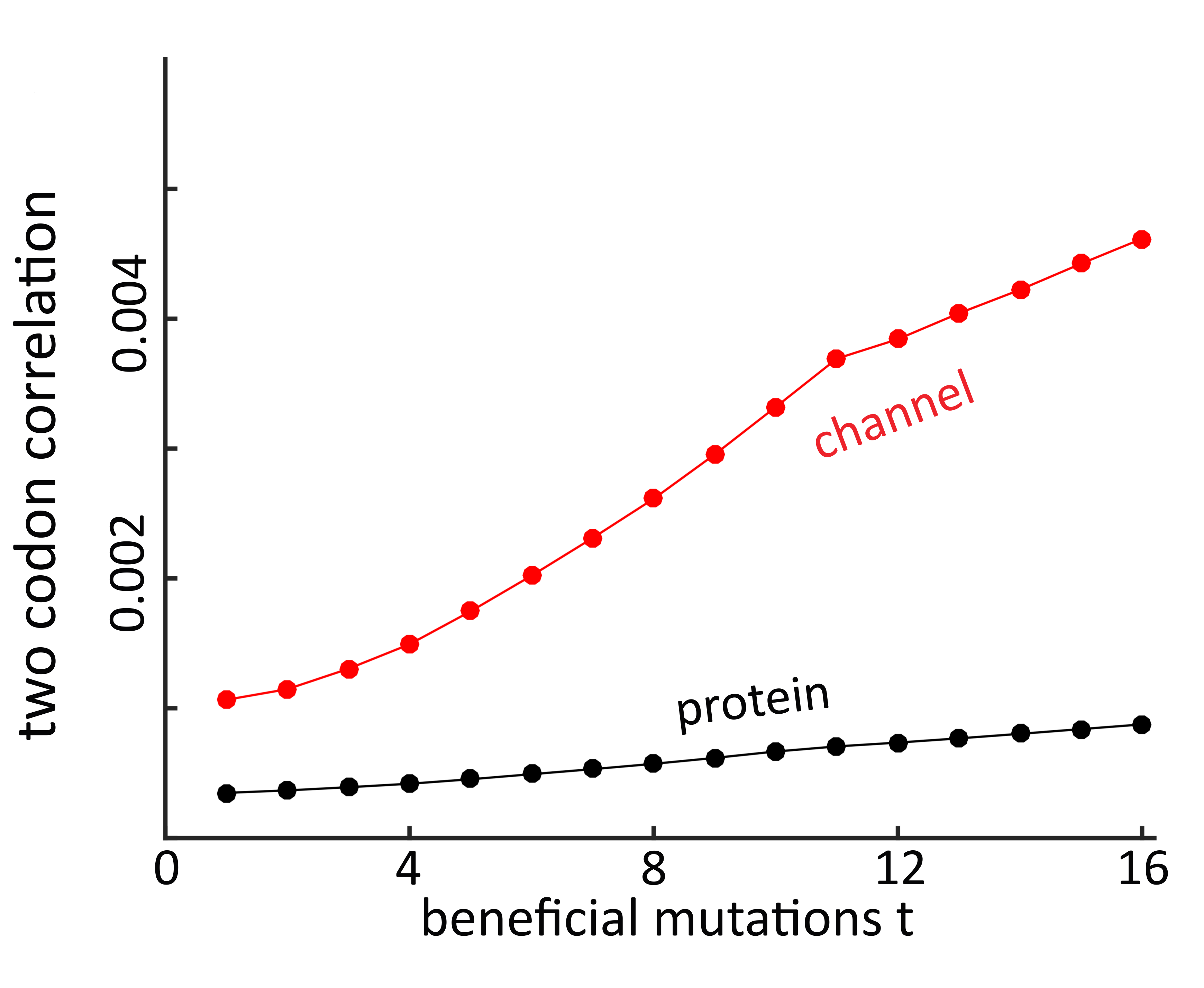}
\caption{The average magnitude of the two-codon
  correlation $|Q_{ij}|$ 
  \eref{eq:correlation}  as a function of the number of beneficial
  mutations, $t$.
The red curve shows $|Q_{ij}|$ in the shear band (AAs in rows $7{-}13$,
of \fref{fig:pinch}). The black curve shows $|Q_{ij}|$ for the whole
protein. The correlations in the channel are clearly larger.}
\label{fig:3c}
\end{figure}

As the shear band (fluid channel) is taking shape, 
the correlation among codons builds up. To see this, 
we align genes from the $10^6$ simulations, in
analogy to sequence alignment of real protein families
\cite{Goebel1994,Marks2011,Jones2012,Lockless1999,
Suel2003,Hopf2017,Poelwijk2017,Halabi2009,Tesileanu2015,Juan2013}. 
At each time step we calculate the two-codon correlation $Q_{ij}$ between
all pairs of codons $c_i$ and $c_j$ ,  
\begin{equ}
\label{eq:correlation}
Q_{ij} \equiv \langle c_i c_j\rangle-\langle c_i \rangle \langle c_j\rangle~,
\end{equ}
where brackets denote ensemble averages. One finds that most of the
correlation is concentrated in the region where the channel will form.
In \fref{fig:3c} one sees that the average correlation is tenfold larger
in the channel than in the whole protein. 
Within the channel, the correlation is long-range, 
and propagates from side to side in the protein
(see \cite{Dutta2018} for a figure).

Analogous correlated domains containing functionally-related
amino acids that co-evolve appear in real protein families 
\cite{Lockless1999,Suel2003,Halabi2009,Tesileanu2015}, 
as well as in coarse-grained models of protein allostery 
\cite{Hemery2015,Flechsig2017,Tlusty2016,Tlusty2017} 
and allosteric matter \cite{Rocks2017,Yan2017}.

We conclude: \textbf{Genetic correlations are significantly larger in the mechanically important regions}.

\subsection{Conserved amino acids}\label{sec:conservation}

In this section, we discuss single mutations (and the lack thereof),
while in the next, we discuss the case where one mutation is
`compensated' by another (this is called epistasis). Both phenomena
are intimately related to those sites on the protein which matter
for the function: These are the sites which are mechanically important.

In the cylinder model (\fref{fig:mutateone}), mutations near 
the top edge and the boundary of the fluid channel have the most deleterious 
effect on the mechanical function (dark regions in the figure).
Therefore, to preserve the functionality of the protein in this model, 
these sensitive amino acids are also conserved more than average
among the solutions. 

In comparison, the center panel of \fref{fig:3gluco} highlights the 
most conserved positions among the 122 aligned homologs of 
the real protein glucokinase. Similar to the model, here also
conservation appears to be correlated with mechanical importance, 
as measured by the magnitude of the shear 
(left panel of \fref{fig:3gluco}).
We conclude: \textbf{Mechanically critical regions of a protein are
  sensitive to mutation}.
If, however, a mutation does occur at such a sensitive 
amino acid, then it should be compensated by another one (or a few). 
This is dealt with in the next section.

%%%%%%%%%%%%%%%%%%%%%%%%%%%%%%%%%%%%%
\subsection{Epistasis links protein mechanics to genetic correlations}\label{sec:epistasis}

The correlations among amino acids in the gene exhibit tight 
correspondence to the pattern of the shear field 
(see \sref{sec:correlation} and \fref{fig:5}). 
We now discuss how to link these genetic correlations among 
mutations to the physical interaction in the amino acid network. 
The procedure will be similar to how the effect of 
a single mutation was interpreted in terms of a 
scattering expansion of Green's function (\sref{sec:scattering}).

In genetics, the term epistasis refers to departure of fitness 
from additivity in the effect of combined mutations owing to 
\emph{inter-genetic} interaction. For example, 
the phenotypic effect of one gene may be masked 
by a different gene \cite{Cordell2002,Phillips2008,Mackay2014}.
In analogy, on the smaller scale of a single gene described here, 
\emph{intra-genetic} epistasis is non-additivity of protein fitness
owing to the non-linear interaction among its amino acids
\cite{Breen2012,Ortlund2007,Harms2013,Clark1997}. 
For example, one mutation can be compensated by another one
in order to keep the protein functional. 
This second mutation can be far away on the gene sequence.
The use of Green's functions allows for a calculable definition of
epistasis in terms of the Dyson series \eref{eq:Dyson}.
\begin{figure}
%%\centering
\includegraphics[width=1.0\columnwidth]{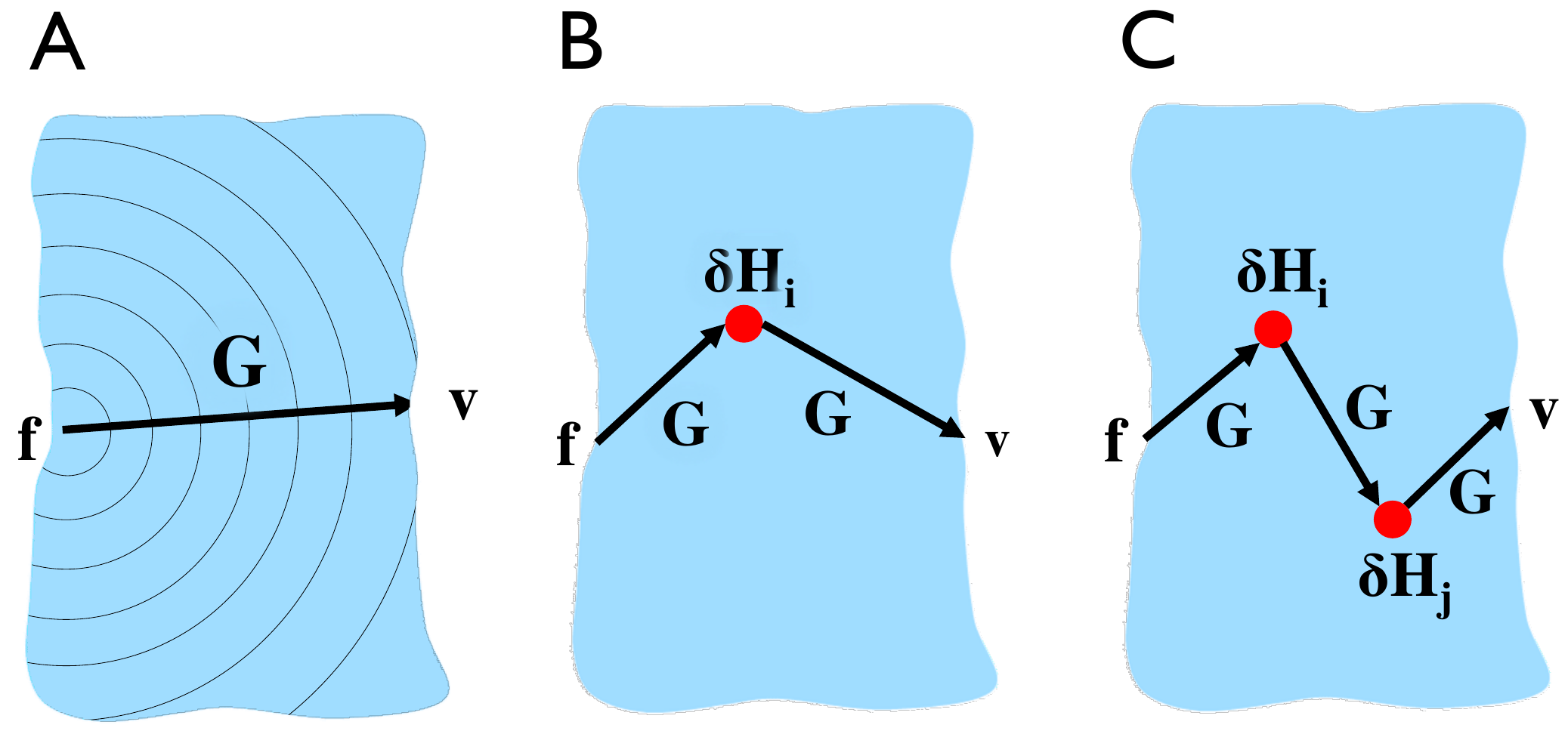}
\caption{\textbf{Force propagation, mutations and epistasis.}
(A) Green's function $\GG$ measures the propagation the mechanical signal across the protein (blue) from the force source $\fb$ (pinch) to the response site $\vb$, depicted as a `diffraction wave'.
(B) A mutation $\DHi$ deflects the propagation of force. The effect of the mutation on the propagator $\DG$ can be described as a series of multiple scattering paths \eref{eq:Dyson}. The diagram shows the first scattering path, $\GG\,\DHi \GG$.
(C) Epistasis is the departure from additivity of the combined fitness change of two mutations. The epistasis between two mutations, $\DHi$ and $\DHj$, is equivalent to a series of multiple scattering paths \eref{eq:multiple}. The diagram shows the path $\GG\, \DHi \GG\,\DHj \GG$.}
\label{fig:2}
\end{figure}%

Algorithmically, one takes one functional solution obtained
from the evolution algorithm and mutates one AA at a site $i$. This
mutation induces a change $\DGi$ as the difference of the new and the
old Green's function.  Then,
\begin{equ}
\DF_i = \vbT \DGi\, \fb~
\end{equ}
is the change of the observable fitness $F$ (which can be computed by
\eref{eq:Woodbury}).
One can similarly perform another, independent mutation at a site
$j\ne i$, producing a second deviation, $\DGj$ and $\DFj$
respectively. Finally, starting again from the original solution,
one mutates both $i$ and $j$ simultaneously, with combined effects
$\DGij$ and $\DFij$. It is then natural to define the epistasis
$\epij$ as the departure of the double mutation from additivity 
of two single mutations, 
\begin{equ}
\label{eq:epistasis}
\epij \equiv \DFij - \DFi - \DFj~.
\end{equ}
%%%%%%%%%%%%%%%%
\begin{figure*}[t]
%%\centering
\includegraphics[width=0.85\textwidth]{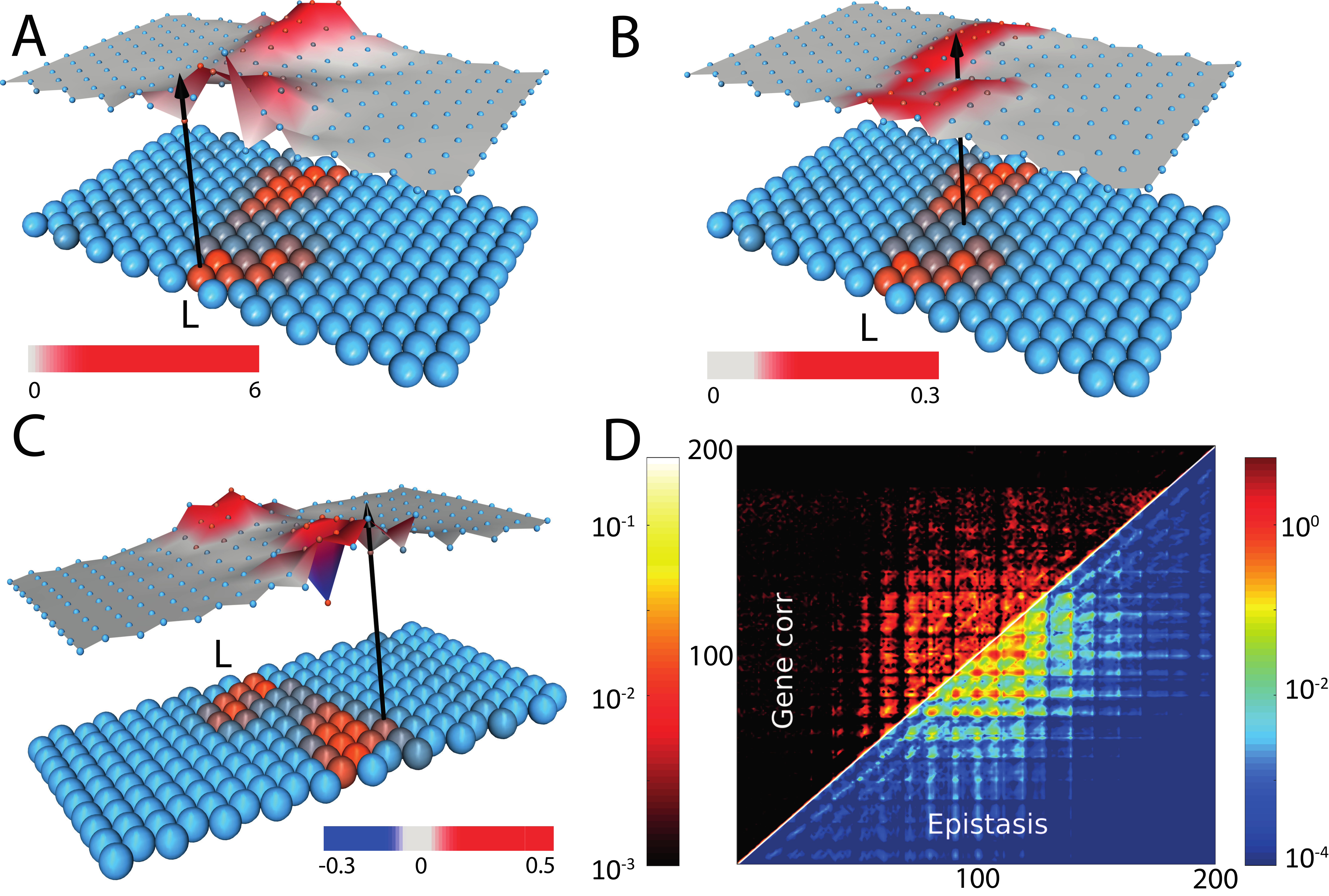}
\caption{\textbf{Mechanical Epistasis.}
The epistasis of \eref{eq:epistasis}, averaged over $ 10^6$ solutions
$\EE_{ij}=\langle \epij \rangle$, between a fixed AA at position $i$
(black arrow) and all other positions $j$. Here, $i$ is located at (A)
the binding site, (B) the center of the channel, and (C) slightly off
the channel. Underneath, the average AA configuration of the protein
is drawn in shades of red ($\aP$) and blue ($\aH$). 
Significant epistasis mostly occurs along the $\aP$-rich channel,
where mechanical interactions are long ranged.   
Though epistasis is predominantly positive, negative values also
occur, mostly at the boundary of the channel (C).
(D) The two-codon correlation function $Q_{ij}$ of
\eref{eq:correlation} measures the coupling between mutations at
positions $i$ and $j$. The epistasis $\EE_{ij}$ and the gene
correlation $Q_{ij}$ show similar patterns. Axes are the positions of
$i$ and $j$ loci. Significant correlations and epistasis occur mostly
in and around the channel region (positions $70{-}130$, rows
$7{-}13$).} 
\label{fig:4}
\end{figure*}%
%%%%%%%%%%%%%%%%%
The epistasis $\epij$ is simply the inner product value of this nonlinearity with the pinch and the response,
\begin{equ}
\label{eq:curvature}
 \epij = \vbT \, \left( \DGij - \DGi - \DGj\right) \, \fb~. 
\end{equ}
\eref{eq:curvature} shows how epistasis is directly related
to mechanical forces among mutated AAs.

To evaluate the average epistatic interaction among amino acids 
in the HP-model, we perform the double mutation calculation 
for all $ 10^6$ solutions and take the ensemble average 
$\EE_{ij}=\langle \epij \rangle$. Landscapes of $\EE_{ij}$ 
show significant epistasis in the channel (\fref{fig:4}). 
AAs outside the high shear region show only small epistasis,
since mutations in the rigid domains hardly change the
elastic response. 
The epistasis landscapes (\fref{fig:4}A-C) are mostly positive 
since the mutations in the channel interact antagonistically
\cite{Desai2007}: after a strongly deleterious mutation, a second
mutation has a smaller effect.  

In the gene, epistatic interactions are manifested in codon
correlations \cite{Hopf2017,Poelwijk2017} shown in \fref{fig:4}D,
which depicts two-codon correlations $Q_{ij}$ of \eref{eq:correlation}
from the alignment of $ 10^6$ functional genes $\cbs$.
We find a tight correspondence between the mean epistasis 
$\EE_{ij}=\langle \epij \rangle$ and the codon correlations $Q_{ij}$. 
Both patterns exhibit strong correlations in the channel region 
with a period equal to channel's length, 10 AAs.
The similarity in the patterns of $Q_{ij}$ and $\EE_{ij}$ indicates
that {\bf a major contribution to the long-range, strong correlations observed among aligned protein sequences stems from the mechanical interactions propagating through the amino acid network}.

\subsection{Epistasis as a sum over scattering paths}
\label{sec:epistasis2}
One can classify epistasis according to the interaction range.
Neighboring AAs exhibit \emph{contact epistasis}
\cite{Goebel1994,Marks2011}, because two adjacent
perturbations, $\DHi$ and $\DHj$, interact nonlinearly via the `and'
gate of the interaction table of \fref{fig:1},  
$\DDHij \equiv \DHij - \DHi - \DHj \neq 0$ 
(where $\DHij$ is the perturbation by both mutations). 
In the case of contact epistasis, the leading term in the Dyson series
\eref{eq:Dyson} of $\DDGij$ is a single scattering from an effective
perturbation with an energy $\DDHij$, which yields the epistasis 
\begin{equation*}
\epij = -\vbT \left( \GG\, \DDHij \GG \right) \fb+\dots ~.
\end{equation*}

\emph{Long-range epistasis} among non-adjacent, non-interacting
perturbations ($\DDHij=0$) is observed along the channel (\fref{fig:4}). 
In this case, \eref{eq:Dyson} expresses the nonlinearity $\DDGij$
as a sum over multiple scattering paths 
which include both $i$ and $j$ (\fref{fig:2}C), 
 \begin{equ}
\epij = \vbT \bigl(
\GG\, \DHi \GG\,\DHj \GG +
\GG\, \DHj \GG\,\DHi \GG \bigr) \fb 
 -\cdots~.
 \label{eq:multiple}
\end{equ}
The perturbation expansion further links long-range epistasis to shear
deformation: Near the transition at which the function emerges, 
Green's function is dominated by the single soft mode, 
$\GG \simeq \ubs\ubsT/\lams$, with fitness $F$ given by
\eref{eq:divergence}. From \eref{eq:Dyson} and \eref{eq:curvature},
one deduces a simple expression for the mechanical epistasis
as a function of the shear,
\begin{equ}
\epij \simeq F \cdot \left(\frac{h_i}{1+h_i}+\frac{h_j}{1+h_j}-\frac{h_i+h_j}{1+h_i+h_j} \right)~.
\label{eq:antagonistic}
\end{equ}
The factor $h_i \equiv \ubsT\DHi\ubs/\lams$ in \eref{eq:antagonistic}
is the ratio of the change in the shear energy due to mutation at $i$ 
(the expectation value of $\DHi$) and the energy $\lams$ 
of the soft mode, and similarly for $h_j$. Thus, $h_i$ and $h_j $ 
are significant only in and around the shear band, where 
the bonds varied by the perturbations are deformed by the soft mode. 

When both sites are outside the channel, $h_i,h_j \ll 1$, 
the epistasis \eref{eq:antagonistic} is small, 
$\epij \simeq 2 h_i h_j F$. It remains negligible even if one of the mutations, $i$, is in the channel, $h_j \ll 1 \ll h_i$, 
and $\epij \simeq h_j F$. 
Epistasis can only be long-ranged along the channel when both
mutations are significant, $h_i\gg 1$ and $ h_j \gg 1$, and 
$\epij \simeq 1$. It follows that \eref{eq:antagonistic} can be roughly approximated as 
\begin{equation}
\epij \simeq F \cdot \min{(1,h_i)} \cdot \min{(1,h_j)}~. 
\label{eq:rank1} 
\end{equation}

We conclude that epistasis is maximal when both sites are at
the start or end of the channel, as illustrated in \fref{fig:4}. 
The nonlinearity of the fitness function gives rise to antagonistic
epistasis since the combined effect of two deleterious mutations is
non-additive as either mutation is enough to diminish the fitness. 

As evident from \eref{eq:rank1}, the epistasis matrix \eref{eq:antagonistic} is approximately a rank-one tensor $\epij \sim \ket{\mathbf{e}}\bra{\mathbf{e}}$,
with a single dominant eigenvector, $\mathbf{e}_i \sim \min{(1,h_i)}$.
The eigenvector $\ket{\mathbf{e}}$ is localized in and around the
shear band. As a result, the epistasis matrix exhibits a `checkered'
pattern visible in \fref{fig:4}D. The rank-one nature of the $\epij$
is verified numerically by spectral decomposition of the epistasis
matrix obtained from the simulation. Interestingly, the genetic
correlation matrix (\eref{eq:correlation}) is also approximately a
rank-one tensor,  
$Q_{ij} \sim \ket{\mathbf{q}}\bra{\mathbf{q}}$, 
with a dominant eigenvector $\ket{\mathbf{q}}$ localized in the channel.
This explains the striking similarity of  the genetic correlation $Q_{ij}$ and the epistasis $\epij$ in \fref{fig:4}D.   

Again, comparing to the real protein glucokinase, 
the rightmost panel of \fref{fig:3gluco} shows that 
the correlation of mutations is concentrated in 
the mechanically critical regions of the protein (left panel). 
Mutations away from these spots seem more independent
and need not be corrected for other mutations.
We conclude: 
\textbf{ mutations correlate near mechanically critical positions}.

\subsection{Multilocus epistasis$^*$}

So far, we examined the interaction between two mutations in terms of
the non-linearity of the double-mutation fitness function  $\epij$
\eref{eq:epistasis}. This two-body interaction can be seen as the
change in the effect of mutation 
$j$ in the presence of another mutation $i$. As an isolated mutation,
$j$ has a fitness effect, $\DFj$, whereas in the presence of $i$ the
effect of $j$ is $\DFij - \DFi$, and the difference defines $\epij
\equiv (\DFij - \DFi )- \DFj~$.

Higher-order epistasis, involving more than two mutations, has a
significant role in shaping the fitness landscape
\cite{Weinreich2013,Poelwijk2017}.    This motivates us to  generalize the
methodology of \sref{sec:epistasis2} to many-body interactions. For
example, the three-loci epistasis, $\epijk$, measures the change in
the two-loci epistasis $\epij$ of the double $i,j$ mutation, induced
by the presence of a third mutation, $k$ \cite{Horovitz1990}: 
\begin{equ}
\label{eq:epistasis3}
 \epijk \equiv \DFijk - \left(\DFij + \DFjk + \DFik \right) + \DFi + \DFj + \DFk~,
\end{equ}
where $\DFijk$ is the phenotypic effect of a triple $i,j,k$ mutation.

In a similar fashion, one derives  the general $N^\mathrm{th}$-order epistasis, among mutations at positions $i_1,\dots,i_N$,
\begin{equ}
\label{eq:epistasisN}
e_{i_1,i_2,\dots, i_N} \equiv \sum_{q=1}^N\left(-1\right)^{N-q} 
\sum_{i_1<\dots <i_q}{\delta\mathbf{F}_{i_1,\dots, i_q}}~,
\end{equ}
 where $\delta\mathbf{F}_{i_1,\dots, i_q}$ is the fitness effect of the $q$-site mutation at positions $i_1,\dots,i_q$. \eref{eq:epistasis} and \eref{eq:epistasis3} are the second- and third-order epistasis terms ($N=2,3$), while the first-order epistasis ($N=1$) is the mutation effect itself, $e_i \equiv \DFi$.
Summing over all orders of epistasis interactions  (\eref{eq:epistasisN}) up to order $N$, one obtains the $N$-site mutation effect 
\begin{equ}
\delta\mathbf{F}_{i_1,i_2,\dots, i_N} = \sum_{q=1}^N
\sum_{i_1<\dots <i_q}{e}_{i_1,\dots, i_q}~.
\end{equ}

To link the multi-locus epistasis to protein mechanics and
deformation, we follow the derivation of \eref{eq:antagonistic}. 
Near the transition at which the function emerges, 
we use the Dyson series \eref{eq:Dyson}, and the  
resulting $N$-site mechanical epistasis is:
\begin{equation}
\label{eq:antagonisticN}
e_{i_1,\dots, i_N} = 
-F \cdot\sum_{q=1}^N{\left(-1\right)^{N-q}  \sum_{i_1<\dots <i_q}
%{\left[ 1 + \left(\sum_{p=1}^q{h_{i_p}}\right)^{-1}\right]{-1}~,}
{ \frac{\sum_{p=1}^q{h_{i_p}}}{1 + \sum_{p=1}^q{h_{i_p}}}}}~,
\end{equation}
where elastic factor $h_{i_p} \equiv \ubsT\delta \mathbf{H}_{i_p}\ubs/\lams$ is the ratio of the change in the shear energy due to mutation at $i_p$ 
and the energy $\lams$ of the soft mode. 

One concludes from \eref{eq:antagonisticN} that the $N$-order epistasis is significant only within and around the shear band, where the bonds are stretched and compressed by the soft mode.
In this region, where all the elastic factors are large. \ie $h_{i_p} \gg 1$,
all orders of epistasis are relevant and are of the same magnitude,   
\begin{equation*}
e_{i_1,i_2,\dots, i_N} \simeq F \cdot \left(-1\right)^N~.
\end{equation*}
We conclude: \textbf{the mechanically critical regions are strongly coupled
with many-body epistatic interactions among the mutations.}

\section{Outlook}
This colloquium has described a method which relates 
biological questions and concepts regarding protein evolution to 
the techniques of theoretical physics. 
Our purpose is to make this approach accessible to a wide community.
While we made an effort to cite some of the current literature, 
there are certainly works which we have incompletely cited. 
We hope that this colloquium will encourage others to build bridges 
between other biological questions and the long
tradition of physics and mathematics.

\begin{acknowledgments}
We thank Albert~Libchaber for inspiring discussions and his 
essential participation in our work on protein.
We thank Sandipan~Dutta who participated in the work on Green's functions. 
We thank Stanislas~Leibler, Michael R.~Mitchell, Elisha~Moses, Giovanni~Zocchi, 
and Olivier~Rivoire for helpful discussions and encouragement.
We are grateful to Karsten~Kruse, Alberto~Morpourgo, Pierre~Collet,
and the referees,
for constructive comments on the manuscript.
JPE was supported by an ERC advanced grant `Bridges', 
and TT by the Institute for Basic Science IBS-R020 
and the Simons Center for Systems Biology of the Institute for Advanced Study, Princeton.
\end{acknowledgments}

%%%% STOPPED

\bibliographystyle{new4}
%\bibliographystyle{apsrmp4-1}
%\small
\bibliography{green}

\end{document}